\documentclass[journal,12pt, onecolumn]{IEEEtran}
\usepackage{lipsum,graphicx,multicol}
\usepackage{xcolor}
\usepackage{color,soul}
\usepackage{mathtools}
\usepackage{amsthm}
\newtheorem{example}{Example}
\usepackage{graphicx}
\usepackage{color,soul}
\usepackage{setspace}
\usepackage{multirow}
\usepackage{tablefootnote}
\usepackage[norelsize, linesnumbered, ruled, lined, boxed, commentsnumbered]{algorithm2e}
\usepackage{amsthm}
\usepackage{amsthm}
\usepackage[normalem]{ulem}
\usepackage{pifont}
\newcommand{\cmark}{\ding{51}} 
\newcommand{\xmark}{\ding{55}} 
\usepackage[font=small,labelfont=bf]{caption}
\usepackage{subcaption}
\usepackage{bm, amsmath, amssymb, amsthm}
\usepackage{amsfonts}
\usepackage {amssymb,amsmath,xcolor}
\usepackage{amsmath}
\usepackage{cite}
\usepackage{array}
\usepackage[acronym]{glossaries}
\usepackage{cases}
\usepackage{float}
\usepackage{amsmath}

\usepackage{adjustbox}
\usepackage[colorlinks = true,
    linkcolor = red,
    urlcolor  = black,
    citecolor = red, 
    anchorcolor = magenta]{hyperref}

\usepackage{cleveref}
\usepackage{mdframed}
\usepackage{xcolor}
\usepackage{comment}
\def\BibTeX{{\rm B\kern-.05em{\sc i\kern-.025em b}\kern-.08em
    T\kern-.1667em\lower.7ex\hbox{E}\kern-.125emX}}  
\newtheorem{theorem}{Theorem}
\newtheorem{Definition}{Definition}
\newtheorem{problem}{Problem}
\newtheorem{corollary}{Corollary}
\newtheorem{proposition}{Proposition}
\newtheorem{lemma}{Lemma}
\newcounter{remark}

\usepackage{xcolor}

\newcommand{\bieee}{\begin{IEEEeqnarray}{rCl}}
\newcommand{\eieee}{\end{IEEEeqnarray}}

\SetKwInput{KwInput}{Input}              
\SetKwInput{KwOutput}{Output}   
 
\begin{document}
\title{Robust Analog Lagrange Coded Computing: Theory and Algorithms via Discrete Fourier Transforms}
\author{\IEEEauthorblockN{Rimpi Borah and J. Harshan}\\
\thanks{This work was presented in part at the IEEE International Symposium on Information
Theory (IEEE ISIT 2024), held at Greece, July 2024 \cite{b4}.}
\IEEEauthorblockA{Department of Electrical Engineering, Indian Institute of Technology Delhi, India}}

\maketitle

\begin{abstract}
Analog Lagrange Coded Computing (ALCC) is a recently proposed computational paradigm wherein certain computations over analog datasets are efficiently performed using distributed worker nodes through floating point representation. While the vanilla version of ALCC is known to preserve the privacy of the datasets from the workers and also achieve resilience against stragglers, it is not robust against Byzantine workers that return erroneous results. Highlighting this vulnerability, we propose a \textcolor{black}{Robust}  ALCC framework that is capable of handling a wide range of integrity threats from the Byzantine workers. As a foundational step, we use error-correction algorithms for Discrete Fourier Transform (DFT) codes to build novel reconstruction strategies for ALCC thereby improving its computational accuracy in the presence of a bounded number of Byzantine workers. Furthermore, capitalizing on some theoretical results on the performance of the DFT decoders, we propose novel strategies for distributing the ALCC computational tasks to the workers, and show that such methods significantly improve the accuracy when the workers’ trust profiles are available at the master server. Finally, we study the robustness of the proposed framework against colluding attacks, and show that interesting attack strategies can be executed by exploiting the inherent precision noise owing to floating point implementation. 
\end{abstract}



\section{Introduction}
\label{sec:introduction}

With the proliferation of data-centric methods for prediction, classification and other analytics, computation over the underlying large-scale dataset necessitates the use of distributed computing \cite{a1}. While a distributed computing framework reaps significant benefits when implemented over a set of trusted servers, it may pose significant challenges when applied over a network with a few untrusted servers. The notion of untrusted servers typically arises either in peer-to-peer networking, or when the underlying networks have poor security features. In such scenarios, distributed computing is vulnerable to (i) privacy issues on the dataset, (ii) distributed denial-of-service (DDoS) attacks, which could stall the overall compute operation, and (iii) integrity attacks, which lead to poor accuracy in the end result \cite{new1,new2}. Although there exists a number of cryptographic approaches to handle privacy, DoS and integrity attacks \cite{new3}, those methods are not feasible for distributed computing owing to high computational loads, and moreover, those methods are not information-theoretically secure as the hardness assumption is based on computational boundedness.

As a promising alternative to the multi-layered cryptographic architecture, distributed coded computing architectures have been studied in untrusted settings, wherein tools from secret sharing schemes and coding-theory have been used to facilitate secure computations in the presence of aforementioned attacks. In particular, such coded computing architectures are known to provide efficient computation in the presence of the so-called \textit{stragglers}\cite{a4,a7,b23}, \textit{Byzantine} workers \cite{a2,a3,a6,a7,a8,a9,a10,b23}, and  \textit{honest-but-curious} workers \cite{a3,a5,a10,a11,a12,b23}. One notable example of such a coded computing paradigm is Lagrange Coded Computing (LCC)\cite{b1}, which is a distributed coded computation framework that provides all the aforementioned features, when the function of interest is an arbitrary multivariate polynomial. Despite these benefits, the major limitation of LCC framework lies in its reliance on finite field dataset. As a result, its computation accuracy significantly degrades because of overflows arising from finite field operations, especially when dealing with large datasets. To address this accuracy issue of LCC, Analog Lagrange Coded Computing (ALCC) has been proposed \cite{b2}, wherein computations are directly performed over analog datasets using distributed worker nodes through floating point implementation. Notably, the accuracy of ALCC remains constant regardless the size of the datasets, and therefore, it outperforms LCC in the large datasets regime.

\begin{figure*}[t]
    \centering
    \begin{subfigure}[b]{0.5\textwidth}
      \centering
 \includegraphics[height=7cm]{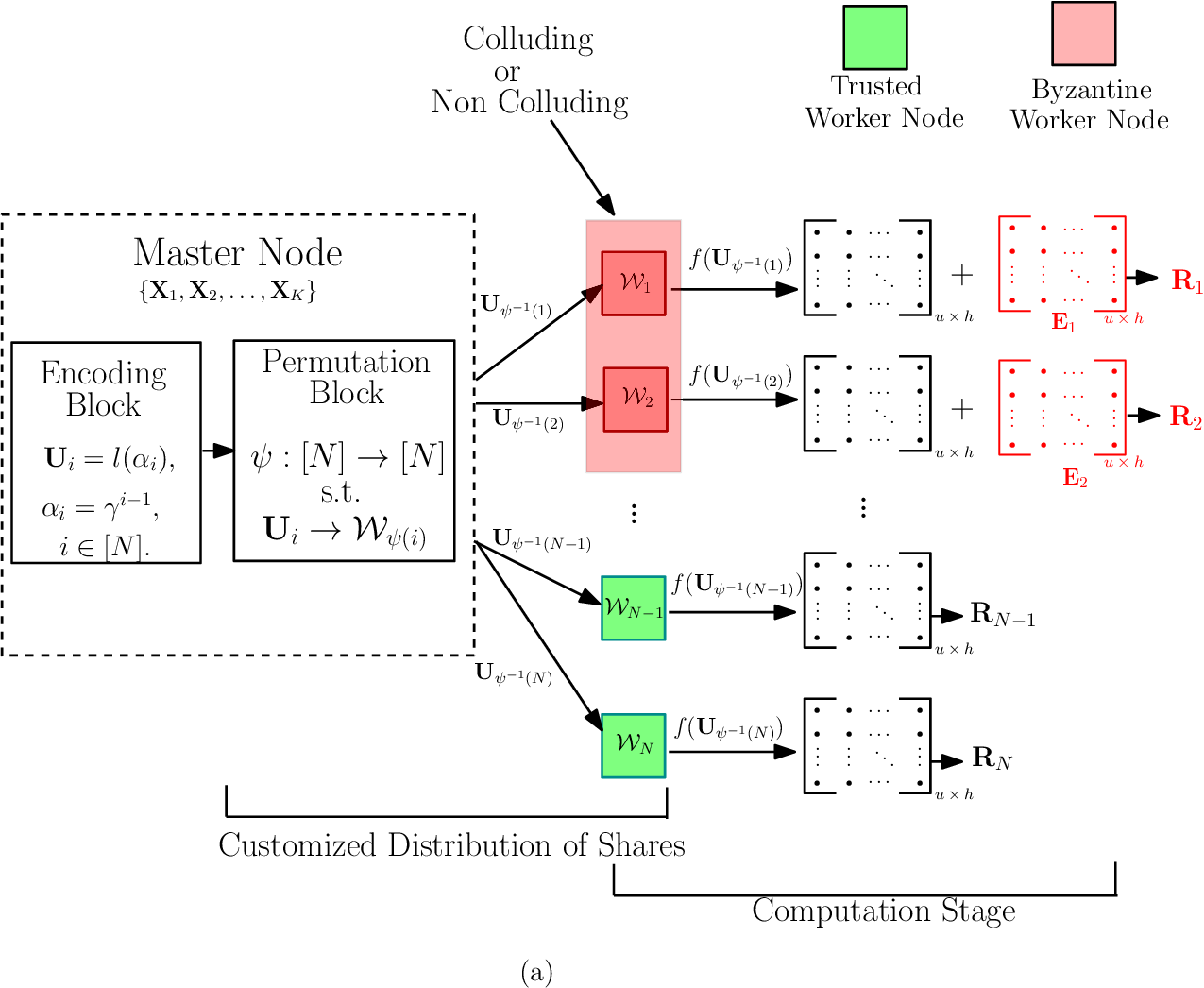}
    \label{fig:1}
    \end{subfigure}
    \begin{subfigure}[b]{0.485\textwidth}
     \centering
  \includegraphics[height=7cm]{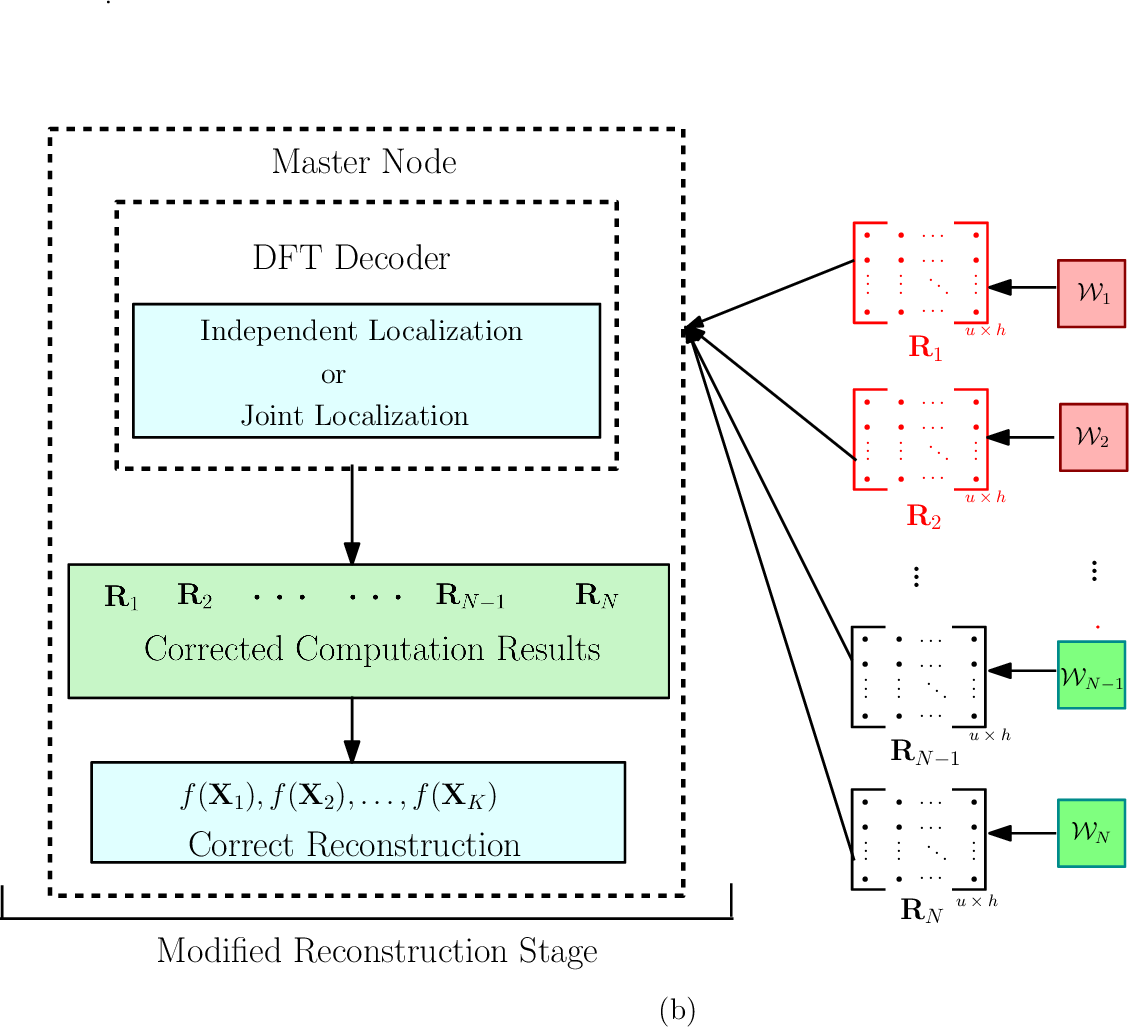}
        \label{fig:2}
    \end{subfigure}
 \caption{\textcolor{black}{Robust  ALCC framework involving Byzantine worker nodes: (a) captures the phases of encoding at the master sever and distributed computation at the workers where Byzantine workers inject noise into their computations. (b) depicts the reconstruction stage at the master server wherein DFT decoders are used to nullify the noise introduced by the Byzantine workers.}} 
 \label{Fig:allc adv}
\end{figure*}

\subsection{Motivation}
\label{subsec:motivation}
Despite the scalability benefits offered by ALCC \cite{b2}, we highlight that it does not provide all the security features offered by LCC. Specifically, while ALCC is known to provide resiliency against stragglers\cite{b2} and preserve the privacy of dataset from the honest-but-curious workers \cite{a11,a12}, it is not resilient against Byzantine workers that return erroneous computation results. In this context, Byzantine workers refer to those worker nodes that intentionally return erroneous computations to the master server, thereby degrading the overall accuracy of ALCC. Identifying this limitation, the authors of \cite{b2} make the following remarks in their conclusions: ``\emph{Another direction is to extend ALCC in order to take into account the presence of Byzantine workers, i.e., the worker nodes that deliberately send erroneous computation results}". To the best of our knowledge, this direction has not been addressed hitherto in a comprehensive manner. Towards bridging this research gap, we investigate the security vulnerabilities of ALCC in the presence of Byzantine worker nodes, and ask a wide range of questions towards building a robust ALCC framework. Specifically, we ask the following questions: (i) Since the ALCC operations are over datasets with floating point representation, can we use coding-theoretic methods at the master server to detect and correct the errors added by the Byzantine workers directly using floating point data? (ii) ALCC typically involves encoding operation of the datasets at the master server, followed by distribution of computation tasks to the worker nodes, and finally the reconstruction stage at the master where the computed results from the workers are combined to get the desired end result. In scenarios where the master server has access to the trust profiles of the workers, can we redesign the encoding and distribution phases of ALCC to improve its accuracy in the presence of Byzantine workers? (iii) Once a robust ALCC framework is designed, based on Kerckhoffs's principle \cite{Kerchoffs} in cryptography, it is typical to expect the Byzantine worker nodes to have the knowledge of the reconstruction mechanism employed at the master node. Under such scenarios, can the floating point operations in ALCC be exploited by the Byzantine worker nodes to design novel colluding attacks?

\subsection{Contributions}
\label{subsec:contribution}

Although erroneous computations at a Byzantine worker node could be due to non-adversarial events such as bugs in software, in this work, we assume that Byzantine nodes are those nodes that have been compromised by external adversarial agents due to poor security features \cite{ransomware}. \textcolor{black}{To capture a more general ALCC setting, unlike \cite{b2}, we assume that trust profiles of the workers are available at the master \cite{axx1,axx2,axx3}. Using these trust profiles, we first categorize the workers into two groups: (i) \textit{reliable} worker nodes, which are
highly trustworthy and do not corrupt their computations, and
(ii) \textit{unreliable} worker nodes, which may corrupt their
computations at any point in time.} Subsequently, we use this trust model to develop the following contributions, which are applicable to ALCC with an arbitrary number of reliable and unreliable worker nodes: 

1) Our preliminary investigation reveals that the vanilla ALCC framework suffers from poor accuracy in the presence of Byzantine worker nodes (see Section \ref{subsec: ALCC reconstruction}). Towards addressing this issue, we identify that, in the absence of Byzantine worker nodes, the computations returned by the worker nodes in the ALCC framework are essentially a set of codewords of a Discrete Fourier Transform (DFT) code \cite{b3,b5,b6,b9,b10,b8} (see Section \ref{sec:secure ALLC}). As a consequence, if Byzantine worker nodes return noisy computations, the introduced errors can be corrected by applying suitable error-correction algorithms for the DFT code before the reconstruction stage of ALCC. Thus, with the use of a DFT decoder, we show that the accuracy of ALCC can be improved in the presence of Byzantine workers (see Section \ref{sec:secure ALLC}).

2) \textcolor{black}{While the DFT decoder improves the accuracy of ALCC in the presence of Byzantine workers, we observe that its accuracy is still affected by precision noise arising from floating-point operations. To alleviate this effect, we first identify that the master server in ALCC receives multiple noisy DFT codewords whose error locations have a common support, thereby motivating the development of new algorithms for error localization. Therefore, we study the localization block of the DFT decoder in the presence of precision noise and derive bounds on the error rate in localizing the errors as a function of various ALCC parameters, including the variance of the precision noise (see Section \ref{subsec:independent localization}).}


3) When employing the standard error-localization algorithms of the DFT code in ALCC, we discover that the algebraic structure of the computations returned by the worker nodes requires the master node to determine the roots of multiple error-locator polynomials, which may have a common set of solutions. Identifying this ALCC-specific structure, we use the derived bounds on the localization error to propose a novel joint-localization algorithm for identifying the locations of Byzantine worker nodes. Subsequently, we use these locations to nullify the erroneous computations, thereby enhancing the accuracy of ALCC. Through experimental results, we show that ALCC with the proposed joint-localization method provides significant accuracy benefits over standard error-localization algorithms in the presence of Byzantine worker nodes. We show that the benefits are remarkable when the number of Byzantine workers equals the error-correction capability of the DFT code (see Section \ref{subsec:joint loc}).

 
\setlength{\tabcolsep}{4pt}
\renewcommand{\arraystretch}{1.1}
\begin{table}[ht]
\centering
 \begin{adjustbox}{max width=0.9\textwidth}
\begin{tabular}{|c|l|c|c|c|c|}
\hline
\textbf{SI} & \textbf{Salient features} & \textbf{LCC} & \textbf{ALCC} & \textbf{DFT decoders} & \textbf{Our work} \\
& & \cite{b1} & \cite{b2} & \cite{b3,b5,b6,b9,b10,b8} & \\
\hline
1 & Straggler resilience & \cmark & \cmark & - & \cmark \\
2 & Security against Byzantine workers & \cmark & \xmark & - & \cmark \\
3 & Privacy against curious workers & \cmark & \cmark & -  & \cmark \\
4 & Analog domain support & \xmark & \cmark & -  & \cmark \\
5 & Error detection and correction capability & \cmark & \xmark & -  & \cmark \\
6 & Customized distribution of shares to workers & \xmark & \xmark & -  & \cmark \\
7 & Design of colluding attack strategies & \xmark & \xmark & -  & \cmark \\
8 & Bounds on the performance analysis of error & - & - & \xmark & \cmark \\
& localization of DFT decoder & & & & \\
\hline
\end{tabular}
\end{adjustbox}
\caption{\label{tab:novelty-table} Novelty of our work with respect to various features.}
\end{table}

4) For the ALCC setting wherein the identities of the unreliable and reliable worker nodes are known to the master server, we identify that the error-localization block of the DFT decoder can exploit this knowledge to improve the accuracy of error localization. In this setting, we use the existing bounds on the localization error to derive special cases, and then use them to design a customized strategy for distributing the encoded shares among unreliable worker nodes. Through experimental results, we show that our strategy provides a low-complexity and tractable method for parameter optimization, and performs close to other exhaustive-search-based methods, which are impractical to implement. We also show that our strategy significantly improves the accuracy of ALCC compared to contiguous share assignment, where no reliability information is exploited at the master (see Section \ref{sec:optimal allocation}).

5) With the support of derived theoretical bounds on the error rate of localization and its behavior with the variance of precision noise, we study the vulnerabilities of ALCC framework in the presence of \textit{colluding} Byzantine worker nodes. In particular, we propose novel attack strategies from the perspective Byzantine workers, aiming at maximizing the error rate of localization, thereby degrading the accuracy of ALCC.  These attack strategies are designed based on the sparsity of the noise matrices introduced by the Byzantine worker nodes and the communication overhead constraint imposed by the colluding channels among them. As the main takeaway, we prove a counter-intuitive result that not all the adversaries should inject noise in their computations in order to optimally degrade the accuracy of the ALCC framework. This is the first work of its kind to address the vulnerability of ALCC against colluding adversaries (see Section \ref{sec:colluding_attacks}).

\subsection{Novelty}
\label{subsec:related work}
As motivated in Section \ref{subsec:motivation}, our contributions are along the lines of LCC \cite{b1} and ALCC \cite{b2}, wherein we have resolved some of the open problems in ALCC by providing resiliency against Byzantine worker nodes \cite{a2,a3,a6,a7,a8,a9,a10,b23} under certain conditions. In particular, we have identified the utility of the DFT decoders \cite{b3,b5,b6,b9,b10,b8} in ALCC, and have proposed novel encoding and decoding methods that are customized to ALCC. Some of the salient features offered by our work in comparison with the existing literature are tabulated in Table \ref{tab:novelty-table}. \textcolor{black}{As shown in Table \ref{tab:novelty-table}, the relevant literature pertaining to our contributions can be broadly classified into two categories: (i) contributions in coded computing and (ii) established results on DFT codes.} With respect to coded computing, several works \cite{f1,f2,f3,f4,f5,f6,f7,f8,f9,f10} have addressed aspects such as numerical stability, privacy, and straggler resilience, both with and without robustness features. However, none of these works addresses the problem of robustness of ALCC against Byzantine worker nodes, which was posed as an open problem in \cite{b2}. Some preliminary results on the problems addressed in this work are available in conference proceedings \cite{b31,b4}, however, these works do not provide analytical guarantees or rigorous proofs for the underlying approaches. 

\textcolor{black}{The established DFT-code based decoders in \cite{b3,b5,b6,b9,b10,b8} address a fundamentally different decoding setting. In particular, they consider the recovery of a single DFT codeword corrupted by errors and present decoding procedures under this model. In contrast, ALCC produces a collection of noisy DFT codewords, all of which are corrupted by the same set of Byzantine workers. Therefore, a key distinction from the established DFT-code literature is that ALCC requires characterizing the reliability of the localization stage under this structured multi-codeword setting. To this end, we derive analytical bounds on the probability of error localization, which, to the best of our knowledge, have not been established in the existing DFT-code literature. These bounds serve as a foundational result for developing several ALCC-specific strategies, including joint error localization across multiple codewords, trust-aware distribution of encoded shares to worker nodes, and collusion-aware decoding strategies. These extensions do not arise from a direct application of existing DFT-code decoding results. The main distinctions between the established DFT-code framework and the proposed ALCC framework are summarized in Table \ref{tab:dft_alcc_contributions}.}

\begin{table}[t]
\centering
\caption{\textcolor{black}{Established DFT-code results\cite{b3,b5,b6,b9,b10,b8} and new contributions in the ALCC setting.}}
\label{tab:dft_alcc_contributions}
\renewcommand{\arraystretch}{1.2}
\begin{small}
\begin{tabular}{|p{3.0cm}|p{6.3 cm}|p{8.0cm}|}
\hline
\textbf{Aspect} &
\textbf{Established DFT-code results} &
\textbf{New in the ALCC setting} \\
\hline

DFT-code structure &
DFT codes and their coding-theoretic properties are established. &
We establish that ALCC worker responses form DFT codewords. \\
\hline

Error correction &
Minimum-distance-based error-correction capability is established. &
We exploit this capability for Byzantine-resilient ALCC. \\
\hline

Multiple codewords &
Conventional decoding considers an individual codeword. &
ALCC naturally produces multiple DFT codewords corresponding to different computation vectors. \\
\hline

Common Byzantine support &
Not part of conventional single-codeword decoding. &
Multiple ALCC codewords share the same Byzantine support. \\
\hline

Finite-precision localization &
Existing DFT algorithms provide localization methods. They do not provide bounds on the rate of localization error. &
We characterize localization error analytically in the ALCC setting. \\
\hline

Joint localization &
No ALCC common-support structure is present in conventional DFT decoding. &
We develop joint localization across multiple codewords by exploiting the common Byzantine support. \\
\hline

Share assignment &
Evaluation points are fixed by the code construction in conventional DFT decoding. &
We treat the assignment of encoded shares/evaluation points to workers as a design parameter and use the localization analysis to develop reliability-aware share-assignment strategies. \\
\hline

Byzantine attacks &
Not considered &
We analyze colluding Byzantine attacks exploiting the localization behavior. \\
\hline

\end{tabular}
\end{small}
\end{table}

\section{ALCC with Byzantine worker nodes}
\label{sec:allc with adv}
The ALCC-based distributed computing setup, as illustrated in Fig. \ref{Fig:allc adv}(a) comprises a master node connected to $N$ worker nodes, denoted by the set $\mathcal{W}=\{\mathcal{W}_{1}, \mathcal{W}_{2}, \ldots, \mathcal{W}_{N}\}$, via dedicated links.  \textcolor{black}{We assume that the master has access to the trust profile of the worker nodes, obtained through an external worker-profiling mechanism. Such a profiling mechanism could be based on historical execution behavior, reliability statistics, or reputation scores, as commonly considered in the distributed systems literature~\cite{axx1,axx2,axx3}. Based on this information, the worker nodes are partitioned into two groups, namely: i) \textit{reliable} worker nodes, denoted by the set $\mathcal{W}_{\rm rel}$, and ii) \textit{unreliable} worker nodes, denoted by the set $\mathcal{W}_{\rm unrel}$. We let $\tau$ and $\mu$ denote the number of reliable and unreliable worker nodes, respectively, such that $\tau+\mu=N$. When such worker reliability information is unavailable, all workers are treated as potentially unreliable, i.e., $\mathcal{W}_{\rm unrel}=\mathcal{W}$, $\mu=N$, and $\tau=0$.} In this context, reliable worker nodes are those that are proven to be honest and accurate in providing the intended computation results to the master node. In contrast, unreliable worker nodes are those that have not been proven honest, i.e., they may be curious about learning the dataset at the master node potentially compromising its privacy, and/or these nodes may return incorrect computation results to the master node. 




Within this framework, the primary objective of the master node is to distribute a computational task $f(\cdot)$ on an underlying dataset among the nodes in $\mathcal{W}$, and subsequently aggregate their computed results, leveraging their collective computing power. For distributed computation, let $\mathbf{X}=(\mathbf{X}_1,\ldots,\mathbf{X}_k)$ be the dataset held by the master node, where each $\mathbf{X}_j \in \mathbb{R}^{m\times n}$ for $j\in[k]$ such that $[k] \triangleq \{1,2,\ldots,k\}$. The master node aims to evaluate a polynomial $f:\mathbb{R}^{m\times n}$ $\rightarrow$ $\mathbb{R}^{u\times h}$ over the dataset $\mathbf{X}$ in a decentralized fashion under the assumption that $f(\cdot)$ is known to all the workers nodes. More specifically, the function $f(\cdot)$ is a $D$-degree polynomial wherein all the entries of the output matrix are multivariate polynomial functions of the entries of the input with a total degree not exceeding $D$. For instance, if $\mathbf{X}_{j} \in \mathbb{R}^{m\times n}$ is such that $x_{\bar{m}\bar{n}}$ is the $(\bar{m},\bar{n})$-th entry of $\mathbf{X}_{j}$, for $\bar{m} \in [m]$ and $\bar{n} \in [n]$, then $\mathbf{Y}_{j} = f(\mathbf{X}_{j})$ refers to the output such that $y_{\bar{u}\bar{h}}$, which is the $(\bar{u},\bar{h})$-th entry of $\mathbf{Y}_{j}$, for $\bar{u} \in [u]$, $\bar{h} \in [h]$ is of the form $y_{\bar{u}\bar{h}} =f_{\bar{u}\bar{h}}(x_{11},x_{12}, \ldots, x_{mn}),$ where $f_{\bar{u}\bar{h}}$ is a multivariate polynomial of degree $D$.

\textcolor{black}{With the above dataset, the master node intends to execute distributed computation of $f(\cdot)$ while ensuring resiliency against $s$ number of straggler nodes in $\mathcal{W}$, and providing privacy from at most $t$ curious worker nodes that intend to collude in order to learn the underlying dataset at the master node. However, in addition to the $s$ stragglers and $t$ curious worker nodes, in this work, we assume the presence of $A$ Byzantine worker nodes who send erroneous computation results to the master node. We assume that the set of $A$ Byzantine worker nodes and the set of $t$ curious worker nodes are subsets of $\mathcal{W}_{unrel}$. In contrast, $s$ stragglers can be from $\mathcal{W}$ irrespective of whether they are reliable or unreliable.} 

In the following subsections, we revisit ALCC, and apply it on our system model, which comprises straggler workers, curious workers and Byzantine worker nodes. To keep our focus on Byzantine worker nodes, we assume the absence of stragglers, i.e., $s = 0$, which implies that all the $N$ worker nodes return their computation results to the master node.\footnote{For exposition, we have used $s = 0$, nonetheless, all the results of this work can be extended to the scenarios involving $s$ stragglers.}  We first outline the encoding strategy of vanilla ALCC followed by the computation and reconstruction stages for obtaining $\mathbf{Y}_{j}=f(\mathbf{X}_{j})$, for $j \in [k]$. Although the encoding strategy is the same as that in \cite{b2}, we modify the computation stage to take into account the presence of Byzantine worker nodes. 



\subsection{Encoding in ALCC}
\label{subsec:Encoding ALCC}

To distribute the computational tasks among the workers, the master node creates $\mathbf{Z}=(\mathbf{X}_1,\mathbf{X}_2,\ldots,\mathbf{X}_k,\\ \mathbf{N}_1,\mathbf{N}_2,\ldots,\mathbf{N}_t)$ where $\{\mathbf{N}_1, \mathbf{N}_2, \ldots, \mathbf{N}_t\}$ are $m \times n$ random matrices with i.i.d. entries drawn from a zero-mean circular symmetric complex Gaussian distribution with standard deviation $\frac{\sigma}{\sqrt{t}}$. Here, the $t$ random matrices are chosen to provide privacy on the dataset against a maximum of $t$ curious worker nodes that may collude to recover the dataset.\footnote{\textcolor{black}{Adding these noise matrices is well known to ensure privacy from the viewpoint of mutual information security (MIS), as long as the parameters $t$ and $\sigma$ are appropriately chosen. Since these privacy guarantees are well studied, we will not delve into them further. However, for the sake of completeness and baseline comparison, we continue to use these parameters in our experiments henceforth.}} Let $\gamma = e^{-\frac{2\pi \iota}{N}}$ and $\omega = e^{-\frac{2\pi \iota}{k+t}}$ be the $N$-th and $(k+t)$-th roots of unity, respectively, with $\iota^2 =-1$. With the above ingredients, Lagrange polynomial is used at the master node to construct the encoded dataset as 
\begin{equation}
\label{eq:u(z)}
l(z)=\sum_{r=1}^{k} \mathbf{X}_r l_r(z) + \sum_{r=k+1}^{k+t} \mathbf{N}_{r-k}l_{r}(z),
\end{equation}
where $l_r(.)$'s are Lagrange monomials defined as
\begin{equation*}
l_r(z)=\prod_{ l\in[k+t] \backslash r} \frac{z-\beta_l}{\beta_r-\beta_l},
\end{equation*}
for all $r\in[k+t]$. Note that $\beta_r$'s are picked to be equally spaced on a circle of radius $\beta$ centered around 0 in the complex plane such that $\beta_r =\beta \omega^{r-1}$ for $\beta \in \mathbb{R}^{+}$.

\subsection{Distribution of Shares among the Worker Nodes}
\label{subsec:distribution of share}

In this step, the master node evaluates $l(z)$ over the $N$-th roots of unity in the complex plane to obtain $\{\mathbf{U}_{i}=l(\alpha_{i})~|~\alpha_i=\gamma^{i-1},i\in[N]\}$. Further, to distribute the shares of the encoded dataset $l(z)$ among the worker nodes, we define a one-to-one mapping $\psi:[N]\rightarrow [N]$, such that the $i$-th evaluation $\mathbf{U}_{i}$ goes to the worker node $\mathcal{W}_{\psi(i)}$, for $i \in [N]$. Although any of the $N!$ mappings, \textcolor{black}{where $N! = N \times (N-1) \times \cdots \times 2 \times 1$,} can be used to distribute the shares, ALCC in \cite{b2} uses the identity mapping, where the $i$-th evaluation $\mathbf{U}{i}$ is assigned to the worker node $\mathcal{W}{i}$.


\subsection{Computation at the Workers}
\label{subsec: computation at workers}
As per ALCC in\cite{b2}, upon receiving $\mathbf{U}_i$ from the master node, $\mathcal{W}_{i}$ computes $f(\mathbf{U}_{i}) \in \mathbb{R}^{u \times h}$ and returns the result to the master node. However, with Byzantine worker nodes, we assume that some nodes return a corrupted version of their computations to the master node, as illustrated in Fig. \ref{Fig:allc adv}(a). Formally, if $i_{1}, i_{2}, \ldots, i_{A}$ represent the indices of the Byzantine worker nodes, a worker node $\mathcal{W}_{i_{a}}$, is supposed to return $f(\mathbf{U}_{i_{a}})$. However, because of its adversarial nature, the corrupted version of $f(\mathbf{U}_{i_{a}})$ can be modelled as $f(\mathbf{U}_{i_{a}})+\mathbf{E}_{i_{a}} \in \mathbb{R}^{u \times h}$ where, $\mathbf{E}_{i_{a}} \in \mathbb{R}^{u \times h}$ is the noise matrix chosen to corrupt the computation. Specifically, since all the computations are performed in floating-point representation, the computations received at the master node are

\begin{equation}
\label{eq:adv and p_noise1}
\mathbf{R}_{i} = f(\mathbf{U}_i) + \mathbf{E}_{i} + \mathbf{P}_{i} \in \mathbb{R}^{u \times h}, 
\end{equation}
\noindent where $\mathbf{E}_{i} = \mathbf{0}$ if $i \notin \{i_{1}, i_{2}, \ldots, i_{A}\},$ and $\mathbf{E}_{i} \neq \mathbf{0}$ if $i \in \{i_{1}, i_{2}, \ldots, i_{A}\},$ such that $\mathbf{P}_{i} \in \mathbb{R}^{u \times h}$ captures the precision errors due to the floating point operations at the worker nodes. The terms ${\mathbf{P}{i}}$ are independent across workers, and we model the entries of ${\mathbf{P}_{i}}$ i.i.d. as $\mathcal{CN}(0,\sigma_q^2)$ random variables. Henceforth, throughout the paper, we refer to $\sigma_q^2$ as the precision-noise variance.
\subsection{Function Reconstruction in ALCC}
\label{subsec: ALCC reconstruction}
Ideally, in the absence of unreliable worker nodes and precision errors, i.e.,  when $\mathbf{E}_{i}=\mathbf{0}$ and $\mathbf{P}_{i} = \mathbf{0}$, upon receiving the set of computation results $\mathbf{R}_i$ for $i\in[N]$ specified in \eqref{eq:adv and p_noise1} from $N$ workers, the master node interpolates the polynomial $f(l(z))$ by using the results returned from at least $(k+t-1)D+1$ workers. Note that, $(k+t-1)D+1$ is the minimum number of returned evaluations needed to ensure a successful interpolation of $f(l(z))$ which results from the fact that the effective degree of $f(l(z))$ is $(k+t-1)D$, where $D$ indicate the degree of the target function $f(\cdot)$ and $k+t-1$ represents the degree of the encoding polynomial $l(z)$ indicated in \eqref{eq:u(z)}. Therefore, if $\tilde D$ is used to represent  $(k+t-1)D$, indicating the effective degree of the polynomial $f(l(z))$, then the number of worker nodes required for successful interpolation of $f(l(z))$ is $K = \tilde D +1$. Finally, to recover $f(\mathbf{X}_{j})$’s, the master node computes $f(l(\beta_r))$ for $r\in[k]$. Note that these steps provide accurate reconstruction of $f(\mathbf{X}_{j})$’s when $\mathbf{E}_{i}=\mathbf{0}$ and $\mathbf{P}_{i} = \mathbf{0}$. However, with precision noise and the noise added by the Byzantine workers, subsequent steps of evaluation and interpolation gets affected. Consequently, the recovered computation is noisy, degrading the overall accuracy. 

To capture the above mentioned errors, let $\mathbf{Y'}$ represent an estmate of $f(\mathbf{X})$ using ALCC. Similarly, let $\mathbf{Y} = f(\mathbf{X})$ represent the centralized computation performed locally at the master node without using ALCC. As a result, the relative error introduced by ALCC with respect to the centralized computation at the master node is computed as,
\begin{equation}
\label{eq:rel_error}
   e_{rel} \triangleq \frac{||\mathbf{Y}-\mathbf{Y'}||}{||\mathbf{Y}||},\
   \end{equation}
where $||.||$ denotes the $l^2$-norm. 

\begin{figure}[ht!]
    \centering
    \begin{subfigure}[b]{0.49\linewidth}
        \centering
        \includegraphics[width=\linewidth]{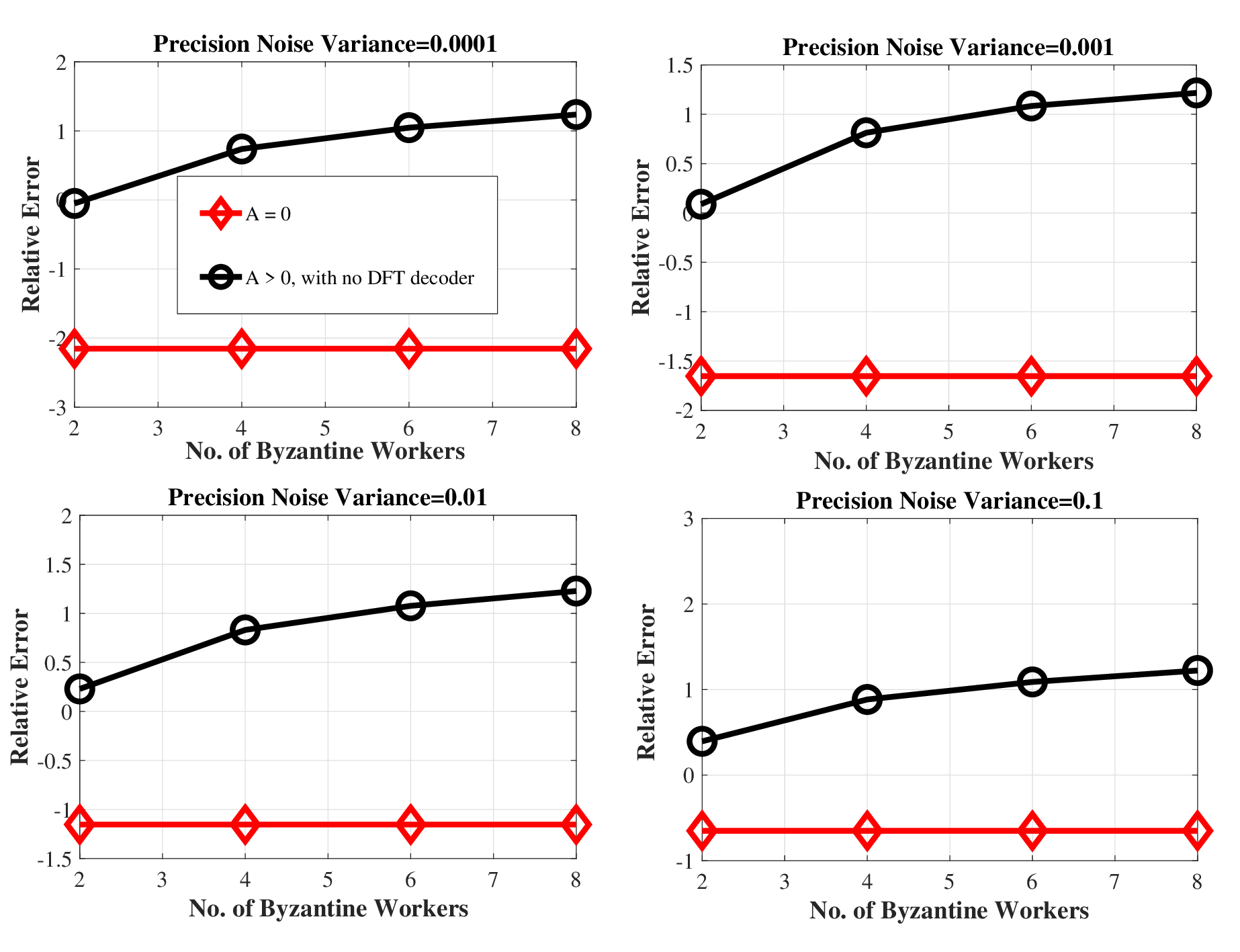}
        \caption{}
    \end{subfigure}
    \hfill
    \begin{subfigure}[b]{0.49\linewidth}
        \centering
        \includegraphics[width=\linewidth]{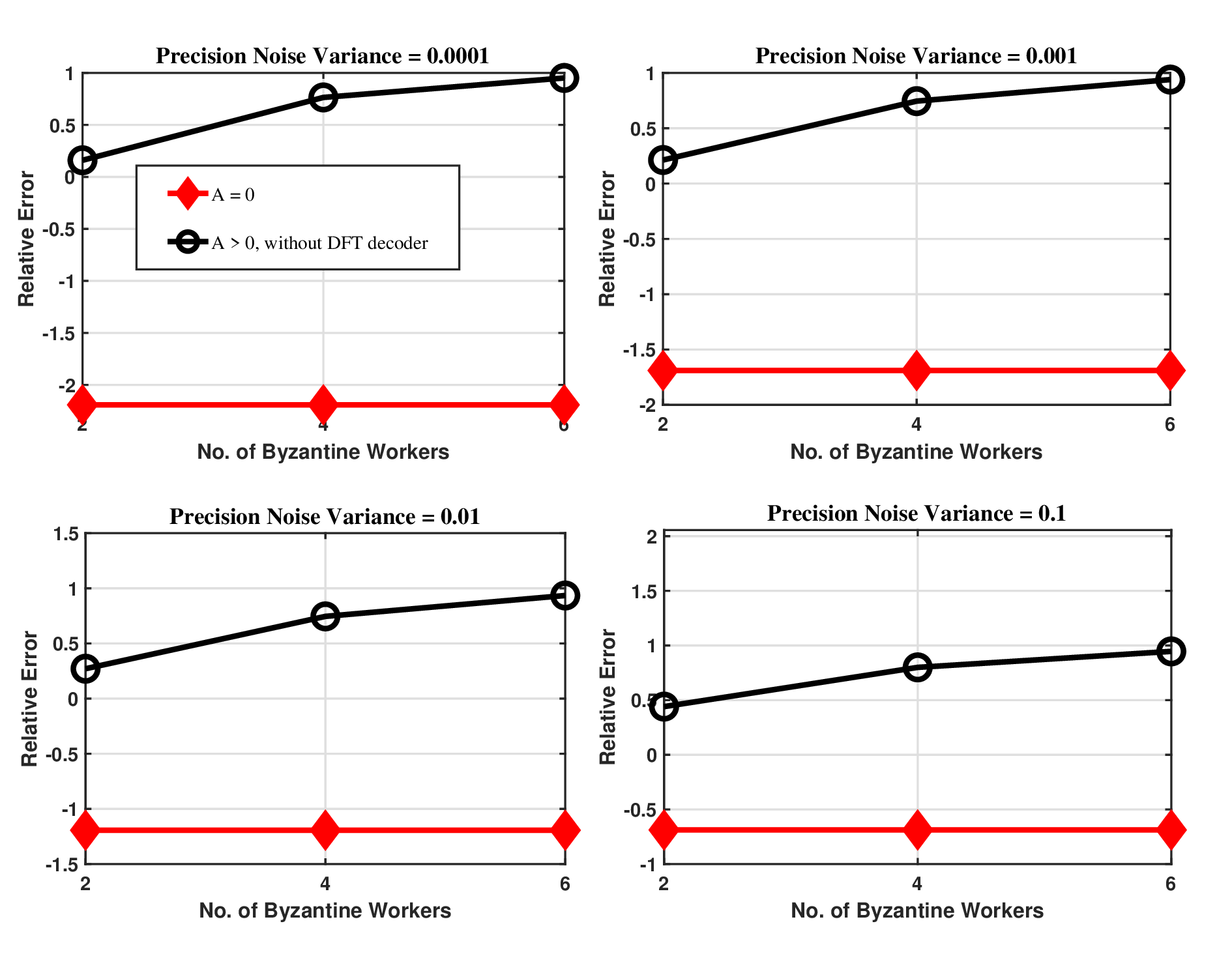}
        \caption{}
    \end{subfigure}

\caption{\textcolor{black}{Average relative error (in $\log_{10}$ scale) of ALCC in the absence of Byzantine workers and in the presence of Byzantine workers without error detection and correction, for the settings $(N,K)=(31,15)$ in (a) and $(N,K)=(27,15)$ in (b), with $k=5$, $t=3$, $\tau=0$, $\beta=1.5$, and $\sigma=10^6$. The non-zero entries of the noise matrices ${\mathbf{E}_{i_a}}$ introduced by the Byzantine worker nodes are i.i.d. as $\mathcal{CN}(10,10^3)$.}}
\label{Fig: alcc in adv plot}
\end{figure}

To demonstrate the impact of Byzantine worker nodes and precision errors on the accuracy of ALCC, we conduct experiments in the presence of Byzantine worker nodes to compute $f(\mathbf{X})=\mathbf{X}^{T}\mathbf{X}$, such that $\mathbf{X}\in \mathbb{R}^{20\times5}$. \textcolor{black}{Specifically, we evaluate the average relative error of ALCC using \eqref{eq:rel_error} for the settings $(N,K)=(31,15)$ and $(27,15)$ at different precision noise variance, while varying the number of Byzantine worker nodes.}  The parameters used for these experiments are
$k=5$, $t=3$, $\beta=1.5$, $\sigma=10^6$ , where $\tau=0$, and computations from $K=15$ worker nodes
are used for reconstruction. These results are presented in Fig.~\ref{Fig: alcc in adv plot}(a) and Fig.~\ref{Fig: alcc in adv plot}(b), corresponding to the $(31,15)$ and $(27,15)$ settings, respectively. The results highlight that the relative error of ALCC increases significantly in the presence of Byzantine worker nodes, thereby confirming that the ALCC framework in \cite{b2} is not resilient against the presence of Byzantine workers. Owing to this limitation, in the next section, we propose a robust ALCC framework that is resilient against the presence of Byzantine worker nodes under certain conditions. \textcolor{black}{We remark that the parameters $\beta=1.5$, $t=3$, and $\sigma=10^6$ used in the above settings are retained from the experimental setup of the original ALCC framework~\cite{b2} to provide a consistent baseline for comparison.}

\section{\textcolor{black}{Robust} ALCC Framework}
\label{sec:secure ALLC}
\textcolor{black}{In this section, we consider a variant of ALCC wherein the trust profiles of the workers are not available. Therefore, as discussed in Section \ref{sec:secure ALLC}. we assume that all the workers are unreliable i.e., $\tau=0$, such that $\mathcal{W}=\mathcal{W}_{unrel}$}.\footnote{The case of $\tau  = 0$ has been considered only to introduce the proposed ideas. However, variant of ALCC with $\tau \neq 0$ will be addressed in Section \ref{sec:optimal allocation}.} Furthermore, we assume the presence of $A$ Byzantine worker nodes, where $0<A<N$. \textcolor{black}{Since the trust profiles of the workers are unavailable, we follow the one-to-one mapping $\psi:[N]\rightarrow[N]$ for encoding the dataset and distributing its shares among the worker nodes. In particular, the $i$-th evaluation $\mathbf{U}_{i}$ is assigned to worker node $\mathcal{W}_{\psi(i)}$, for $i\in[N]$, which is consistent with the original ALCC framework~\cite{b2}}. However, in order to provide robustness to the framework against Byzantine worker nodes, we propose a modification to the reconstruction stage. Recall the set of computation results returned by the worker nodes, denoted by, $\{\mathbf{R}_{i} \in \mathbb{R}^{u \times h} ~|~ i \in [N]\}$, where $\mathbf{R}_{i}$ is specified in \eqref{eq:adv and p_noise1}. Based on the structure of $\mathbf{R}_{i}$, the following proposition establishes the foundational result of this work. 


\begin{proposition}
\label{prop:1}
For an ALCC setting with parameters N and $K = (k+t-1)D + 1$, the erroneous computations returned by the $A$ Byzantine worker nodes can be nullified as long as $A \leq v \triangleq \lfloor \frac{N-K}{2}\rfloor$ and the floating-point operations of ALCC have infinite precision.  
\end{proposition}

\begin{IEEEproof}
     We refer the readers to the proof given in Appendix \ref{proof:1}.\\
\end{IEEEproof}

For DFT codes, both coding-theoretic and subspace-based error-correction algorithms are well known \cite{b3},\cite{b5},\cite{b6}, \cite{b9}, \cite{b10}. In general, their implementation involves the following steps: (i) computing the syndrome vector, (ii) estimating the number of errors, (iii) identifying the location of errors, and (iv) estimating the error values. Consequently, in the presence of Byzantine worker nodes, these algorithms can be independently applied on each noisy DFT codeword $\mathbf{r}_{\bar{u}\bar{h}}$ to recover an estimate of $f(\mathbf{U}_{i})$ for $i \in [N]$. Subsequently, the reconstruction method discussed in Section \ref{subsec: ALCC reconstruction} can be used to recover $f(\mathbf{X}_{j})$, for $j\in[k]$, as shown in Fig. \ref{Fig:allc adv}(b). In the next subsection, we present the typical steps of the DFT decoder \cite{b3},\cite{b5},\cite{b6} when $\mathbf{P}_{i}=\mathbf{0}$ for $i\in [N]$.

\subsection{DFT Based Error Correction Algorithm for ALCC}
\label{subsec:basics of DFT}
Given the set of $M=u\times h$ noisy codewords at the master node as captured by $\mathbf{R}_{eff}$, the first step is to compute the syndrome vector for each noisy codeword. Specifically, for 
 $\mathbf{r}_{\bar{u}\bar{h}}$, the syndrome vector is computed as
\begin{equation}
\label{eq:lin eq}
\mathbf{{s}}_{\bar{u}\bar{h}}=\mathbf{r}_{\bar{u}\bar{h}} \mathbf{H}^{\dagger}=\mathbf{e}_{\bar{u}\bar{h}}\mathbf{H}^{\dagger},
\end{equation}

\noindent where $\mathbf{{s}}_{\bar{u}\bar{h}}$ represents a row vector of length $N-K$. Here, $\dagger$ denotes the Hermitian operator, and $\mathbf{H}$ denotes the parity check matrix of the $(N,K)$ DFT code such that $\mathbf{G}\mathbf{H}^{\dagger}=\mathbf{0}$, where $\mathbf{G}$ is the generator matrix. The vector $\mathbf{e}_{\bar{u}\bar{h}}$ represents the error vector of length $N$ corresponding to the noisy codeword $\mathbf{r}_{\bar{u}\bar{h}}$ with Hamming weight $A$, where $\mathbf{e}_{\bar{u}\bar{h}} =[\mathbf{E}_{1}(\bar{u}, \bar{h}) ~\mathbf{E}_{2}(\bar{u}, \bar{h}) ~\ldots~ \mathbf{E}_{N}(\bar{u},\bar{h})]$. More specifically, the vector $\mathbf{e}_{\bar{u}\bar{h}}$ denotes the error values introduced by the $A$ Byzantine worker nodes at the $(\bar{u},\bar{h})$ entry of each noise matrix $\mathbf{E}_{i}$. The DFT-based error-correction algorithm consists of the following three sequential steps: (i) estimation of the number of errors, (ii) localization of the errors, and (iii) estimation of the error values.


\textbf{i) Estimation of number of errors:} Using $\mathbf{{s}}_{\bar{u}\bar{h}}$ defined in \eqref{eq:lin eq}, the next step is to estimate the number of errors introduced by the Byzantine worker nodes on each noisy codeword $\mathbf{r}_{\bar{u}\bar{h}}$. Note that, the noise injected by the Byzantine worker nodes into each entry of their respective noise matrices may be independent and vary in sparsity across different entries. As a result, the number of errors can differ across codewords, depending on the specific entries corrupted by the Byzantine worker nodes. Therefore, it is essential to estimate the number of errors for each noisy codeword $\mathbf{r}_{\bar{u}\bar{h}}$ individually before proceeding with error localization and correction. With infinite precision, the number of errors for a given noisy codeword $\mathbf{r}_{\bar{u}\bar{h}}$, can be estimated by finding the rank of the syndrome matrix $\mathbf{{S}}_{m}$, where the $i$-th row of the matrix $\mathbf{{S}}_{m}$ is given by, $[ \mathbf{{s}}_{\bar{u}\bar{h}}(1+i-1),\mathbf{{s}}_{\bar{u}\bar{h}}(2+i-1),\ldots,\mathbf{{s}}_{\bar{u}\bar{h}}(v+i-1)]$ for $i=1,2,\ldots,v$. Here, $v=\lfloor \frac{N-K}{2}\rfloor$, denotes the maximum error correction capability of the code, and $\mathbf{s}_{\bar{u}\bar{h}}(i)$ represents the $i$-th element of the syndrome vector $\mathbf{{s}}_{\bar{u}\bar{h}}$ corresponding to the noisy codeword $\mathbf{r}_{\bar{u}\bar{h}}$.

\textbf{ii) Localization of errors:}  After estimating the number of errors introduced by the Byzantine worker nodes, denoted by $A_{\bar{u}\bar{h}}$, on the noisy codeword $\mathbf{r}_{\bar{u}\bar{h}}$, the next step is to compute the error locator polynomial ${g}_{\bar{u}\bar{h}}(z)$, which has degree $A_{\bar{u}\bar{h}}$, corresponding to the noisy  codeword $\bar{r}_{\bar{u}\bar{h}}$. The coefficients of the error locator polynomial ${g}_{\bar{u}\bar{f}}(z)$ are computed by solving the following system of linear equations,
\begin{IEEEeqnarray}{rcl} 
\label{eq:err loc coeff}
\mathbf{{s}}_{\bar{u}\bar{h}}(i)g_{A_{\bar{u}\bar{h}}}+\mathbf{{s}}_{\bar{u}\bar{h}}(i+1)g_{A_{\bar{u}\bar{h}-1}}+\ldots +\mathbf{{s}}_{\bar{u}\bar{h}}(i+A_{\bar{u}\bar{h}}-1)g_{0} 
=-\mathbf{{s}}_{\bar{u}\bar{h}}(i+A_{\bar{u}\bar{h}}),
\end{IEEEeqnarray}
\noindent for $i=1,2,3\ldots,2v-A_{\bar{u}\bar{h}}$, where $g_{0},g_{1}\ldots,g_{A_{\bar{u}\bar{h}}}$, and $g_{0}=1$ represents the coefficients of the error locator polynomial ${g}_{\bar{u}\bar{h}}(z)$. Further, the roots of error locator polynomial ${g}_{\bar{u}\bar{h}}(z)$ indicate the location of the errors.

\textbf{iii) Estimation fo error values:} After identifying the location of the errors, the next step is to compute the error vector $\mathbf{e}_{\bar{u}\bar{h}}$ for each noisy codeword $\mathbf{r}_{\bar{u}\bar{h}}$ by solving the system of linear equations specified in \eqref{eq:lin eq}. Once the error vector for a noisy codeword is determined, these values are subtracted from its corresponding error locations to correct those errors. The corrected codewords are then used for the subsequent reconstruction stage.

All the internal blocks of DFT decoder discussed above assume $\mathbf{P}_{i}=\mathbf{0}$ on each $\mathbf{R}_{i}$ for $i\in [N]$. However, in practice, the decoder suffers from performance degradation due to inherent precision noise introduced by the worker nodes when computing $f(\mathbf{U}_{i})$, arising from floating point operations. As a result, $\mathbf{P}_{i}\neq \mathbf{0}$ for each $\mathbf{R}_{i}$, $i\in [N]$. To address this, several variations of the above mentioned steps of DFT decoder are available in the literature \cite{b3,b5,b6,b9,b10,b8}, and those can be adopted to operate under such finite-precision scenarios.

\textcolor{black}{Building on the existing approaches \cite{b3,b5,b6,b9,b10,b8}, we investigate whether the specific structure of ALCC can be exploited to develop decoding methods that improve Byzantine error correction in the ALCC setting. Towards this direction, we observe that, in the presence of Byzantine workers, the master node in ALCC receives multiple noisy DFT codewords corresponding to different entries of the resulting computation matrices ${\mathbf{R}_i}$, for ${i\in[N]}$. Importantly, although the error values introduced by the Byzantine workers may vary across these codewords, their error locations have a common support determined by the identities of the Byzantine workers. This common-support structure is specific to the ALCC setting and is not present when decoding individual DFT codewords independently. This structure, however, cannot be directly exploited in every block of the DFT decoder. In particular, the task of estimating the number of errors must be performed separately for each codeword. This is because a Byzantine worker may inject an error into some entries of its computation result but not others, resulting in different numbers of errors across codewords. Similarly, the task of estimating the error values must be performed separately for each codeword, since the error values introduced by the Byzantine workers may differ across entries. Therefore, these two blocks do not provide an opportunity to exploit the common-support structure across multiple codewords.} \textcolor{black}{In contrast, the error-localization block can exploit this common-support structure, since the error locations across the multiple codewords correspond to the same set of Byzantine workers. Motivated by this observation, we focus our analysis on the error-localization block of the DFT decoder. To this end, in the next section, we analyze the performance of the error-localization block under precision noise and use this analysis to develop novel localization methods.} 

\section{Error Localization for \textcolor{black}{Robust} ALCC}
\label{sec:err localization}

Recall that the master node retrieves $M=u\times h$ noisy DFT codewords $\{\mathbf{r}_{\bar{u}\bar{h}}\}$, each perturbed by precision noise. In this context, we apply a variant of the error correction algorithm discussed in the previous section on each noisy codeword, and present the performance of the error localization block. 

\subsection{Independent Error Localization in ALCC}
\label{subsec:independent localization}
In order to localize the errors introduced by the Byzantine worker nodes, we apply DFT decoder on each noisy codeword $\mathbf{r}_{\bar{u}\bar{h}}$ independently, assuming the number of errors on the corresponding noisy codeword $\mathbf{r}_{\bar{u}\bar{h}}$ has been accurately estimated as $A_{\bar{u}\bar{h}}$, where $A_{\bar{u}\bar{h}}$ is within the error correcting capability of the code. Since the DFT decoder is applied independently on each noisy codeword $\mathbf{r}_{\bar{u}\bar{h}}$, for the ease of notation, in the rest of the section, we will use $\mathbf{r}$ in place of $\mathbf{r}_{\bar{u}\bar{h}}$, $A$ in place of $A_{\bar{u}\bar{h}}$, $\mathbf{e}$ in place of $\mathbf{e}_{\bar{u}\bar{h}}$, and ${g}(z)$ in place of ${g}_{\bar{u}\bar{h}}(z)$,
With finite precision, the noisy version of the error locator polynomial ${g}(z)$ is given by, $\bar{g}(z) = g(z) + e(z),$ where $g(z)$ denotes the error-locator polynomial with infinite precision and $e(z)$ is the error polynomial due to precision errors. \textcolor{black}{We note that the coefficients of $e(z)$ are in general correlated due to the standard decoder operations on the syndrome vector. However, for analytical tractability, we model the coefficients as random variables which are i.i.d. as $\mathcal{CN}(0,\sigma_p^2)$. Henceforth, we refer to $\sigma_p^2$ as the error-locator noise variance. This modeling assumption provides analytical tractability, enables us to derive a bound on the probability of localization error, and allows us to analyze the localization performance as a function of $\sigma_p^2$ and the number of Byzantine workers $A$.\footnote{\textcolor{black}{In the subsequent sections, we demonstrate that this modeling assumption preserves the analytical behavior in a manner that is consistent with the empirically observed probability of localization error. More importantly, it retains the dependence of the localization error on the underlying design parameters, enabling principled design choices that translate into tangible performance gains. }}} In the absence of precision noise, the roots of the error-locator polynomial accurately provide information about the positions of the errors. For example, if $X_{q}$ is a root of the polynomial ${g}(z)$, where $X_{q} =\gamma^{q-1}$ and $\gamma = e^{-\frac{2\pi \iota}{N}}$, with $q\in[N]$, then the index $q$ corresponds to the location of the error. However, with precision errors, one needs to evaluate $||\bar{g}(X_{q})||^2$ for each $q\in[N]$ and arrange the evaluations in the ascending order as $||\bar{g}(X_{\hat{i}_{1}})||^2 \leq||\bar{g}(X_{\hat{i}_{2}})||^2\leq \ldots \leq||\bar{g}(X_{\hat{i}_{N}})||^2,$ where \{$\hat{i}_1,\hat{i}_2, \ldots, \hat{i}_N\}$ is the ordered set based on the evaluations. Subsequently, the $A$ smallest evaluations and its corresponding set of indices $\hat{\mathcal{L}} = \{\hat{i}_1,\hat{i}_2, \ldots, \hat{i}_A\}$ are treated as the detected error locations \cite{b5,b9,b10,b8}. Further, the detected error locations are used to subtract the error vectors from the noisy codewords, to recover an estimate of $f(\mathbf{U}_{i})$. Finally, these estimates of $f(\mathbf{U}_{i})$ are used for the reconstruction stage. 

In the rest of this section, we present an analysis on the error localization block. When recovering the roots of ${g}(z)$ of degree $A$, let the true locations of the errors be $\mathcal{L} = \{i_1,i_2,...,i_A\}$, however, let the recovered roots of $\bar{g}(z)$ be $\hat{\mathcal{L}} = \{\hat{i}_1,\hat{i}_2, \ldots, \hat{i}_A\}$. Conditioned on this information, the localization step is in error if $\hat{\mathcal{L}} \neq \mathcal{L}$. Therefore, for a given choice of the error vector $\mathbf{e}$ on codeword $\mathbf{r}$, the error rate of localization due to precision noise is 
\begin{equation}
 \label{eq:first union  bound on pep}
 P_{error}= \text{Prob}(\hat{\mathcal{L}}\neq \mathcal{L}) = \text{Prob} (\bigcup\limits_{a=1}^{A} i_{a} \notin \hat{\mathcal{L}}).
 \end{equation} Furthermore, for $i_{a}$ to not appear in $\hat{\mathcal{L}}$, there should be another index $j_{b} \in \mathcal{V}\triangleq[N] \backslash \mathcal{L}$, where $b\in\{1,2,\ldots,N-A\}$, such that $||\bar{g}(X_{j_{b}})||^2\leq ||\bar{g}(X_{i_{a}})||^2$. Denoting $\mathcal{V}=\{j_{1},j_{2},\ldots,j_{N-A}\}$, we rewrite \eqref{eq:first union  bound on pep} and bound it as
\begin{IEEEeqnarray}{rcl}
\label{eq:lowerboundstep}
   \text{Prob} (\bigcup\limits_{b=1}^{N-A} j_{b} \in \hat{\mathcal{L}}) \leq \sum_{b=1}^{N-A} \text{Prob} \left( j_{b} \in \hat{\mathcal{L}}\right).
\end{IEEEeqnarray}
Further, if $ j_{b}\in \hat{\mathcal{L}}$, this implies $ ||\bar{g}(X_{j_{b}})||^2 \leq||\bar{g}(X_{i_{a}})||^2$ for some $a\in[A]$. Therefore, \eqref{eq:lowerboundstep} can be expressed as
\begin{IEEEeqnarray}{rCl}
\label{eq:last union bound}
    \leq \sum_{b=1}^{N-A} \text{Prob} \Big( ||\bar{g}(X_{j_{b}})||^2 \leq ||\bar{g}(X_{i_{1}})||^2 \cup  ||\bar{g}(X_{j_{b}})||^2 \leq ||\bar{g}(X_{i_{2}})||^2 \cup \dots \cup 
    ||\bar{g}(X_{j_{b}})||^2 \leq ||\bar{g}(X_{i_{a}})||^2 \Big).
\end{IEEEeqnarray}
Applying union bound on \eqref{eq:last union bound}, we obtain
\begin{IEEEeqnarray}{rcl}
\label{eq:union bound2}
   P_{error}\leq \sum_{b=1}^{N-A} \sum_{a=1}^{A}  \text{Prob}\Big(||\bar{g}(X_{j_{b}})||^2 \leq ||\bar{g}(X_{i_{a}})||^2\Big),
\end{IEEEeqnarray}
\begin{IEEEeqnarray}{rcl}
\label{eq:union bound9}
   =\sum_{b=1}^{N-A} \sum_{a=1}^{A} PEP_{j_{b},i_{a}},
\end{IEEEeqnarray}
where, $ PEP_{j_{b},i_{a}}$ is given by
\begin{equation}
\label{eq:PEP}
 PEP_{j_{b},i_{a}}= \text{Prob} \Big(||\bar{g}(X_{j_{b}})||^2\leq ||\bar{g}(X_{i_{a}})||^2\Big),
\end{equation}

\noindent represents the pairwise error probability between $i_{a}$ and $j_{b}$ in \eqref{eq:union bound2}. Therefore, in order to derive an expression for \eqref{eq:union bound9}, we need to find an expression for the pairwise error probability between any two $A(N-A)$ pairs.

Since closed-form expressions on $ PEP_{j_{b},i_{a}}$ are intractable to obtain, we propose a lower bound on it.
\begin{theorem}
\label{thm:corr}
For a given $A, N$, $\sigma^{2}_{p}$, and $\mathbf{e}$, a lower bound on the pairwise error probability in \eqref{eq:PEP} can be expressed as
\begin{equation}
\label{eq:lower_bound}
 PEP_{j_{b},i_{a}}\geq PL_{j_{b},i_{a}} \triangleq \frac{\kappa}{(1+\kappa)}e^{-\frac{\eta (c_I^2 +c_Q^2)\kappa}{4\sigma^2_p(1+\kappa)}},
\end{equation}
where $\eta, c_I$, $c_{Q}$ are constants which depend on the vector $\mathbf{e}$ and the indices $j_{b}$ and $i_{a}$, and \\ $\kappa =\frac{2}{\eta \sum_{l = 1}^{A} 1 - \mbox{cos}\big(l\frac{2\pi}{N}|j_{b}-i_{a}|\big)}$.
\end{theorem}

\begin{IEEEproof}
    We refer the readers to the proof given in Appendix \ref{proof:2}.\\
\end{IEEEproof}
\noindent Based on the results in Theorem \ref{thm:corr}, we are now ready to make inferences on the behavior of the pairwise error probability as a function of $\sigma_p^2$ and present the following corollary.
\begin{corollary}
\label{lemma:sigma_beh}
For a given $A$ such that $A \leq \lfloor \frac{N-K}{2}\rfloor$, the lower bound in Theorem \ref{thm:corr} is a non-decreasing function of $\sigma^2_p$.
\end{corollary}

\begin{figure}[ht]
\begin{center}
\includegraphics[scale = 0.30]{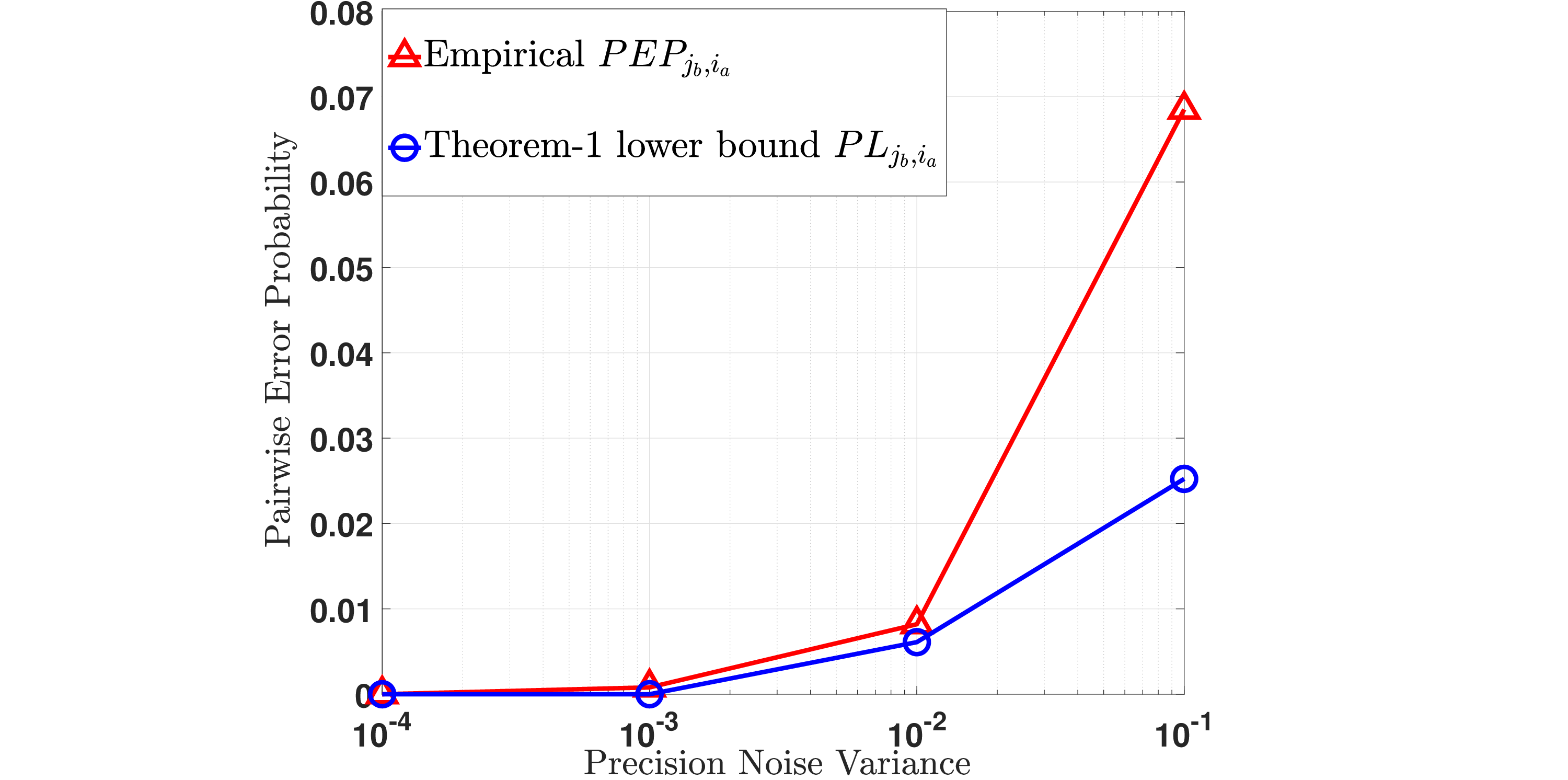}
\caption{Effect of precision noise variance on the pairwise error probability for a fixed non-Byzantine and Byzantine index pair $(j_b,i_a)$, with parameters  $N=31$, $k=7$, $t=3$, $K=19$, $v=6$, $\beta=1.5$, $\sigma=10^6$, $A=6$, $\eta=50$, and $f(\mathbf{X})=\mathbf{X}^{T}\mathbf{X}$, with $\mathbf{X}\in\mathbb{R}^{28\times5}$. The empirical pairwise error probability $PEP_{j_b,i_a}$ is obtained from $10^4$ Monte Carlo realizations and compared with the analytical lower bound $PL_{j_b,i_a}$ from Theorem~\ref{thm:corr}.}
\label{Fig:fixed_pair_pep}
\end{center}
\end{figure}

\noindent \textcolor{black}{The above corollary establishes that the analytical lower bound on the pairwise error probability is a non-decreasing function of $\sigma_p^2$ for every pair $(j_b,i_a)$, where $i_a\in\mathcal{L}$ and $j_b\in\mathcal{V}$. However, this result alone does not imply that the actual pairwise error probability is non-decreasing with $\sigma_p^2$. We therefore empirically examine the behavior of both the actual pairwise error probability and its analytical lower bound for a fixed pair of Byzantine and non-Byzantine locations. Specifically, we empirically compare the pairwise error probability $PEP_{j_b,i_a}$ with its analytical lower bound $PL_{j_b,i_a}$ from Theorem~\ref{thm:corr} for a fixed non-Byzantine and Byzantine pair $(j_b,i_a)$, with $j_b\in\mathcal{V}$ and $i_a\in\mathcal{L}$, while keeping the set of Byzantine locations $\mathcal{L}$ fixed across all Monte Carlo realizations, over different values of precision noise variance. For each precision noise variance, we perform $10^4$ Monte Carlo realizations of the complete error-localization block of the proposed Robust ALCC setting using the DFT decoder. In each realization, a pairwise error event is declared when $\left|\bar{g}(X_{j_b})\right|^2\leq\left|\bar{g}(X_{i_a})\right|^2$, i.e., when the non-Byzantine location $j_b$ is preferred over the Byzantine location $i_a$. The empirical $PEP_{j_b,i_a}$ is obtained as the fraction of realizations in which this event occurs, while $PL_{j_b,i_a}$ is evaluated for the same pair using the corresponding empirically computed error-locator noise variance $\sigma_p^2$. For each precision-noise variance, we fix the location of the Byzantine workers, and then generate $10^4$ realizations of the precision noise matrices $\{\mathbf{P}_{i}, i = 1, 2, \ldots, N\}$. Subsequently, we use these realizations to generate an ensemble of $10^4$ polynomials representing $e(z)$, through the standard error-localization block of the DFT decoder. Finally, we empirically compute the variance of each of the coefficients of $e(z)$ using the $10^4$ realizations, and then average them across the number of coefficients to obtain $\sigma_p^2$. The resulting comparison for the above setting is shown in Fig.~\ref{Fig:fixed_pair_pep}, wherein the empirical $PEP_{{j_b,i_a}}$ remains above $PL_{{j_b,i_a}}$ over the considered range of precision noise variance, consistent with the lower-bound relation established in Theorem \ref{thm:corr}.
Importantly, we also observe that the lower bound is not tight, or lie very close to the empirical values of pairwise error probability $PEP_{{j_b,i_a}}$. These results
provide empirical support for the relevance of $PL_{{j_b,i_a}}$ as a
tractable characterization of the corresponding pairwise error probability.}

 \textcolor{black}{At the overall localization level, the union bound in \eqref{eq:union bound2} decomposes the localization error probability $P_{\mathrm{error}}$ into $A(N-A)$ pairwise error events. Since the individual pairwise error probabilities $PEP_{j_b,i_a}$ in \eqref{eq:PEP} do not admit a tractable closed-form expression, we use the lower bound in Theorem~\ref{thm:corr} to form the following aggregate quantity as a tractable surrogate for $P_{{error}}$. Accordingly, we define
\begin{equation}
\label{eq:surrogate PL} 
P_{\mathrm{SLE}}
\triangleq
\sum_{a=1}^{A}\sum_{b=1}^{N-A}PL_{j_b,i_a}.
\end{equation}
We refer $P_{\mathrm{SLE}}$ as the surrogate localization error probability and use it as a tractable proxy for $P_{{error}}$ in the subsequent analysis and design decisions throughout this paper.}

\begin{figure}[H]
\begin{center}
\includegraphics[scale = 0.29]{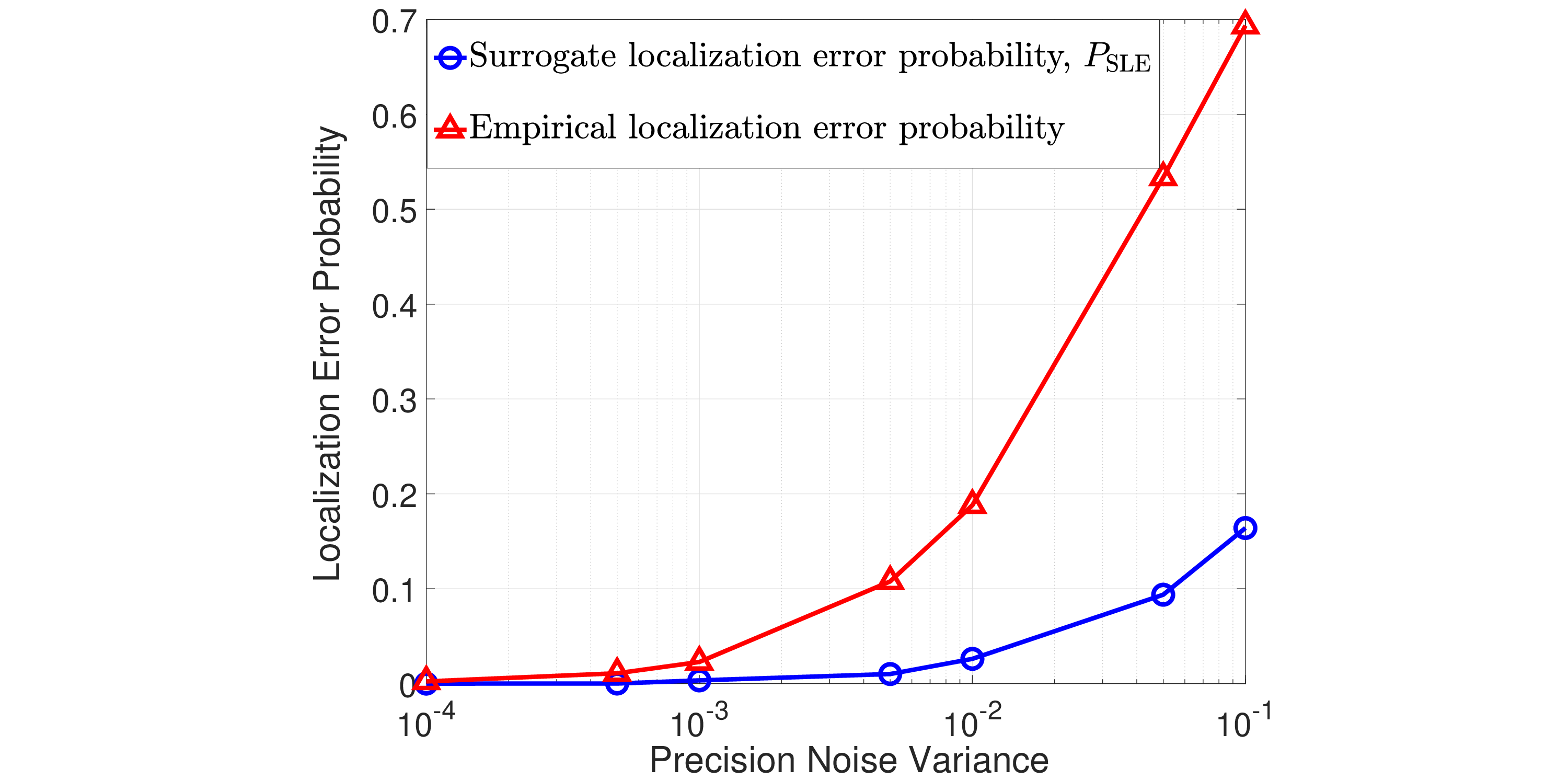}
\caption{Comparison of the empirical localization error probability with the surrogate localization error probability $P_{\mathrm{SLE}}$ at different precision noise variance. The parameters used in the experiments are $N=31$, $k=7$, $t=3$, $K=19$, $v=6$, $\beta=1.5$, $\sigma=10^6$, $A=6$, $\eta=50$, and $f(\mathbf{X})=\mathbf{X}^{T}\mathbf{X}$, with $\mathbf{X}\in\mathbb{R}^{28\times5}$. The non-zero entries of the matrices $\{\mathbf{E}_{i_a}\}$ are i.i.d. according to $\mathcal{CN}(10,10^3)$.}
\label{Fig:localization_trends1}
\end{center}
\end{figure}

\textcolor{black}{We emphasize that the surrogate for localization error  probability $P_{\mathrm{SLE}}$ in \eqref{eq:surrogate PL} is not claimed to be a tight numerical approximation of the exact localization error probability. Rather, it provides a tractable characterization of its variation with the system parameters. To further examine this behavior, we compare values of empirical localization error probability with $P_{\mathrm{SLE}}$ for different values of precision noise variance. The empirical localization error probability is obtained by comparing the complete set of locations identified by the DFT decoder with the actual set of Byzantine locations in each Monte Carlo realization, with a realization declared erroneous whenever the two sets are not identical. The empirical localization error probability is then obtained by averaging these error indicators over the Monte Carlo realizations. Throughout this experiment, the set of Byzantine locations is kept
fixed across all Monte Carlo realizations. Further, for each precision-noise variance, the error-locator noise variance $\sigma_p^2$ is empirically computed using the same procedure as that used to generate the results in Fig.~\ref{Fig:fixed_pair_pep}. The corresponding values of $PL_{j_b,i_a}$ are then evaluated using this empirically computed $\sigma_p^2$ and used to compute the surrogate localization error probability $P_{\mathrm{SLE}}$ defined in \eqref{eq:surrogate PL} by aggregating the contributions from all $A(N-A)$ pairs. These results are shown in Fig. \ref{Fig:localization_trends1}. As shown in Fig. \ref{Fig:localization_trends1}, $P_{\mathrm{SLE}}$ exhibits an increasing trend with precision noise variance, consistent with the observed increase in the empirical localization error  probability. Since a closed-form expression for the overall localization error probability is not available, we use the observed behavior of $P_{\mathrm{SLE}}$, together with its consistency with the empirical localization error probability, as the basis for studying the dependence of localization performance on the precision-noise variance in the subsequent analysis. These observations provide empirical support for the use of $P_{\mathrm{SLE}}$ as a tractable characterization of the variation of localization performance with the precision-noise variance over the considered range.}

\begin{figure}[ht!]
    \centering
    \begin{subfigure}[b]{0.49\linewidth}
        \centering
        \includegraphics[width=\linewidth]{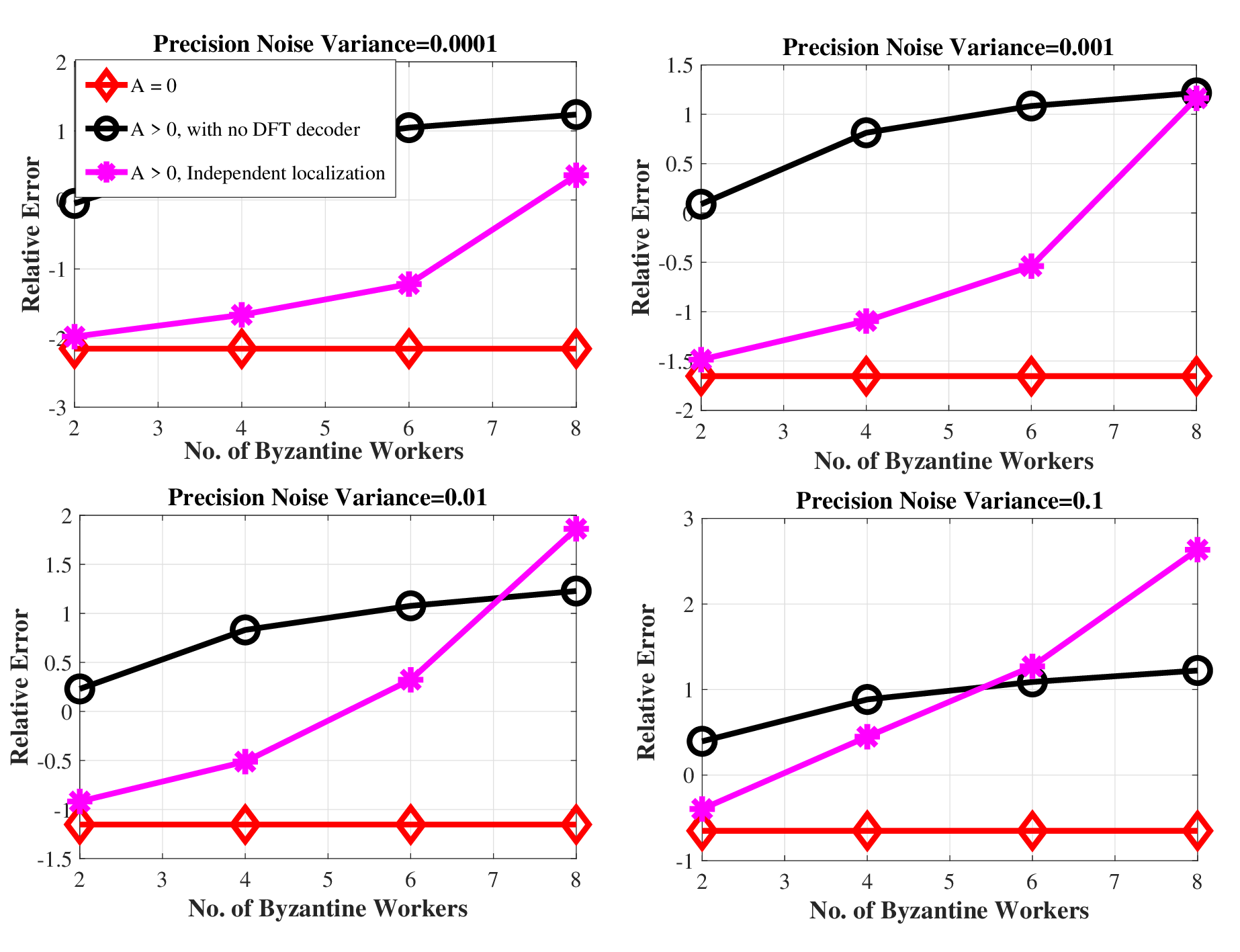}
           \caption{}
    \end{subfigure}
    \hfill
    \begin{subfigure}[b]{0.49\linewidth}
        \centering
        \includegraphics[width=\linewidth]{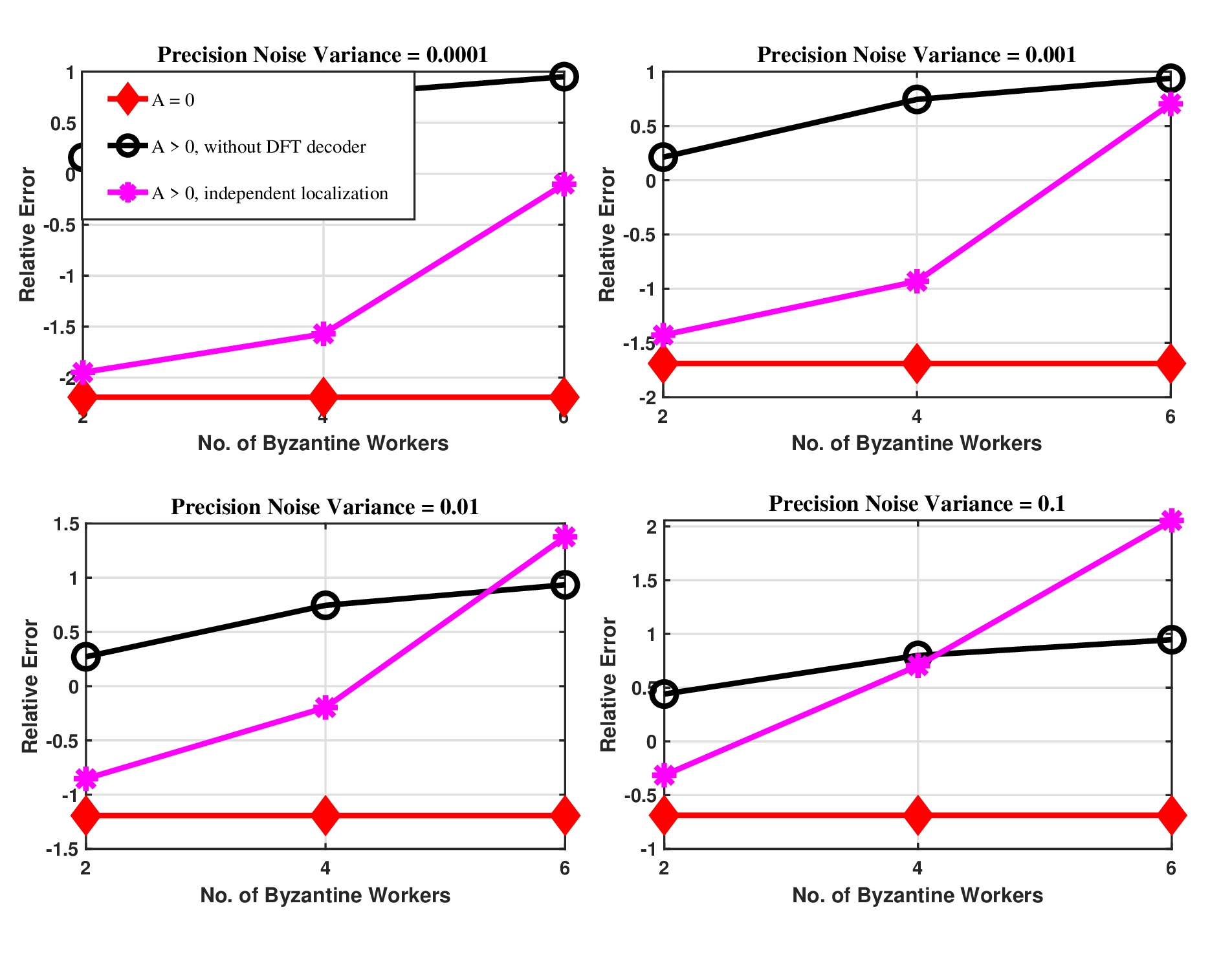}
        \caption{}
    \end{subfigure}

\caption{\textcolor{black}{Average relative error (in $\log_{10}$ scale) of ALCC in the presence of Byzantine worker nodes, with and without DFT decoder, for $\tau=0$, $\beta=1.5$, $t=3$, and $\sigma=10^6$. In (a), $(N,K)=(31,15)$, and in (b), $(N,K)=(27,15)$. The non-zero entries of the noise matrices ${\mathbf{E}_{i_a}}$ are i.i.d. as $\mathcal{CN}(10,10^3)$.}}
\label{Fig: ind localization}
\end{figure}

\subsubsection{Experimental Results on Independent Error Localization in ALCC}
\label{subsub:experimental results on Ind}

\textcolor{black}{In order to showcase the effect of DFT decoder and to understand its behavior as a function of the precision noise variance, we conduct experiments using $(N, K) = (31,15)$ and $(N, K) = (27,15)$ DFT decoders in the ALCC setting.} The parameters and the function used for the experiments are the same as those used to generate the results in Fig.~\ref{Fig: alcc in adv plot}. For the above experiment setup, we compute the average relative error of ALCC when using the DFT decoder at different precision noise variance by varying the number of Byzantine worker nodes up to the corresponding error-correction capability. \textcolor{black}{For $(N, K) = (31,15)$ and $(N, K) = (27,15)$, the corresponding error-correction capabilities are $v=8$ and $v=6$, respectively. The results on relative error are presented in Fig.~\ref{Fig: ind localization}(a) and Fig.~\ref{Fig: ind localization}(b), respectively.} The plots in Fig.~\ref{Fig: ind localization} confirm that ALCC with the DFT decoder is robust to a few Byzantine worker nodes. \textcolor{black}{However, when the precision noise variance is high, the error localization performance degrades as the number of Byzantine worker nodes increases.} This degradation in localization performance degrades the relative error of computation, which in turn renders the DFT decoder ineffective when used as an off-the-shelf block.


To address this limitation, in the next subsection, we show that the error-localization block of DFT decoder can be customized to improve the accuracy of ALCC.
\subsection{ALCC Based Joint Error Localization}
\label{subsec:joint loc}
Recall that at the reconstruction phase of ALCC framework master node receives a total set of $M=u\times h$ codewords $\mathbf{r}_{\bar{u}\bar{h}}$ for $\bar{u} \in [u], \bar{h} \in [h]$. Consequently, when the DFT decoder is employed within the ALCC framework, the error-localization block produces a collection of $M$ error-locator polynomials corresponding to the noisy codewords $\{\mathbf{r}_{\bar{u}\bar{h}}\}$. Along with each polynomial, the decoder also estimates its degree, which equals the number of corrupted evaluations $A_{\bar{u}\bar{h}}$ introduced into $\mathbf{r}_{\bar{u}\bar{h}}$ by the Byzantine workers. Since the corruption patterns introduced by different Byzantine workers can vary, the master node generally receives error-locator polynomials of different degrees.
 
\textcolor{black}{To characterize these corruption patterns, let $\mathbf{B}_{i_{a}} \in \{0, 1\}^{u \times h}$ specify the binary matrix that captures the positions of the non-zero entries of $\mathbf{E}_{i_{a}}$. These matrices are henceforth referred to as the \emph{base matrices} for the Byzantine worker nodes. More specifically, a non-zero entry in a base matrix $\mathbf{B}_{i_{a}}$ for $a\in[A]$ signifies that the corresponding component of the returned matrix $f(\mathbf{Y}_{i_{a}})$ is corrupted with an additive noise introduced by the worker node $i_{a}$. As a consequence, depending on the sparsity of the base matrices $\mathbf{B}_{i_{a}}$ of the Byzantine worker nodes, the degrees of the $M$ error locator polynomials vary. We first consider a canonical corruption pattern that exposes the key design principle behind the proposed localization method.}

\textbf{Joint localization for all-one base matrices}: \textcolor{black}{Suppose every Byzantine worker corrupts all entries of its returned matrix, i.e., $\mathbf{B}_{i_a}=\mathbf{J}_{u\times h}$ for every $a\in[v]$, where $\mathbf{J}_{u\times h}$ denotes the all-one matrix. In this case, every returned codeword is affected by all Byzantine workers, and therefore every error-locator polynomial has the maximum degree $v=\lfloor\frac{N-K}{2}\rfloor$. More importantly, all $M$ error-locator polynomials correspond to the same set of Byzantine worker locations. To recover these common roots, we use the analysis of the DFT decoder presented in Section~\ref{subsec:independent localization}. In particular, the surrogate localization error probability $P_{\mathrm{SLE}}$ derived in \eqref{eq:surrogate PL} characterizes the effect of precision noise on the coefficients of the error-locator polynomial. As shown in Fig.~\ref{Fig:localization_trends1}, the surrogate predicts that localization performance improves as the precision noise variance decreases. Furthermore, the surrogate preserves the same qualitative dependence on the underlying system parameters as the empirical localization error probability. This consistency enables us to use $P_{\mathrm{SLE}}$ as a design surrogate for constructing improved localization strategies.}

\textcolor{black}{The above observation motivates exploiting the availability of multiple error-locator polynomials corresponding to the same root locations. Since all $M$ polynomials estimate the same error-locator polynomial but with different realizations of precision noise, averaging their coefficients reduces the perturbation introduced by finite-precision arithmetic while preserving the underlying roots. Accordingly, given the $M$ error-locator polynomials $\{\bar{g}_{\bar{u}\bar{h}}(z), \bar{u} \in [u], \bar{h} \in [h]\}$ of degree $v$, the master node constructs the averaged polynomial
$$
\bar{g}_{\max}(z)=\frac{1}{M}\sum_{\bar{u}\in[u]}\sum_{\bar{h}\in[h]}\bar{g}_{\bar{u}\bar{h}}(z),
$$
and localizes the Byzantine workers by solving the roots of $\bar{g}_{\max}(z)$. The following proposition formalizes the benefit of this averaging operation.}

\begin{proposition}
\label{prop:all_one_matrix}
\textcolor{black}{When $\sigma_p^2>0$ and $M>1$, if $\mathbf{E}_{i_a}$ has non-zero
entries in all its locations for every $a\in[v]$, then the proposed
joint localization method offers smaller values of $P_{\mathrm{SLE}}$ in \eqref{eq:surrogate PL} compared to that obtained using any one of $M$ error locator polynomials under independent localization.}
\end{proposition}

\begin{IEEEproof}
     We refer the readers to the proof given in Appendix \ref{proof:3}.
\end{IEEEproof}

\textcolor{black}{Proposition~\ref{prop:all_one_matrix} establishes the central insight behind the proposed decoder. When multiple error-locator polynomials correspond to the same set of Byzantine workers, coefficient averaging suppresses precision noise and simultaneously improves localization accuracy without increasing the complexity of solving for the roots. This result motivates extending the averaging idea to the more general setting where the base matrices are arbitrarily sparse.}

\textbf{Joint localization for arbitrary base matrices}: When the base matrices are arbitrary, not every Byzantine worker corrupts every entry of the returned matrix. Consequently, the decoder receives error-locator polynomials whose degrees can vary between $1$ and $v$. In this setting, only the degree-$v$ polynomials correspond to the complete set of Byzantine worker locations, whereas lower-degree polynomials correspond to subsets of Byzantine workers and therefore cannot be averaged directly. Nevertheless, all the error-locator polynomials are generated by the same set of at most $v$ Byzantine workers. Therefore, despite having different degrees, the roots of all the polynomials must belong to a common set of at most $v$ worker locations. This observation naturally extends the averaging principle into a joint localization problem that simultaneously exploits information from all available error-locator polynomials.

Let $\mathcal{G}_{v}$ and $\mathcal{G}_{v^{-}}$ denote the set of polynomials with degree $v$ and degree less than $v$, respectively. Then, we obtain a new set of polynomials, denoted by $\mathcal{G}_{joint} = \{\bar{g}_{max}(z)\} \cup \mathcal{G}_{v^{-}}$, where $\bar{g}_{max}(z) = \frac{1}{|\mathcal{G}_{v}|} \sum_{\bar{g}_{\bar{u}\bar{h}}(z) \in \mathcal{G}_{v}} \bar{g}_{\bar{u}\bar{h}}(z)$ is the polynomial obtained by averaging the degree $v$ polynomials. Furthermore, let $\bar{M} \leq M$ represent the cardinality of $\mathcal{G}_{joint}$. We now propose a method to jointly solve for the roots of $\mathcal{G}_{joint}$. Assuming that the polynomials of $\mathcal{G}_{joint}$ are sorted in some order, let $t_{m}(x) \in \mathcal{G}_{joint}$ be the $m$-th polynomial of degree $d(m)$, where $1 \leq d(m) \leq v$. We propose to solve for the roots of every polynomial in $\mathcal{G}_{joint}$ independently and then take the union of the roots. Let this set of roots be denoted by $\mathcal{S}_{I}$, which is of cardinality $v''$. Furthermore, let $\mathcal{S}_{total} = \mathcal{S}_{I}^{v}$ be the set of all $v$-tuples over the set $\mathcal{S}_{I}$. For a given $\mathcal{S} \in \mathcal{S}_{total}$, we evaluate $t_{m}(z)$ at all the entries of $\mathcal{S}$, and obtain its first $d(m)$ smallest evaluations. Let the sum of the first $d(m)$ smallest evaluations of $t_{m}(z)$ be denoted by $E_{t_{m}(z)}(\mathcal{S})$, and the corresponding roots be $R_{t_{m}(z)}(\mathcal{S}) \subset \mathcal{S}$. Thus, the sum of the first $d(m)$ smallest evaluations of polynomials over the entire set $\mathcal{G}_{joint}$ corresponding to $\mathcal{S}$ is $\mathcal{E}_{\mathcal{S}} \triangleq \sum_{m = 1}^{\bar{M}} E_{t_{m}(z)}(\mathcal{S})$. Also, the set of the roots of all the polynomials in $\mathcal{G}_{joint}$ corresponding to $\mathcal{S}$ is denoted by $\mathcal{R}_{\mathcal{S}} \triangleq \{R_{t_{1}(z)}(\mathcal{S}), R_{t_{2}(z)}(\mathcal{S}), \ldots, R_{t_{\bar{M}}(z)}(\mathcal{S})\}$. Having obtained $\mathcal{E}_{\mathcal{S}}$ and $\mathcal{R}_{\mathcal{S}}$, we propose Problem \ref{opt_Problem}.

\begin{mdframed}  
\begin{problem} 
\label{opt_Problem}
For a given $N$, $\sigma_{p}^2$, $\mathcal{G}_{joint}$ , and $v$ solve $\hat{\mathcal{S}} = \arg \min_{\mathcal{S} \in \mathcal{S}_{total}} \mathcal{E}_{\mathcal{S}},$ and then obtain $\mathcal{R}_{\hat{\mathcal{S}}}$ as its solution. 
\end{problem}
\end{mdframed}

Note that as $v''$ (the cardinality of $\mathcal{S}_{I}$) increases, solving the optimization problem in Problem \ref{opt_Problem} becomes more computationally complex. However, in the practical scenario, when, $\sigma^{2}_{p}$ is small, the difference $v'' - v$ is expected to be small, which makes the optimization problem feasible to solve with computational complexity proportional to  $\binom{v''}{v}$, even if  $v'' \geq v$. In order to further decrease the complexity of solving the optimization problem, we substitute $\mathcal{S}_{I}$ by $\mathcal{S}_{II}$, where $\mathcal{S}_{I}$ is a randomly selected subset with the cardinality of $\mathcal{S}_{II}$ is $v' \leq v''$ such that $\mathcal{S}_{II} \subseteq \mathcal{S}_{I}$. The parameter $v'$ is referred to as the set constraint length for the proposed joint localization step, and this parameter can be adjusted to balance the computational complexity and error rate of joint localization. Therefore, increasing $v'$ leads to improvement in accuracy, at the expense of higher computational complexity. \textcolor{black}{The overall joint error localization method is summarized in
Algorithm \ref{alg:joint_localization}, while
Example \ref{example:exp1} illustrates its different steps through a numerical example.}
\begin{algorithm}[t]
\caption{\textcolor{black}{Joint Error Localization from Error-Locator Polynomials}}
\label{alg:joint_localization}
\KwIn{Error-locator polynomials
$\mathcal{G}=\{\bar{g}_{\bar{u}\bar{h}}(z)\}$,
maximum degree $v$, set constraint length $v\leq v'\leq v''$,
and $\mathcal{G}_{v}\neq\emptyset$.}

\KwOut{Estimated Byzantine-worker locations
$\mathcal{R}_{\hat{\mathcal{S}}}$.}

Partition $\mathcal{G}$ into
$\mathcal{G}_{v}$ (degree $v$) and
$\mathcal{G}_{v^-}$ (degree $<v$).

Compute
$\bar{g}_{\max}(z)
=
\frac{1}{|\mathcal{G}_{v}|}
\sum_{\bar{g}(z)\in\mathcal{G}_{v}}
\bar{g}(z)$.

Construct
$\mathcal{G}_{\rm joint}
=
\{\bar{g}_{\max}(z)\}\cup\mathcal{G}_{v^-}$.

\ForEach{$t_m(z)\in\mathcal{G}_{\rm joint}$}{
Identify the roots of $t_m(z)$ among the $N$ evaluation points
and denote them by $\mathcal{R}(t_m(z))$.
}

Form the union
$\mathcal{S}_{I}
=
\bigcup_{m=1}^{\bar{M}}
\mathcal{R}(t_m(z))$,
where $|\mathcal{S}_{I}|=v''$.

\eIf{$v'<v''$}{
Select a subset
$\mathcal{S}_{II}\subseteq\mathcal{S}_{I}$
such that $|\mathcal{S}_{II}|=v'$, and replace $\mathcal{S}_{I}$ by $\mathcal{S}_{II}$.
}{
Retain $\mathcal{S}_{I}$.
}

Construct
$\mathcal{S}_{\rm total}
=
\{\mathcal{S}\subseteq\mathcal{S}_{I}:|\mathcal{S}|=v\}$.

\ForEach{$\mathcal{S}\in\mathcal{S}_{\rm total}$}{
    \For{$m=1,\ldots,\bar{M}$}{
        Evaluate $t_m(z)$ at all elements of $\mathcal{S}$.
        
        Compute $E_{t_m(z)}(\mathcal{S})$
        as the sum of the smallest $d(m)$ squared magnitudes
        of the evaluations.
        
        Record the corresponding roots
        $R_{t_m(z)}(\mathcal{S})$.
    }
    
    Compute
    $\mathcal{E}_{\mathcal{S}}
    =\sum_{m=1}^{\bar{M}}E_{t_m(z)}(\mathcal{S})$.
    
    Store
    $\mathcal{R}_{\mathcal{S}}
    =\{R_{t_1(z)}(\mathcal{S}),\ldots,
    R_{t_{\bar{M}}(z)}(\mathcal{S})\}$.
}

Obtain
$\hat{\mathcal{S}}
=\arg\min_{\mathcal{S}\in\mathcal{S}_{\rm total}}
\mathcal{E}_{\mathcal{S}}$.

\Return
$\mathcal{R}_{\hat{\mathcal{S}}}$.
\end{algorithm}

\begin{example}
\label{example:exp1}
\textcolor{black}{Consider $N=7$ and $K=3$, so that $v=\left\lfloor\frac{N-K}{2}\right\rfloor
=\left\lfloor\frac{7-3}{2}\right\rfloor=2.$ Let $u=h=2$, giving $M=uh=4$ noisy codewords $\mathbf r_{\bar{u}\bar{h}}$ and hence four error-locator polynomials at the DFT decoder. Suppose there are two Byzantine workers, indexed by $i_1=1$ and $i_2=2$, whose base matrices are
$\mathbf B_{i_1}
=
\begin{bmatrix}
1&1\\
1&0
\end{bmatrix},$ and $
\mathbf B_{i_2}
=
\begin{bmatrix}
1&1\\
0&1
\end{bmatrix}.$
Here, the $(\bar{u},\bar{h})$-th entry of $\mathbf B_{i_a}$ indicates whether Byzantine worker $i_a$ introduces a nonzero error at the $(\bar{u},\bar{h})$-th entry of its error matrix $\mathbf E_{i_a}$. Hence, the error pattern across the four noisy codewords is
\[
\begin{array}{c|cccc}
&\mathbf r_{11}&\mathbf r_{12}&\mathbf r_{21}&\mathbf r_{22}\\
\hline
i_1=1&1&1&1&0\\
i_2=2&1&1&0&1
\end{array}.
\] Since $v=2$, the first two polynomials have degree $v$ and therefore
belong to $\mathcal{G}_{v}$, while the remaining two have degree less
than $v$ and therefore belong to $\mathcal{G}_{v^-}$. Hence,
$\mathcal{G}_{v}=\{\bar{g}_{1}(z),\bar{g}_{2}(z)\}$ and
$\mathcal{G}_{v^-}=\{\bar{g}_{3}(z),\bar{g}_{4}(z)\}$. Since
$|\mathcal{G}_{v}|=2$, the degree-$v$ polynomials are averaged as
specified in the proposed joint localization procedure to obtain
$\bar{g}_{\max}(z)
=
\frac{1}{|\mathcal{G}_{v}|}
\sum_{\bar{g}(z)\in\mathcal{G}_{v}}\bar{g}(z).$
Consequently, the set of polynomials used in the joint localization
step is $\mathcal{G}_{\mathrm{joint}}
=
\{\bar{g}_{\max}(z)\}\cup\mathcal{G}_{v^-}
=
\{\bar{g}_{\max}(z),\bar{g}_{3}(z),\bar{g}_{4}(z)\}.$
Thus, $\bar{M}=3$ polynomials are considered in the joint localization
step.}

\textcolor{black}{Let $t_1(z)=\bar{g}_{\max}(z)$, $t_2(z)=\bar{g}_{3}(z)$, and
$t_3(z)=\bar{g}_{4}(z)$. Suppose that independently applying the
DFT-based localization step to these three polynomials gives the
candidate error-location sets
$\mathcal{R}(t_1)=\{1,2\}$,
$\mathcal{R}(t_2)=\{3\}$, and
$\mathcal{R}(t_3)=\{2\}$, respectively. Following the proposed joint
localization procedure, we first collect all candidate locations
identified from the individual polynomials. Thus, the resulting
candidate set is $\mathcal{S}_{I}
=\bigcup_{m=1}^{3}\mathcal{R}(t_m)
=\{1,2,3\},$ with $v''=|\mathcal{S}_{I}|=3>v=2$. For this example, we take the set constraint length as $v'=v''=3$,
so that no further reduction of $\mathcal{S}_{I}$ is performed. Since the joint localization step seeks a set of $v$ candidate
locations, the algorithm constructs $\mathcal{S}_{\mathrm{total}}$ by
considering every $v$-element subset of $\mathcal{S}_{I}$. In this
example, $\mathcal{S}_{\mathrm{total}}
=
\left\{
\mathcal{S}\subseteq\mathcal{S}_{I}:|\mathcal{S}|=v
\right\}
=
\left\{
\{1,2\},\{1,3\},\{2,3\}
\right\},$ and $|\mathcal{S}_{\mathrm{total}}|
=
\binom{3}{2}
=
3.$}

\textcolor{black}{For each candidate set $\mathcal{S}\in\mathcal{S}_{\mathrm{total}}$,
the algorithm evaluates each polynomial in $\mathcal{G}_{\mathrm{joint}}$
at every candidate location $s\in\mathcal{S}$. For illustration, suppose the squared
magnitudes of the evaluations for the three candidate sets are as
follows:}
\begin{table}[H]
\centering
\label{tab:...}
\begin{tabular}{|c|c|c|}
\hline
$\mathcal{S}$ & $t_m(z)$ & $\{|t_m(s)|^2:s\in\mathcal{S}\}$ \\
\hline
$\{1,2\}$ & $t_1,\ d(1)=2$ & $\{0.02,0.03\}$ \\
          & $t_2,\ d(2)=1$ & $\{0.01,0.20\}$ \\
          & $t_3,\ d(3)=1$ & $\{0.15,0.02\}$ \\
\hline
$\{1,3\}$ & $t_1,\ d(1)=2$ & $\{0.02,0.40\}$ \\
          & $t_2,\ d(2)=1$ & $\{0.01,0.03\}$ \\
          & $t_3,\ d(3)=1$ & $\{0.15,0.04\}$ \\
\hline
$\{2,3\}$ & $t_1,\ d(1)=2$ & $\{0.03,0.40\}$ \\
          & $t_2,\ d(2)=1$ & $\{0.20,0.03\}$ \\
          & $t_3,\ d(3)=1$ & $\{0.02,0.04\}$ \\
\hline
\end{tabular}
\end{table}

\textcolor{black}{For each candidate set, $t_1(z)$ has degree $2$, and hence both
evaluations are included in $E_{t_1(z)}(\mathcal{S})$. In contrast,
$t_2(z)$ and $t_3(z)$ have degree $1$, and hence only the smaller of
the two squared-magnitude evaluations is included. Therefore, the
objective values $\mathcal{E}_{\mathcal{S}}$ for the candidate sets
$\mathcal{S}\in\mathcal{S}_{\mathrm{total}}$ are
\begin{align*}
\mathcal{E}_{\{1,2\}}
&=(0.02+0.03)
  +\min\{0.01,0.20\}
  +\min\{0.15,0.02\} =0.05+0.01+0.02
=0.08,\\[2mm]
\mathcal{E}_{\{1,3\}}
&=(0.02+0.40)
  +\min\{0.01,0.03\}
  +\min\{0.15,0.04\} =0.42+0.01+0.04
=0.47,\\[2mm]
\mathcal{E}_{\{2,3\}}
&=(0.03+0.40)
  +\min\{0.20,0.03\}
  +\min\{0.02,0.04\} =0.43+0.03+0.02
=0.48.
\end{align*}}
\noindent \textcolor{black}{Thus,
$\mathcal{E}_{\{1,2\}}<\mathcal{E}_{\{1,3\}}$ and
$\mathcal{E}_{\{1,2\}}<\mathcal{E}_{\{2,3\}}$, and the optimization in
Problem~\ref{opt_Problem} gives
\[
\hat{\mathcal{S}}
=
\arg\min_{\mathcal{S}\in\mathcal{S}_{\mathrm{total}}}
\mathcal{E}_{\mathcal{S}}
=
\{1,2\}.
\]
The corresponding stored root sets
$\mathcal{R}_{\hat{\mathcal{S}}}
=
\{R_{t_1(z)}(\hat{\mathcal{S}}),\ldots,
R_{t_{\bar{M}}(z)}(\hat{\mathcal{S}})\}$
are then returned as the estimated locations of the Byzantine workers.}

\end{example}

\begin{figure}[ht!]
    \centering
    \begin{subfigure}[b]{0.49\linewidth}
        \centering
        \includegraphics[width=\linewidth]{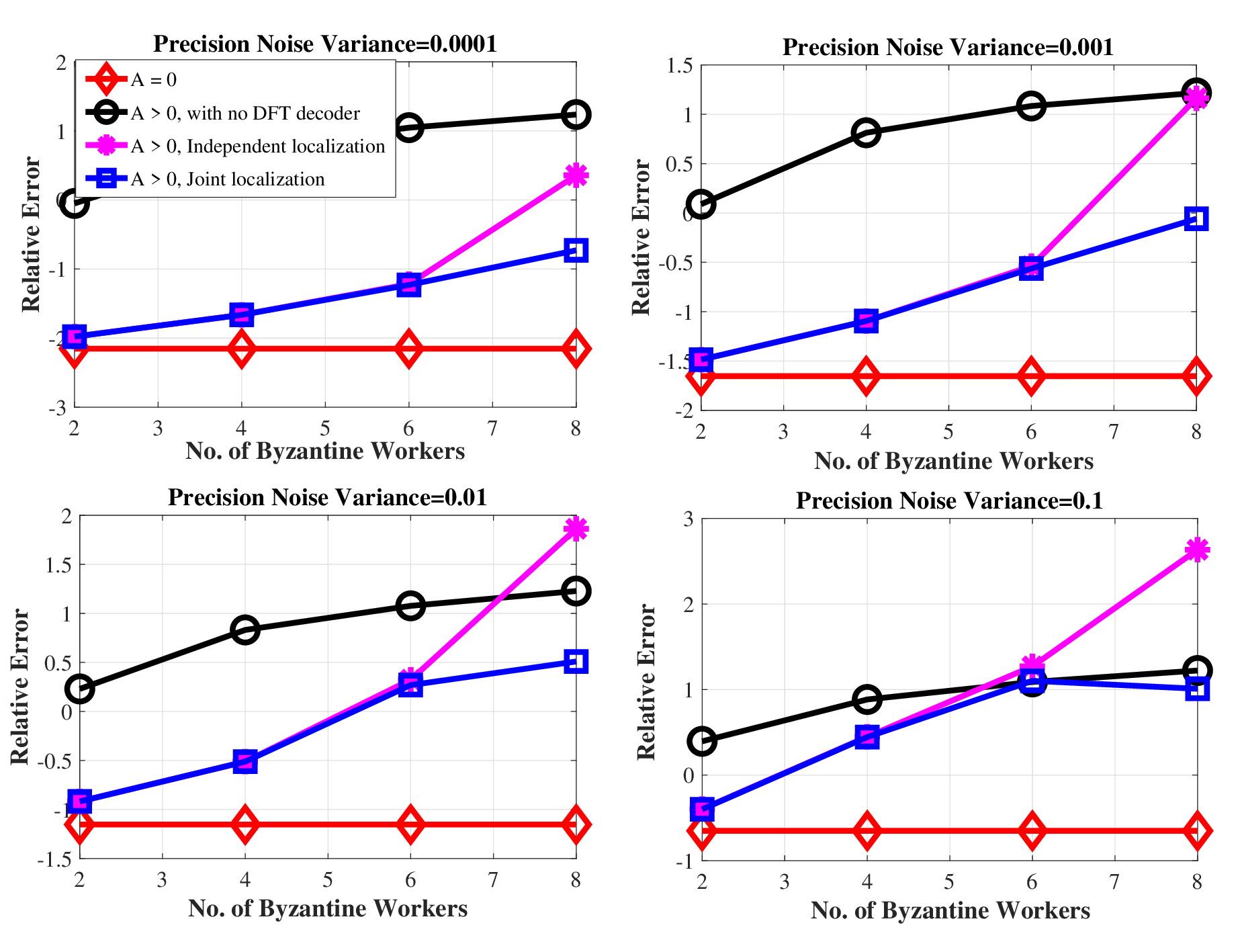}
        \caption{}
        \label{fig:joint_loc_31_15}
    \end{subfigure}
    \hfill
    \begin{subfigure}[b]{0.49\linewidth}
        \centering
        \includegraphics[width=\linewidth]{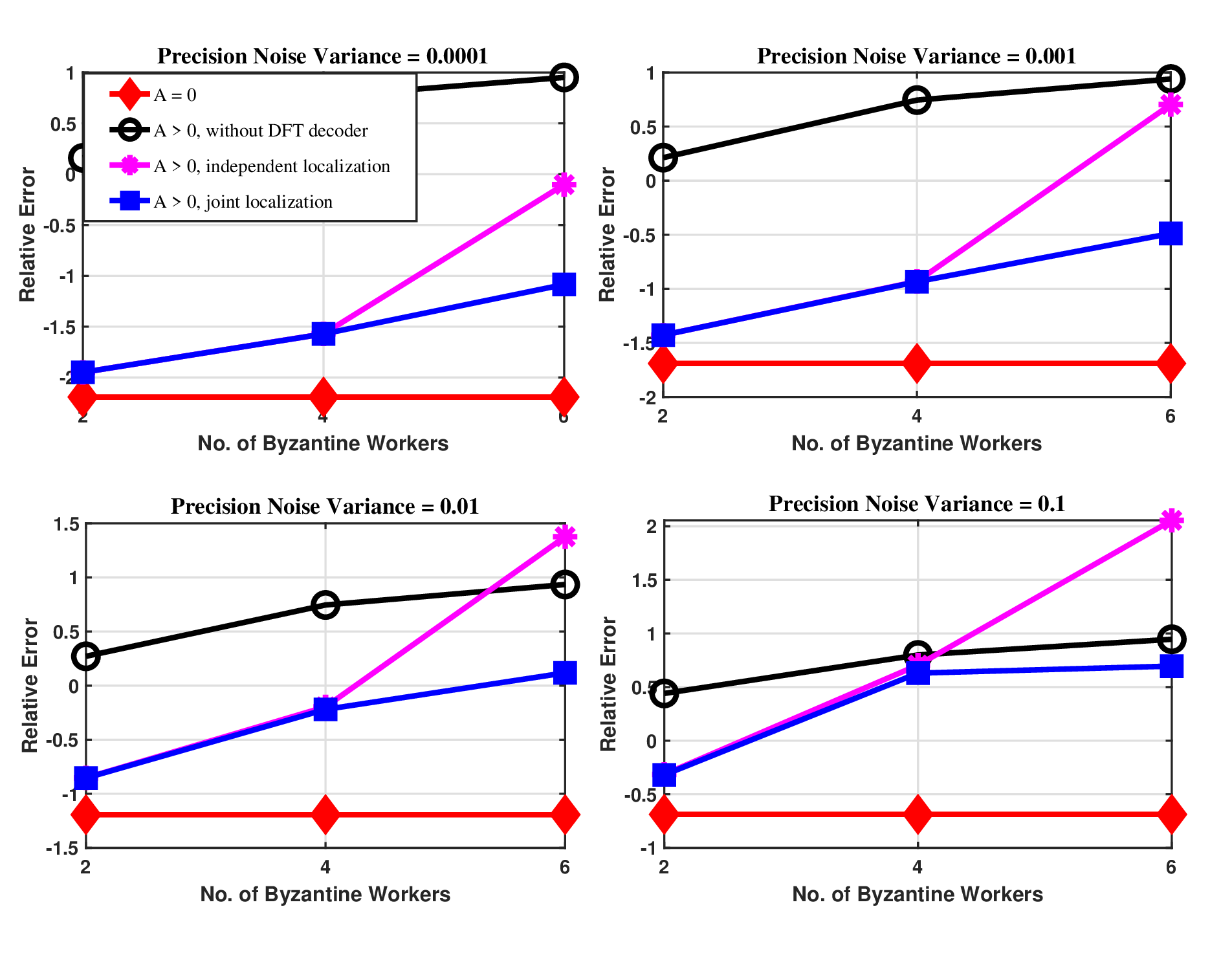}
        \caption{}
        \label{fig:joint_loc_27_15}
    \end{subfigure}

 \caption{\textcolor{black}{Average relative error (in $\log_{10}$ scale) of ALCC with and without DFT decoding and the proposed joint error localization in the presence of Byzantine worker nodes. In (a), $N=31$, $K=15$, and $v=8$, while in (b), $N=27$, $K=15$, and $v=6$. The remaining system parameters are $t=3$, $\tau=0$, $\beta=1.5$, and $\sigma=10^6$. The non-zero entries of the noise matrices ${\mathbf{E}_{i_a}}$ are i.i.d. as $\mathcal{CN}(10,10^3)$.}}
    \label{Fig: Joint localization}
\end{figure}

\subsubsection{Experimental Results on Joint Error Localization in ALCC}
\label{subsub:experimental results on joint loc}
\textcolor{black}{To demonstrate the benefits of joint error localization over independent localization in terms of accuracy of ALCC, we perform additional experiments using the $(N,K)=(31,15)$ and $(27,15)$ DFT decoders. For these experiments, we compute the average relative error at different precision noise variances, while varying the number of Byzantine worker nodes up to the corresponding error-correction capability. For the $(31,15)$ and $(27,15)$ DFT codes, the corresponding
error-correction capabilities are $v=8$ and $v=6$, respectively.} The underlying function $f(\cdot)$ and the parameters used for the experiments are the same as those used to generate the results in Fig \ref{Fig: ind localization}. The set of noise matrices $\{\mathbf{E}_{i_{a}}\}$ added by the Byzantine worker nodes are non-zero at each entry, i.e., $\mathbf{B}_{i_{a}} = \mathbf{J}_{u \times h}$, for $a\in[A]$. Furthermore, for the joint localization method, we use $\mathcal{S}_{total} =\mathcal{S}_{I}^{v}$. \textcolor{black}{With this setup, the results on average relative error are presented in Fig. \ref{Fig: Joint localization}(a) and Fig. \ref{Fig: Joint localization}(b), respectively.} From the plots shown in Fig. \ref{Fig: Joint localization}(a) and Fig. \ref{Fig: Joint localization}(b), we observe that, for small precision noise variance, the performance of independent localization and joint localization is similar. As the number of Byzantine workers increases to the error-correcting capability of the code, the performance difference gradually increases, and joint localization performs better in terms of accuracy.

\textcolor{black}{We remark that the same set of parameters is used in Figs. \ref{Fig: alcc in adv plot}, \ref{Fig: ind localization}, and \ref{Fig: Joint localization}. Maintaining the same system parameters across these figures allows the observed performance differences to be attributed to the successive introduction of the DFT decoder, independent localization, and the proposed joint localization, rather than to changes in the underlying system configuration.}

\subsection{\textcolor{black}{Computational Complexity of Independent and Proposed Joint Error Localization}}
\label{subsec:complexity_localization}
\textcolor{black}{In this section, we analyze the computational complexity of the error-localization stage of the DFT decoder, which is modified by the proposed joint localization approach. Since the proposed method operates only on the recovered error-locator polynomials, the complexity of the preceding DFT decoding operations is not included in this comparison. For independent localization, each of the $M$ error-locator polynomials, whose degree is at most $v$, is evaluated at the $N$ evaluation points corresponding to the $N$-th roots of unity. Thus, the computational complexity of independent localization is $O(MNv)$.}

\textcolor{black}{For the proposed joint localization, the $M$ error-locator polynomials are first partitioned into $\mathcal{G}_{v}$ and $\mathcal{G}_{v^{-}}$, containing the degree-$v$ and degree-less-than-$v$ polynomials, respectively. Since the degree of each polynomial is available at the master, this requires $O(M)$ operations. Let $L=|\mathcal{G}_{v}|$. Averaging the $L$ degree-$v$ polynomials requires $O(Lv)$ operations. The resulting set $\mathcal{G}_{\rm joint}=\{\bar{g}_{\max}(z)\}\cup\mathcal{G}_{v^{-}}$ has cardinality $\bar{M}=M-L+1$. Obtaining the roots of the $\bar{M}$ polynomials by evaluating them at the $N$ evaluation points requires $O(\bar{M}Nv)$ operations. Let $v''=|\mathcal{S}_{I}|$ denote the resulting number of locations in $\mathcal{S}_{I}$. The set $\mathcal{S}_{I}$ can be restricted to a subset $\mathcal{S}_{II}\subseteq\mathcal{S}_{I}$ of cardinality $v'\leq v''$, as mentioned in the proposed joint localization approach. Thus, the number of $v$-element subsets considered in the optimization is $\binom{v''}{v}$ when $v'=v''$, and $\binom{v'}{v}$ when $v'<v''$.
For each subset $\mathcal{S}$ and polynomial $t_m(z)\in\mathcal{G}_{\rm joint}$, computing $E_{t_m(z)}(\mathcal{S})$ requires $O(v^2)$ operations, since polynomial evaluation dominates the subsequent magnitude computation and selection of the $d(m)$ smallest values. Therefore, evaluating $\mathcal{E}_{\mathcal{S}}=\sum_{m=1}^{\bar{M}}E_{t_m(z)}(\mathcal{S})$ for one subset requires $O(\bar{M}v^2)$ operations. Consequently, exhaustive evaluation requires $O\!\left(\binom{v''}{v}\bar{M}v^2\right)$ operations when $v'=v''$, and $O\!\left(\binom{v'}{v}\bar{M}v^2\right)$ operations when $v'<v''$.}

\begin{table}[t]
\centering
\caption{\textcolor{black}{Computational complexity of independent and
proposed joint error localization.}}
\label{tab:localization_complexity}
\renewcommand{\arraystretch}{1.1}
\color{black}{
\begin{tabular}{|p{8.5cm}|p{5.5cm}|}
\hline
\centering\textbf{Error Localization method} &
\centering\textbf{Computational complexity}\arraybackslash \\ \hline

Independent localization &
$O(MNv)$ \\ \hline

Proposed joint localization, $L=M$ &
$O(Mv+Nv)$ \\ \hline

Proposed joint localization, $0<L<M$, $v'=v''$ &
$O\!\left(Mv+\bar{M}Nv+
\binom{v''}{v}\bar{M}v^2\right)$ \\ \hline

Proposed joint localization, $0<L<M$, $v'<v''$ &
$O\!\left(Mv+\bar{M}Nv+
\binom{v'}{v}\bar{M}v^2\right)$ \\ \hline

\end{tabular}}
\end{table}

\textcolor{black}{Further, when $L=M$, all recovered error-locator polynomials have degree $v$. In this case, the $M$ polynomials are replaced by their coefficient-wise average, and only the resulting polynomial needs to be evaluated at the $N$ evaluation points. Thus, the proposed joint localization has complexity $O(Mv+Nv)$, compared with $O(MNv)$ for independent localization. When $0<L<M$, the proposed joint localization retains the $M-L$ lower-degree polynomials in addition to the averaged degree-$v$ polynomial, giving $\bar{M}=M-L+1$. Therefore, the overall complexity is $O\!\left(Mv+\bar{M}Nv+\binom{v''}{v}\bar{M}v^2\right)$ when $v'=v''$, and $O\!\left(Mv+\bar{M}Nv+\binom{v'}{v}\bar{M}v^2\right)$ when $v'<v''$. The resulting computational complexities of independent and proposed joint error localization are summarized in Table~\ref{tab:localization_complexity}. In particular, when $L=M$, the proposed joint localization requires only one polynomial evaluation over the $N$ evaluation points after coefficient averaging, compared with $M$ polynomial evaluations for independent localization. When $0<L<M$, the combinatorial search over subsets introduces an additional computational cost, which can be reduced by decreasing the set constraint length from $v''$ to $v'$.}

\subsection{\textcolor{black}{Localization under Estimation of the Number of Errors}}
\label{subsec:localization_estimated_errors}
\textcolor{black}{Note that the preceding sections on the error-localization block, including the independent localization and the proposed joint localization algorithms, assume that the number of errors $A_{\bar{u}\bar{h}}$ is available at the decoder for $\bar{u}\in[u]$ and $\bar{h}\in[h]$. In practice, however, $A_{\bar{u}\bar{h}}$ must also be estimated in the presence of precision noise. Since the Byzantine workers may corrupt different entries of the matrix-valued polynomial, the number of errors can differ across the received codewords. Therefore, the number of errors must be estimated independently for each received codeword. In particular, due to finite precision, the syndrome vector in \eqref{eq:lin eq} is perturbed by the precision noise. In this context, let $\bar{\mathbf{s}}_{\bar{u}\bar{h}}$ denote the resulting perturbed syndrome vector, and let $\bar{\mathbf{S}}_m$ denote the corresponding syndrome matrix constructed from $\bar{\mathbf{s}}_{\bar{u}\bar{h}}$ in the same manner as $\mathbf{S}_m$. Unlike the infinite-precision case, $\bar{\mathbf{S}}_m$ is generally no longer rank deficient because the precision noise introduces additional components into the syndrome. Consequently, its rank cannot be directly used to determine the number of Byzantine errors. Thus, the estimation of the number of errors is also vulnerable to finite-precision perturbations.}


  \begin{figure}[H]
\centering
\includegraphics[scale = 0.3]{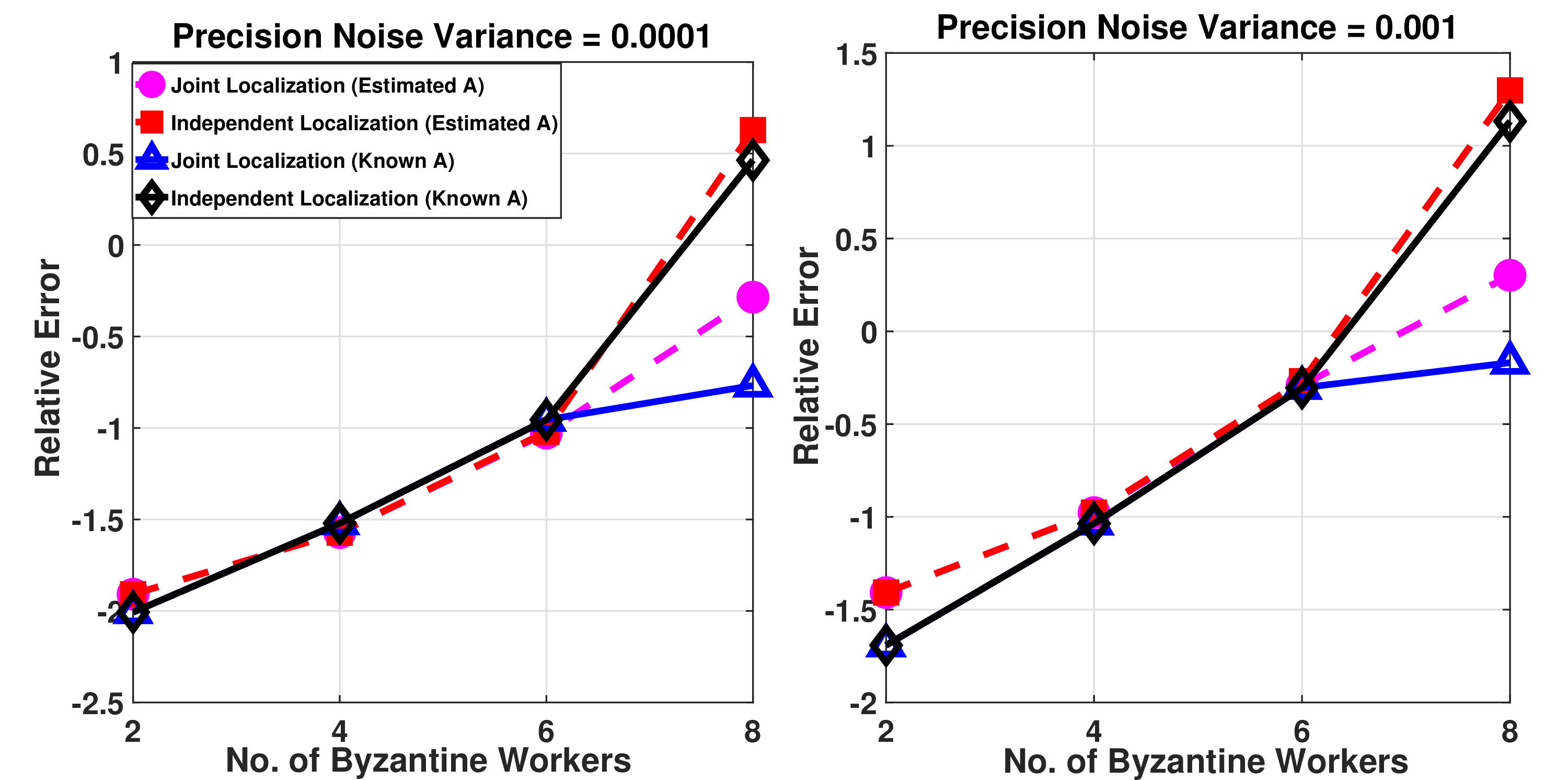}
\caption{\textcolor{black}{Average relative error (in $\log_{10}$ scale) of ALCC with parameters $N=31$, $k=5$, $t=3$, $K=15$, $\beta=1.5$, and $\sigma=10^6$, in the presence of Byzantine workers and for different precision-noise variance.The results are shown for two cases: (i) the number of errors $A$ is estimated at the DFT decoder (Estimated $A$), and (ii) the number of errors $A$ is correctly known at the DFT decoder (Known $A$). Here, the non-zero entries of the adversarial matrices $\{\mathbf{E}_{i_a}\}$ are i.i.d. according to $\mathcal{CN}(10,10^3)$.}}
\label{fig:estimation}
\end{figure}

\textcolor{black}{To examine the practical effect of estimating the number of errors on the subsequent localization and the overall ALCC reconstruction performance, we perform additional Monte Carlo experiments. Specifically, we compare the average relative error of ALCC using the DFT decoder under two settings: (i) the number of errors is correctly known at the DFT decoder, corresponding to the assumption used in the preceding analysis and experiments of our original manuscript, and (ii) the number of errors is estimated independently for each received noisy codeword using the empirical eigenvalue-thresholding procedure considered for comparison in \cite{b6}. A commonly used approach for estimating the number of large errors in DFT codes is to examine the eigenvalues of the syndrome covariance matrix and apply a threshold to distinguish the eigenvalue components associated with large errors from those primarily introduced by small precision noise. Specifically, following the thresholding procedure used in \cite{b6}, we compare the magnitudes of the eigenvalues of the syndrome covariance matrix with a threshold of the form $\gamma v\sigma_q^2$, where $\sigma_q^2$ denotes the precision noise variance and $v=\left\lfloor \frac{N-K}{2}\right\rfloor$. The number of eigenvalues exceeding this threshold is then used as the estimated number of errors. We select $\gamma$ from a
larger range to maximize the estimation performance, consistent with the
experimental comparison procedure in \cite{b6}. We use the same ALCC
setting as in the original experiments, with $N=31$, $K=15$,
$\beta=1.5$, $t=3$, and $\sigma=10^6$, and the non-zero entries of the
adversarial matrices are generated i.i.d. according to
$\mathcal{CN}(10,10^3)$. The resulting average relative errors are shown
in Fig.~\ref{fig:estimation} for two values of precision-noise variance
and varying numbers of Byzantine workers up to the error-correction
capability $v=8$.}

\textcolor{black}{As shown in Fig.~\ref{fig:estimation}, for a small or moderate number of Byzantine workers, the
relative error obtained when the number of errors is estimated remains
close to that obtained when the number of errors is correctly known at
the decoder. However, as the number of Byzantine workers
increases toward the error-correction capability $v=8$, the gap between
the relative error corresponding to the cases of estimated-number and known-number becomes more pronounced,
with a larger gap observed at the higher precision-noise variance.
This indicates that the proposed joint localization method is sensitive to incorrect estimation of the number of errors when $A = v.$ Despite this degradation, Fig.~\ref{fig:estimation} shows that the
proposed joint localization method continues to outperform independent
localization at $A = v$ when both are operated with the same estimated number of
errors. Thus, although the overall reconstruction accuracy degrades in
this regime due to imperfect estimates on the number
of errors, the proposed joint localization method still provides a substantial
improvement over the independent localization method.}

\section{\textcolor{black}{Robust}  ALCC with Profiled Worker Nodes}
\label{sec:optimal allocation}
\textcolor{black}{In contrast to the ALCC setting discussed in Section \ref{sec:secure ALLC}, in this section, we consider a setting where the worker reliability information described in Section \ref{sec:allc with adv} is available at the master, and thereby the master partitions the worker nodes into reliable $\mathcal{W}_{\rm rel}$ and unreliable $\mathcal{W}_{\rm unrel}$ worker nodes. However, we assume that the identities of the Byzantine workers within $\mathcal{W}_{\rm unrel}$ remain unknown to the master. More specifically, in this setting, $\tau>0$ and $\mu>0$, respectively, such that $\tau+\mu=N$ and $\mathcal{W}_{\rm unrel}$ is nonempty.} Furthermore, the case when $\tau\geq(\tilde{D}+1)$ is irrelevant because $(\tilde{D}+1)$ reliable worker nodes can always be chosen for reconstruction. Therefore in this setup, the relevant range for $\tau$ is $0<\tau<\tilde{D}+1$.

Within this framework, we use the same encoding strategy employed in the framework discussed in Section \ref{sec:secure ALLC}, when $\tau=0$. Further, while distributing the shares of the encoded data among the worker nodes any one-to-one mapping i.e., $\psi :[N]\rightarrow [N]$ can be used such that the $i$-th evaluation $\mathbf{U}_{i}$ goes to the worker node $\mathcal{W}_{\psi(i)}$, for $i\in[N]$. However, in the subsequent discussion, we show that in the scenario where  $0<\tau<\tilde{D}+1$, providing the identity mapping i.e., $\psi(i)=i$ may not be optimal in terms of relative error of the framework. 
In the reconstruction stage of ALCC, when using  the DFT decoder with $\tau>0$, we modify the block of error localization, particularly while computing the roots of the error locator polynomial $\bar{g}_{\bar{u}\bar{h}}(z)$. In this context, recall that the Byzantine worker nodes are a subset of the set of unreliable worker nodes. This implies that the evaluation of $\bar{g}_{\bar{u}\bar{h}}(z)$ should be restricted to those $N$-th roots of unity corresponding to the indices of unreliable worker nodes. Let the set of indices corresponding to $\mu$ unreliable worker nodes be $\mathcal{U}$ = $\{i_{u_{1}},i_{u_{2}},\ldots, i_{u_{\mu}}\}$, where $\mathcal{U}\subseteq [N]$ and $|\mathcal{U}|=\mu$, such that the indices of the Byzantine worker nodes ${\mathcal{L}} = \{{i}_1,{i}_2, \ldots, {i}_A\}\subseteq \mathcal{U}$. Formally, in this variant of ALCC framework, with one-to-one mapping, we suggest evaluating $\bar{g}_{\bar{u}\bar{h}}$ at those $N$-th roots of unity corresponding to the indices present in the set $\mathcal{U}$ instead of all $N$ indices in the set $[N]$. More specifically, we evaluate $||\bar{g}_{\bar{u}\bar{h}}(X_{q})||^2$, where $X_{q} =\gamma^{q-1}$ and $q\in \mathcal{U}$. We then arrange the evaluations in the ascending order in order to obtain $$||\bar{g}_{\bar{u}\bar{f}}(X_{\hat{i}_{u_{1}}})||^2 \leq||\bar{g}_{\bar{u}\bar{f}}(X_{\hat{i}_{u_{2}}})||^2\leq \ldots \leq||\bar{g}_{\bar{u}\bar{f}}(X_{\hat{i}_{\mu}})||^2,$$ where \{$\hat{i}_{u_{1}},\hat{i}_{u_{2}}, \ldots, \hat{i}_{u_{\mu}}\}$ is the ordered set based on the evaluations. Subsequently, the $A$ smallest evaluations and its corresponding set of indices $ \hat{\mathcal{L}} = \{\hat{i}_{u_{1}},\hat{i}_{u_{2}}, \ldots, \hat{i}_{u_{A}}\}$ are treated as the detected error locations. Further, the following proposition captures the error rate of localization for ALCC with $\tau>0$.

\begin{proposition}
    For an ALCC framework with $N$ worker nodes, where $\mu>0$, $0<\tau<\tilde D+1$, $\mathcal{U}$, and $A$ Byzantine worker nodes with $\mathcal{L}$, an upper bound on error rate of localization can be expressed as
\begin{IEEEeqnarray}{rcl}
\label{eq:error rate tau>0}
 \sum_{i=1}^{\mu-A} \sum_{a=1}^{A}  \text{Prob}\bigg(  ||\bar{g}_{\bar{u}\bar{f}}(X_{i_{u_{i}}})||^2 \leq ||\bar{g}_{\bar{u}\bar{f}}(X_{i_{a}})||^2 \bigg),
\end{IEEEeqnarray}
 where $\mathcal{U} = \{i_{u_{1}},i_{u_{2}},\ldots, i_{u_{\mu}}\}$, and $\mathcal{L}=\{i_{1},i_{2},\ldots,i_{A}\}$.
\end{proposition}
\begin{IEEEproof}
This bound can be obtained along the similar lines of analysis in Section \ref{subsec:independent localization}, except that the index of the erroneous detected location belongs to $\mathcal{U}$, instead of $[N] \backslash \mathcal{L}$. 
 \end{IEEEproof}
Since the indices of Byzantine worker nodes are chosen randomly over  $\mathcal{U}$, for a given $\mu>0$, $0<\tau<\tilde D+1$ and $A$ Byzantine worker nodes, an upper bound on average error rate of localization, denoted by $\Bar{P}_{error}= \mathbb{E}_{\mathcal{L}\subseteq \mathcal{U}} \big[P_{error}\big]$, is

\begin{small}
\begin{IEEEeqnarray}{rcl}
\label{eq:pl3}
    \mathbb{E}_{\mathcal{L}\subseteq \mathcal{U}}\bigg[\sum_{i=1}^{\mu-A} \sum_{a=1}^{A}  \text{Prob}\bigg(||\bar{g}_{\bar{u}\bar{f}}(X_{i_{u_{i}}})||^2 \leq ||\bar{g}_{\bar{u}\bar{f}}(X_{i_{a}}||^2 \bigg)\bigg].
\end{IEEEeqnarray}
\end{small}

In the next section, we discuss the possibility of identifying a customized mapping $\psi(\cdot)$ in order to minimize the above average error rate of localization. 

\subsection{Assignment of Shares Based on Error Rate of Localization}
\label{subsec:error rate based assignment}
Consider an ALCC setting where the unreliable worker nodes are such that $\mathcal{U}$ consists of consecutive indices, such as $\{i,i+1,\ldots,i+\mu-1\}$, for $i\in\{1,2,\ldots,N-\mu+1\}$, and the identity mapping is used for $\psi(\cdot)$ such that $\mathbf{U}_i$ is assigned $\mathcal{W}_{i}$, for all $i \in [N]$. This implies that the unreliable worker nodes receive evaluations of the encoding polynomial $l(z)$ using the $N$-th roots of unity that are contiguous on the unit circle. As a result, this set of contiguous $N$-th roots of unity is used as candidate solutions when solving for the roots of the error locator polynomial $\bar{g}_{\bar{u}\bar{h}}(z)$ at the DFT decoder. Further, since the roots of $\bar{g}_{\bar{u}\bar{h}}(z)$ are computed by ordering its evaluations due to precision noise, an unreliable node, which is not a Byzantine worker node, may be wrongly detected as Byzantine due to its proximity to a true Byzantine node. Thus, using contiguous $N$-th roots of unity to the set of unreliable worker nodes may degrade the average error rate of localization. To address this problem, we believe that the subset of $N$-th roots of unity for assigning the evaluations to the unreliable workers must be carefully chosen at the master node. In this context, let $\mathcal{P} \subseteq [N]$ be the set of $\mu$ distinct indices used to pick the roots of unity for the evaluation of $l(z)$. Using this, we can construct the mapping $\psi(\cdot)$ such that for each index $j \in \mathcal{U}$, there exists a unique $i \in \mathcal{P}$ with $\psi(i) = j$. This ensures that, the unreliable worker node $\mathcal{W}_j$ receives $\mathbf{U}_i$, where $i\in\mathcal{P}$. The remaining evaluations $\{\mathbf{U}_i : i \notin \mathcal{P}\}$ can be assigned to the reliable worker nodes $\mathcal{W}_{{rel}}$ in an arbitrary manner. Note that the master node has a total $\binom{N}{\mu}$ ways of selecting the set $\mathcal{P}$. Therefore, to achieve a lower error rate of localization, one can choose the optimal set $\mathcal{P}^{*}\subseteq [N]$, which minimizes $\Bar{P}_{error}$ in \eqref{eq:pl3}. However, since there is no closed form expression for $\Bar{P}_{error}$, in the rest of the section, we present a lower bound on it and use this as an objective function to formulate a minimization problem.

Note that any two pairs in the expression  of $\Bar{P}_{error}$ in \eqref{eq:pl3} can be lower bounded using Theorem \ref{thm:corr}. Therefore, instead of using $\Bar{P}_{error}$ as the objective function, its lower bound, which is obtained by applying the results of Theorem \ref{thm:corr}, can be used to formulate an optimization problem as

\begin{small}
  \begin{IEEEeqnarray}{rcl}\
\label{eq:surro_objective}
\arg\min_{\substack{\mathcal{P} \subseteq [N] \\ \text{s.t.} |\mathcal{P}|=\mu}}  \Bigg\{ \mathbb{E}_{\mathcal{L}\subseteq \mathcal{P}} \Bigg[ \sum_{i=1}^{\mu-A} \sum_{a=1}^{A}  
    \frac{\kappa_{ia}}{(1+\kappa_{ia})} e^{-\frac{\eta H(\mathcal{L},i)\kappa_{ia}}{4\sigma^2_p(1+\kappa_{ia})}} \Bigg]\Bigg\},
  \end{IEEEeqnarray}
\end{small}

\noindent where $\mathcal{P}=\{j_{u_{1}},j_{u_{2}},\ldots, j_{u_{\mu}}\}\subseteq[N]$, denotes the set of evaluation indices for the unreliable worker nodes, and
\begin{IEEEeqnarray}{rcl} 
\frac{\kappa_{ia}}{(1+\kappa_{ia})}= \frac{2}{\eta \sum_{l = 1}^{A}\big( 1 - \mbox{cos}\big(l\frac{2\pi}{N}|j_{u_{i}}-i_{a}|\big)\big)},
\end{IEEEeqnarray}
\begin{IEEEeqnarray}{rcl}
\label{eq:f(L,i)}
H(\mathcal{L},i)=\big|\big|\prod_{a=1}^{A}(j_{u_{i}}-i_{a})\big|\big|^2.
\end{IEEEeqnarray}
Here, $i_{a}\in\mathcal{L}$ for $a\in[A]$ denotes the evaluation indices for the Byzantine worker nodes, such that $\mathcal{L}\subseteq \mathcal{P}$. Given that the objective function in \eqref{eq:surro_objective} is a tractable expression, the corresponding minimization problem can be solved through computational tools. 

Further, we notice that instead of using $(\mu - A)A$ terms to compute the objective function, we can lower bound it again using the bound $\gamma_{min}\leq \frac{\kappa_{ia}}{(1+\kappa_{ia})}\leq \gamma_{max}$, $\forall i, a$, to formulate the following optimization problem 

\begin{small}
\begin{IEEEeqnarray*}{rcl}
\label{eq:bound3}
\arg\min_{\substack{\mathcal{P} \subseteq [N] \\ \text{s.t.} |\mathcal{P}|=\mu}}\mathbb{E}_{\mathcal{L}\subseteq \mathcal{P}} \Bigg[ \sum_{i=1}^{\mu-A} \gamma_{min}  e^{-\frac{\eta H(\mathcal{L},i)\gamma_{max}}{4\sigma^2_p}} \Bigg],
\end{IEEEeqnarray*}
\end{small}

\noindent where $\gamma_{min}$ and $\gamma_{max}$ are constants, which can be computed offline. Although the above problem can be solved using numerical methods, we propose a variant of it in Problem \ref{optprob:11}, wherein the dominant term of the summation is used instead of the sum of the $\mu - A$ terms. Thus, for a given choice of $N$, $A$, $\mu>0$, $0<\tau<\tilde D+1$, $\sigma_{p}^2$, $\eta$, the master node in ALCC can obtain $\mathcal{Q}^{*}$ by solving Problem \ref{optprob:11}, and then use it for assigning the shares among the unreliable worker nodes.

\begin{mdframed}
 \begin{problem}
 \label{optprob:11}
 For ALCC with $\mu>0$, $0<\tau<\tilde D+1$, and $A$ Byzantine worker nodes, solve
\begin{IEEEeqnarray*}{rcl}
    \mathcal{Q}^{*} = \arg\min_{\substack{\mathcal{P} \subseteq [N] \\ \text{s.t.} |\mathcal{P}|=\mu}}{\mathbb{E}_{\mathcal{L}\subseteq \mathcal{P}}}
\Bigg[ \max_{i\in \{1,2,\ldots, \mu-A\}} e^{-\frac{\eta H(\mathcal{L},i)\gamma_{max}}{4\sigma^2_p}} \Bigg].
\end{IEEEeqnarray*}
 \end{problem}
  \end{mdframed}



\subsection{Experimental Results}
\label{subsec:experimenatal_placement}

\begin{figure}[H]
\centering
\includegraphics[scale = 0.3]{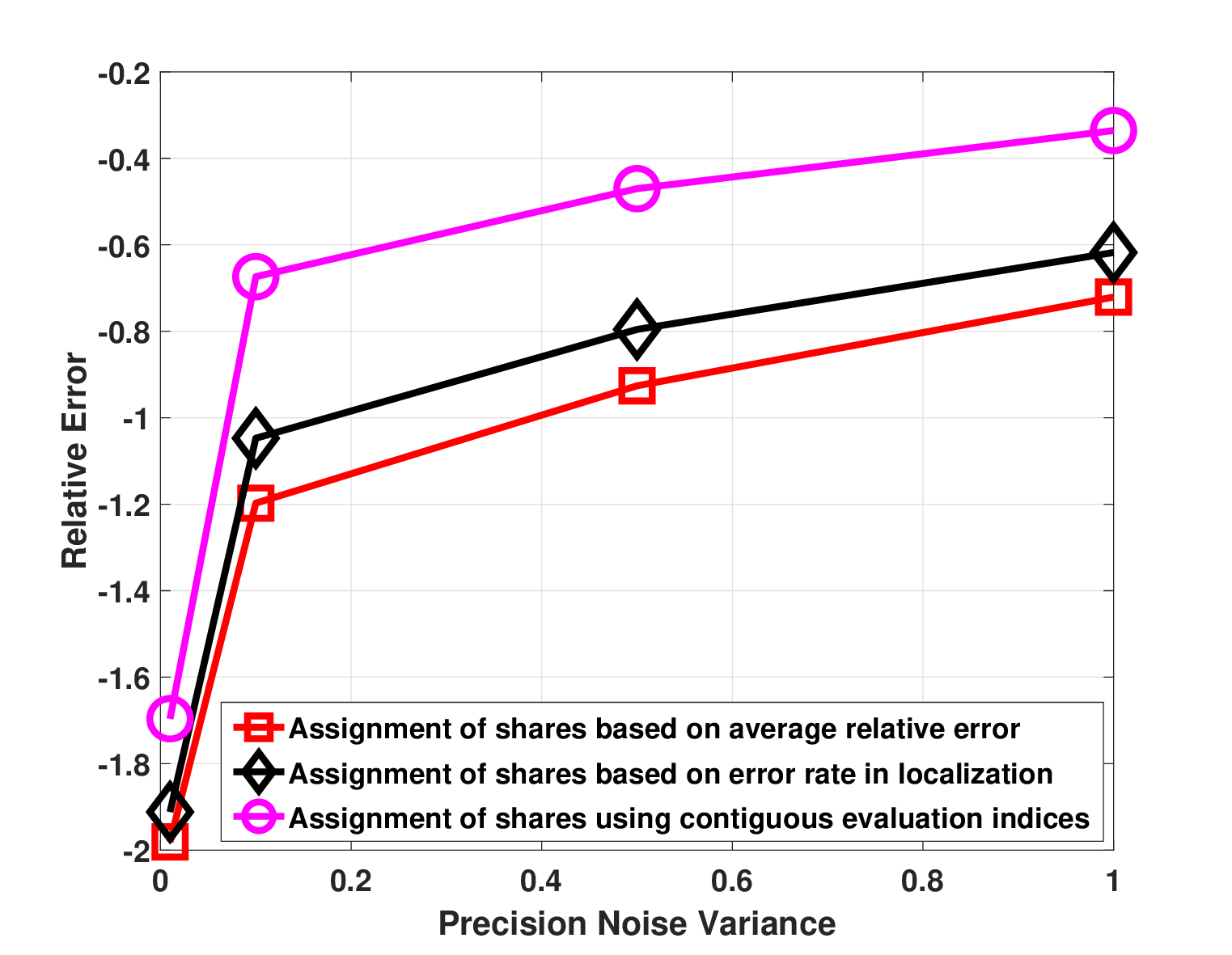}
\vspace{-0.1cm}
\caption{Average relative error (in $\log_{10}$ scale) when (i) indices are obtained by minimizing the average relative error of the framework, (ii) indices are obtained by solving Problem \ref{optprob:11}, and (iii) using contiguous indices.} 
\label{fig:accuracy of optimal vs suboptimal}
\end{figure}

In order to showcase the effectiveness of our proposed method for assigning evaluation indices, we consider a baseline method wherein the set of evaluation indices for unreliable worker nodes is chosen by minimizing the average relative error of the ALCC framework. The solution to such a problem is denoted by $\mathcal{R}^*$. Note that this method is computationally infeasible as there is no closed form expression on relative error as a function of the system parameters, and therefore, experiments must be conducted to evaluate the relative error for each $\mathcal{P}$. Further, we conduct experiments within ALCC to obtain two sets of evaluation indices for unreliable worker nodes, one using the baseline approach, and the other by solving Problem \ref{optprob:11}. Further, we compute the average relative error using the index sets from both methods, at their respective precision noise variances as presented in Table \ref{tab:optimal and suboptimal}. Additionally, we perform experiments when the evaluation indices assigned to the unreliable worker nodes are contiguous. Specifically, we compute the average relative error at different precision noise variances, while computing the function $\mathbf{f(X)}=\mathbf{X}^T\mathbf{X}$, such that $\mathbf{X}\in \mathbb{R}^{20\times 5}$, using i) contiguous index sets, and ii) the index sets $\mathcal{R}^*$ and $\mathcal{Q^*}$ from Table \ref{tab:optimal and suboptimal}. Note that the average relative error is computed over $10^3$ iterations, where the Byzantine worker nodes are chosen uniformly at random over the the set of $\mu$ unreliable worker nodes in each iteration.
The parameters used for the experiments are  $N = 11,  k=3, t = 1, \beta = 1.5,  \mu=5, A=2, K = 7$, $\eta=10$, $\sigma = 10^6$. Here, the non-zero entries of noise matrices $\{\mathbf{E}_{i_{a}}\}$ are i.i.d. as $\mathcal{CN}(10, 10^3)$. These results are presented in Fig. \ref{fig:accuracy of optimal vs suboptimal}. As illustrated in Fig. \ref{fig:accuracy of optimal vs suboptimal}, assigning contiguous evaluation indices to the unreliable worker nodes leads to higher relative error compared to our proposed method of assigning the evaluation indices and the method that aims to minimize the average relative error of ALCC. Furthermore, the plots in Fig. \ref{fig:accuracy of optimal vs suboptimal} confirm that our proposed method of assigning evaluation indices to the unreliable worker nodes based on minimizing error rate of localization is not significantly away from the method that aims to minimize the average relative error, in terms of accuracy of the framework.

\begin{table}[htbp]
\centering
\begin{small}
\begin{tabular}{|l|l|l|ll}
\cline{1-3}
\begin{tabular}[c]{@{}l@{}}Precision \\ Noise Variance\end{tabular} & \begin{tabular}[c]{@{}l@{}}$\mathcal{R}^{*}$\end{tabular} & \begin{tabular}[c]{@{}l@{}}$\mathcal{Q}^{*}$\end{tabular} &  &  \\ \cline{1-3}
1                                                                   & 1, 2, 5, 8 ,9                                                & 1, 3, 6, 9, 11                                                  &  &  \\ \cline{1-3}
0.5                                                                 & 1, 4, 5, 8, 9                                                & 1, 3, 6, 9, 11                                                  &  &  \\ \cline{1-3}
0.1                                                                 & 2, 5, 6, 9, 10                                               & 1, 3, 6, 8, 11                                                  &  &  \\ \cline{1-3}
0.01                                                                & 2, 4, 6, 8, 11                                               & 1, 2, 4, 5, 7                                                   &  &  \\ \cline{1-3}
\end{tabular}
\end{small}
\caption{Evaluation indices for unreliable worker nodes obtained through minimizing the relative error of the framework and error rate of localization. Parameters used for the experiments are  $N = 11,  k=3, t = 1, \beta = 1.5,  \mu=5, A=2, K = 7$, $\eta=10$, $\sigma = 10^6$.}
\label{tab:optimal and suboptimal}
\end{table}

When using ALCC with $\tau>0$, note that the joint localization method in Section \ref{subsec:joint loc}, can further reduce the error rate of localization compared to independent localization. 

\begin{remark}
\textcolor{black}{Note that, to obtain the sets $\mathcal{R}^{*}$ and $\mathcal{Q}^{*}$ corresponding to the unreliable workers, as presented in Table~\ref{tab:optimal and suboptimal}, we use smaller system parameters than those in Figs.~\ref{Fig: alcc in adv plot}, \ref{Fig: ind localization}, and \ref{Fig: Joint localization}. This is because evaluating the baseline solution $\mathcal{R}^{*}$ requires computing the average relative error of ALCC through Monte Carlo simulations of the complete ALCC framework for each of the $\binom{N}{\mu}$ candidate worker subsets. For each subset, we perform $10^3$ Monte Carlo iterations, where the $A$ Byzantine workers are selected uniformly at random from the $\mu$ unreliable workers in each iteration. Thus, the computational cost grows exponentially with the system size.}
\end{remark}

\section{\textcolor{black}{Robust}  ALCC with Colluding Byzantine Worker Nodes}
\label{sec:colluding_attacks}
In this section, we address some potential collusion attacks by the Byzantine worker nodes on ALCC. For exposition, we consider an ALCC setting comprising $N$ worker nodes, where $\tau=0$, such that $\mathcal{W}=\mathcal{W}_{unrel}$, implying that all the $N$ worker nodes are unreliable.\footnote{However, these attacks are also applicable to scenarios when $\tau>0$.} Therefore, we adopt the same encoding strategy, identity mapping for distributing the shares, and the same reconstruction strategy as discussed in Section \ref{sec:secure ALLC} for the ALCC setting with $\tau=0$. However, in contrast to the scenario discussed in Section \ref{sec:secure ALLC}, in this setting we assume the presence of $v$ Byzantine worker nodes, where $v= \lfloor \frac{N-K}{2}\rfloor$, denotes the error correcting capability of the DFT decoder. Further, we assume that the $v$ Byzantine worker nodes have access to a dedicated colluding channel for communication.

Before proposing an optimized framework for colluding attacks, we look at a possible strategy through which the Byzantine worker nodes can inject noise matrices in order to maximize relative error of ALCC. Since there is no closed-form expression on relative error as a function of system parameters, we believe that the upper bound on the error rate of localization specified in \eqref{eq:union bound2} can be used as an objective function for maximization from the perspective of the Byzantine worker nodes. Although the nature of collusion is broadly applicable to decide the noise matrices $\mathbf{E}_{i_{1}}, \mathbf{E}_{i_{2}}, \ldots, \mathbf{E}_{i_{v}}$, we focus primarily on a scenario where the Byzantine worker nodes collude to decide on the sparsity of their noise matrices. This is because, the sparsity of the noise matrices will decide the degree of the error locator polynomials, which in turn affects the error rate localization. Therefore, once the sparsity and the positions of the non-zero entries are chosen, we assume that non-zero entries are injected on those positions in a statistically independent manner.

Recall that for each $\mathbf{E}_{i_{a}}$, such that $a \in [v]$, $\mathbf{B}_{i_{a}} \in \{0, 1\}^{u \times h}$ specifies the binary matrix that captures the positions of the non-zero entries of each $\mathbf{E}_{i_{a}}$. Therefore, before formally addressing the problem on selecting the base matrices $\{\mathbf{B}_{i_{1}}, \mathbf{B}_{i_{2}}, \ldots, \mathbf{B}_{i_{v}}\}$ in order to maximize the error rate of localization, Proposition \ref{prop:all_one_matrix} highlights an important consideration on how not to choose them.  More specifically, from Proposition \ref{prop:all_one_matrix}, if $\mathbf{B}_{i_{a}} = \mathbf{J}_{u \times h}$, $\forall a \in [v]$, then the master node receives a total of $M=u\times h$ polynomials of degree $v$ which can be used to reduce the effective variance of precision noise across the coefficients of error locator polynomials, thereby minimizing the error rate of localization. Therefore, to counter this, Byzantine worker nodes must select their base matrices in order to forbid the master node from averaging the coefficients of the error locator polynomials. Therefore, to define a problem statement on the choice of the base matrices $\{\mathbf{B}_{i_{1}}, \mathbf{B}_{i_{2}}, \ldots, \mathbf{B}_{i_{v}}\}$, we first define the effective base matrix, denoted by $\mathbf{B}_{eff} \in \{0, 1\}^{uh \times v}$, as 

\begin{small}
\begin{equation}
\mathbf{B}_{eff} = \begin{bmatrix}
\mathbf{B}_{i_{1}}(1,1) & \mathbf{B}_{i_{2}}(1, 1) & \ldots & \mathbf{B}_{i_{v}}(1, 1)\\
\mathbf{B}_{i_{1}}(1,2) & \mathbf{B}_{i_{2}}(1, 2) & \ldots & \mathbf{B}_{i_{v}}(1, 2)\\
\vdots & \vdots & \vdots & \vdots\\
\mathbf{B}_{i_{1}}(1,h) & \mathbf{B}_{i_{2}}(1, h) & \ldots & \mathbf{B}_{i_{v}}(1, h)\\
\mathbf{B}_{i_{1}}(2,1) & \mathbf{B}_{i_{2}}(2, 1) & \ldots & \mathbf{B}_{i_{v}}(2, 1)\\
\vdots & \vdots & \vdots & \vdots\\
\mathbf{B}_{i_{1}}(u,h) & \mathbf{B}_{i_{2}}(u, h) & \ldots & \mathbf{B}_{i_{v}}(u, h)\\
\end{bmatrix}.
\end{equation}
\end{small}

The above representation is such that the rows represent the $M$ error locator polynomials, and the number of ones in a given row captures the degree of the corresponding error locator polynomial. Note that, for $1 \leq a \leq v$,  the $a$-th column of $\mathbf{B}_{eff}$ matrix corresponds to the entries of base matrices $\mathbf{B}_{i_{a}}$ of $a$-th Byzantine worker which in turn implies that $a$-th column of $\mathbf{B}_{eff}$ matrix is completely in control of the $a$-th Byzantine worker, but not the elements across the columns. As a result, the Byzantine worker nodes must collude in order to control the number of ones in a row, which in turn controls the degree of the error locator polynomials. Given that the master node has a total of $M$ error locator polynomials captured by the rows of $\mathbf{B}_{eff}$, each of degree at most $v$, the relevant objective function from the perspective of Byzantine worker nodes is average error rate of localization across all $M$ error locator polynomials. With this background, a problem for the Byzantine workers is to choose $\mathbf{B}_{eff}$, such that average error rate of the localization across $M$ error locator polynomials is maximized. Based on this, we present Problem \ref{opt:B_{eff1}}.

\begin{mdframed}
\begin{problem}
\label{opt:B_{eff1}}
    For a given $\sigma_{p}^2$, $N$, $M$ and $v$, solve\\    
    \begin{IEEEeqnarray}{rcl}
    \label{eq: opt B_eff}
    \mathbf{B^*}_{eff} = \arg \max_{\mathbf{B}_{eff} \in \{0, 1\}^{uh \times v}} \frac{1}{M }\left(\sum _{c=1}^{M}P_{error_{c}}\right) 
    \end{IEEEeqnarray}
     wherein $P_{error_{c}}$ denotes the error rate of localization for $c$-th error locator polynomial captured by $c$-th row of $\mathbf{B}_{eff}$, where $1\leq c\leq M.$
\end{problem}
 \end{mdframed}

Although the master node can either perform independent localization or joint localization in order to solve for the roots of $M$ error locator polynomials, the combinatorial nature of Problem \ref{opt_Problem}, makes it intractable to derive an exact expression on the error rate of joint localization for various set constraint lengths. To address this, we target an analytically-tractable variant of the joint localization step, and show that the Byzantine worker nodes can choose $\mathbf{B^*}_{eff}$, which maximizes the error rate of  joint localization method, wherein independent localization is performed on each polynomial in $\mathcal{G}_{joint}$, after averaging all the $v$ degree polynomials. 

Furthermore, the error rate of localization $P_{error_{c}}$ for $c$-th error locator polynomial, appearing on the right hand side of the expression in \eqref{eq: opt B_eff} can be upper bounded using the expression specified in \eqref{eq:union bound2}, where $c\in [M]$. Specifically, if the error locator polynomial corresponding to the $c$-th row of $\mathbf{B}_{eff}$ has degree $A_{c}\leq v$, then the bound on the error rate in \eqref{eq:union bound2}, involves a total of $A_{c}(N - A_{c})$ pairwise error probabilities terms denoted by, $PEP_{j_{b}i_{a}}$ between each index $i_{a}\in \mathcal{L}$ and an index $j_{b}\in \mathcal{V}$. However, to simplify the bound analytically, we consider only the most dominant pairwise error term for each index $i_{a}\in\mathcal{L}$, i.e., the term with maximum $PEP_{j_{b}i_{a}}$ over all $j_{b}\in \mathcal{V}$, instead of accounting all $N-A_{c}$ terms in $\mathcal{V}$. More specifically, we consider the dominant event when $j_{b}$ is the nearest neighbour of $i_{a}$, i.e., $j_{b}=i_{a}+1$ modulo $N$ or $j_{b}=i_{a}-1$ modulo $N$. As a result, the upper bound on error rate of localization can be approximated as $2 \times \sum_{a=1}^{A_{c}} PEP_{dom}(i_{a})$, where $PEP_{dom}(i_{a})$ represents the pairwise error probability between the index $i_{a}$ and its nearest neighbour. Also, the factor of $2$ arises due to the symmetry in the $N$-th roots of unity. 
In this context, to keep our analysis tractable we consider the most dominant pair among these $A_{c}$ distinct pairs. Hence, we approximate the upper bound on error rate of localization in \eqref{eq:union bound2} for $c$-th error locator polynomial of degree $A_{c}$ as
 \begin{IEEEeqnarray}{rcl}
 \label{eq:error rate of M polynomial}
 P_{error_{c}} \leq A_{c}\times PEP_{dom}(i_{a}^{*},c),
\end{IEEEeqnarray}
 where $PEP_{dom}(i_a^*, c) = \max_{1 \leq a \leq A_c} {PEP}_{dom}(i_a)$, and $1 \leq c\leq M$. However, since the right-hand side of the error rate expression in \eqref{eq:error rate of M polynomial} is not analytically tractable, we use the lower bound provided in Theorem \ref{thm:corr} as a surrogate function for the pairwise error probability term $PEP_{dom}(i_{a}^{*},c)$. Therefore, substituting the lower bound from Theorem \ref{thm:corr} into \eqref{eq:error rate of M polynomial}, error rate of localization can be computed for each individual error locator polynomial of degree $A_{c}\leq v$ captured by the rows of $\mathbf{B}_{eff}$. Subsequently, the average error rate of localization across all $M$ error locator polynomials can be obtained which serves as an analytically tractable objective function for solving the Problem \ref{opt:B_{eff1}}. Hence, instead of solving Problem \ref{opt:B_{eff1}}, we propose to solve Problem \ref{opt:B_{eff}}.

  \begin{mdframed}
\begin{problem}
\label{opt:B_{eff}}
    For a given $\sigma_{p}^2$, $N$, $M$ and $v$, solve\\
    \begin{small}
    \begin{IEEEeqnarray}{rcl}
    \label{eq: opt B_eff_modify}
    \mathbf{B^{\dagger}}_{eff} = \arg \max_{\{0, 1\}^{uh \times v}} \frac{1}{M }\left(\sum _{c=1}^{M} A_{c} \times PL_{(i_{a}^{*},i_{a}^{*}+1,c)}\right)
    \end{IEEEeqnarray}
    \end{small}
    where $PL_{(i_{a}^{*},i_{a}^{*}+1,c)}$ denotes the lower bound on pairwise error probability between the dominant
    index $i_{a}^{*}$ among $A_{c}$ distinct terms and its nearest neighbour $i_{a}^{*}+1$ for $c$-th error locator polynomial captured by $c$-th row of $\mathbf{B}_{eff}$, where $1\leq c\leq M.$
\end{problem}
 \end{mdframed}

Given that the objective function of Problem \ref{opt:B_{eff}} is available in closed-form, the Byzantine workers can obtain $\mathbf{B^{\dagger}}_{eff}$ using numerical methods. However, to reduce the computational complexity, we believe that the following lemma can be used to directly identify the number of ones in a row of $\mathbf{B^{\dagger}}_{eff}$.

 
\begin{lemma}
\label{lemma:inc_A}
For a given $N$, when $\sigma^{2}_{p}$ is small, for an error locator polynomial of degree $A_{c}$, such that $A_{c}\leq v$, captured by a row of $\mathbf{B}_{eff}$, the expression $A_{c} \times PL_{(i_{a}^{*},i_{a}^{*}+1,c)}$ in \eqref{eq: opt B_eff_modify} is a non-decreasing function of $A_{c}$.
\end{lemma}

\begin{IEEEproof}
   We refer the readers to the proof given in  Appendix \ref{proof:4}.\\
\end{IEEEproof}

Although Lemma \ref{lemma:inc_A} suggests to choose an all-one vector for a row of $\mathbf{B}_{eff}$, choosing all-one vectors for multiple rows of $\mathbf{B}_{eff}$ can aid the master node to average the corresponding error-locator polynomials thereby reducing the error rate of localization. Therefore, the following proposition uses Proposition \ref{prop:all_one_matrix} and Lemma \ref{lemma:inc_A} to present a desired structure on $\mathbf{B}_{eff}$.

\begin{proposition}
\label{prop:strog_coll}
When $\sigma_{p}^{2}$ is small, a bound on the objective function in \eqref{eq: opt B_eff_modify} can be maximized by (i) picking the all-one row as the first row of $\mathbf{B}_{eff}$, (ii) picking $v-1$ ones at random positions with uniform distribution for the other rows.  
\end{proposition}

\begin{IEEEproof}
    We refer the readers to the proof given in  Appendix \ref{proof:5}.\\
\end{IEEEproof}
The strategy proposed in Proposition \ref{prop:strog_coll} ensures that only one error locator polynomial out of $M$ polynomials has the maximum degree $v$, whereas the rest of the $M - 1$ polynomials have degree $v-1$. Since degree $v-1$ polynomials can have $v-1$ distinct locations of the $v-1$  Byzantine workers, the master node, without exact information on the locations, cannot perform the averaging operation among them. 
We highlight that $\mathbf{B}_{eff}$ obtained from Proposition \ref{prop:strog_coll} may not give us $\mathbf{B^{\dagger}}_{eff}$, however, it is an effective way to directly generate the base matrices without computational burden.

In the rest of this section, we propose two strategies using which the Byzantine workers can locally generate their base matrices, satisfying a global structure on $\mathbf{B}_{eff}$. In particular, since their colluding channels have limited capacity, the proposed variants are based on the strength of the colluding channels.

\subsection{Strongly Colluding Attack}
\label{subsec:strong colluding}
In this threat model, we assume that the Byzantine workers have access to a high-capacity colluding channel in order to jointly arrive at $\mathbf{B}_{eff}$. In particular, we assume that one of the Byzantine worker nodes, is a master Byzantine worker, which possesses the knowledge of all the underlying parameters of ALCC. Subsequently, it chooses $\mathbf{B}_{eff}$ appropriately (this could be a solution of Problem \ref{opt:B_{eff}} or Proposition \ref{prop:strog_coll}) and distributes it to the other Byzantine worker nodes. Finally, all the Byzantine workers design the entries of their noise matrices $\mathbf{E}_{i_{1}}, \mathbf{E}_{i_{2}}, \ldots, \mathbf{E}_{i_{v}}$ according to the columns of $\mathbf{B}_{eff}$.

\begin{Definition}
An attack model is referred to as a strongly colluding model when the Byzantine worker nodes explicitly communicate $\mathbf{B}_{eff}$ in their colluding channel.
\end{Definition}

\begin{figure}[ht!]
    \centering
    \begin{subfigure}[b]{0.49\linewidth}
        \centering
        \includegraphics[width=\linewidth]{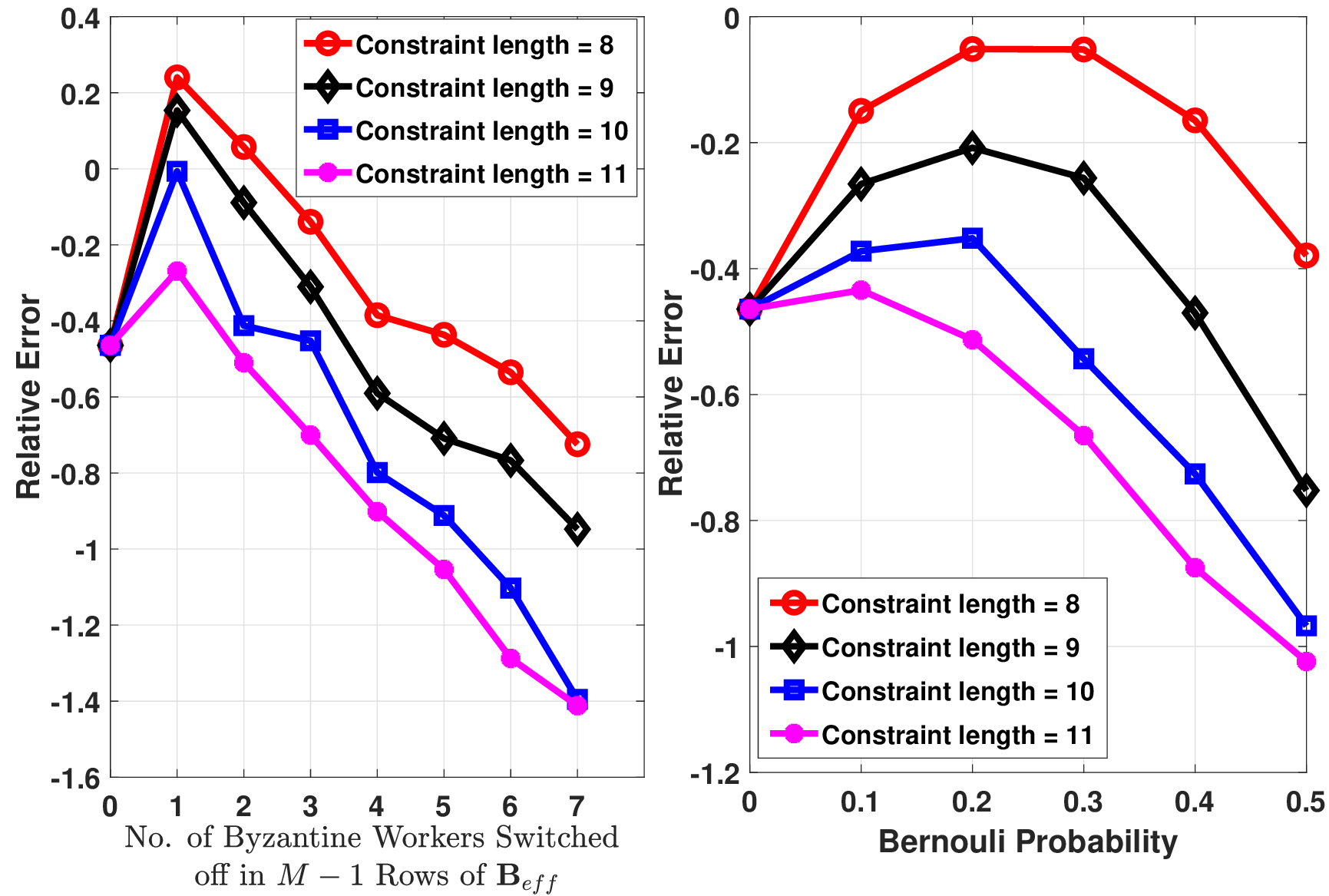}
        \caption{}
    \end{subfigure}
    \hfill
    \begin{subfigure}[b]{0.49\linewidth}
        \centering
        \includegraphics[width=\linewidth]{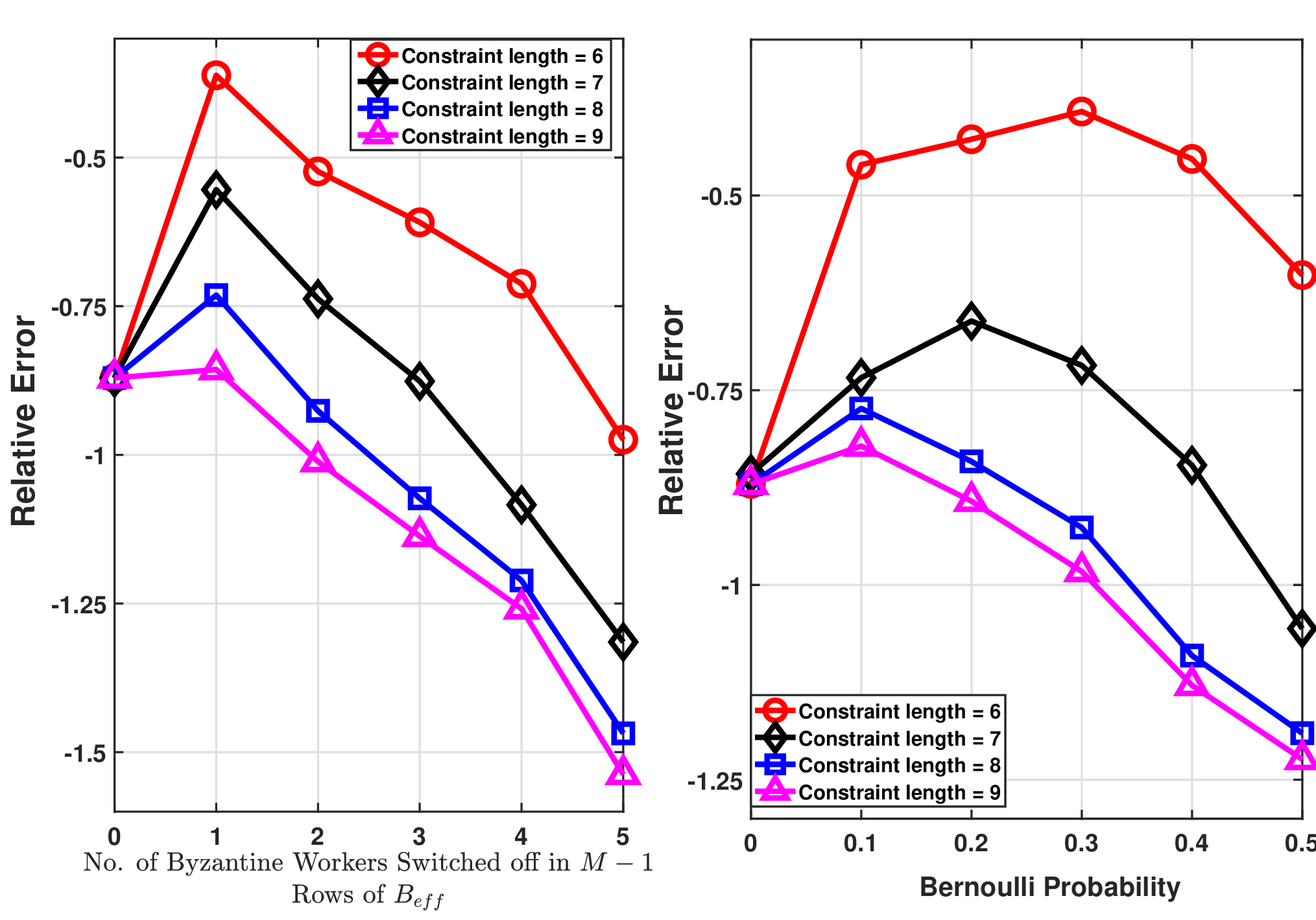}
        \caption{}
    \end{subfigure}

    \caption{\textcolor{black}{Average relative error (in $\log_{10}$ scale) of ALCC against
    strongly and weakly colluding attacks. In (a),
    $(N,K)=(31,15)$ and $v=8$, in (b),
    $(N,K)=(27,15)$ and $v=6$. In both cases,
    $\tau=0$, $\beta=1.5$,  $k=5$, $t=3$, and $\sigma=10^6$, and the
    non-zero entries of the noise matrices
    $\{\mathbf{E}_{i_a}\}$ are generated i.i.d. according to
    $\mathcal{CN}(10,10^3)$.}}
    \label{fig:colluding}
\end{figure}
For the scenarios when the colluding channel between the Byzantine worker nodes cannot support high communication overhead, we propose an alternative model, wherein the master Byzantine worker broadcasts a \emph{common seed}, enabling the other Byzantine workers to generate their column of $\mathbf{B}_{eff}$.
\subsection{Weakly Colluding Attack}
\label{subsec:weak colluding}
Unlike the strongly colluding attack, in this model, leveraging its knowledge of the ALCC parameters the master Byzantine worker  selects a probability value $p \in (0, 1)$ instead of $\mathbf{B}_{eff}$. Subsequently, a suitably quantized version of $p$ is broadcast to the other Byzantine workers, ensuring compatibility with communication constraints. Thereafter, each Byzantine worker constructs its column in $\mathbf{B}_{eff}$ by generating its entries in a statistically independent manner using a Bernoulli process with probability $p$. Since identical $p$ is used at all the Byzantine workers, this process is equivalent to generating $\mathbf{B}_{eff} \sim \mbox{Bernoulli}(p)$, where $p$ denotes the probability associated with bit 0. While this threat model incurs less communication overhead over the strongly colluding model, it may not generate $\mathbf{B}_{eff}$ with the structure of Proposition \ref{prop:strog_coll} at each instant, thus justifying its name. 
To execute an attack under this model, the master Byzantine worker would need to choose an appropriate value of $p$ to degrade the probability of localization error.
Although we do not have optimality conditions on the choice of $p$, we propose Problem \ref{prob:prop_weak}, as a practical way to synthesize $p$. The rationale behind proposing Problem \ref{prob:prop_weak}, is presented in Appendix \ref{proof:6}.


\begin{mdframed}
\begin{problem}
\label{prob:prop_weak}
\textcolor{black}{Let $\mathbf{B}_{eff} \sim \mbox{Bernoulli}(p)$, such that $N_{0}$ denotes the number of zero entries in $\mathbf{B}_{eff}$ and $N_{{1}}$ denotes the number of rows in $\mathbf{B}_{eff}$ that have all ones, then solve}
\begin{equation}
\label{eq:pval}
p^{*} = \arg \min_{\ p \in (0, 1)} \frac{1}{Mv}\mathbb{E}[N_{0}] + \frac{1}{M}\mathbb{E}[N_{{1}}].
\end{equation}
\end{problem}
\end{mdframed}

\vspace{0.2cm}
\textcolor{black}{To showcase the impact of both colluding attacks, we conduct experiments while computing the function $f(\mathbf{X})=\mathbf{X}^T\mathbf{X}$, such that $\mathbf{X}\in\mathbb{R}^{20\times5}$, using the $(N,K)=(31,15)$ and $(N,K)=(27,15)$ DFT codes. For the $(31,15)$ setting, $v=8$, whereas for the $(27,15)$ setting, $v=6$. In both settings, we consider $v$ Byzantine worker nodes and use joint localization with varying constraint lengths. Since the error-correction capability $v$ varies across the two settings, the corresponding constraint lengths are adjusted accordingly. Specifically, the $(31,15)$ setting with $v=8$ uses constraint lengths $8,9,10,11$, while the $(27,15)$ setting with $v=6$ uses constraint lengths $6,7,8,9$. Note that we retain the same parameter setting as in Figs.~\ref{Fig: alcc in adv plot}, \ref{Fig: ind localization}, and \ref{Fig: Joint localization} to evaluate the effect of the proposed colluding attacks on joint localization under the same experimental setting.} For the strongly colluding case, the first row of
$\mathbf{B}_{eff}$ has all ones, whereas the other rows have variable
number of ones. For the weakly colluding case, we generate the rows of
$\mathbf{B}_{eff}$ using different probability values. The other
parameters used for both settings are $\tau=0$, $\beta=1.5$, $t=3$, $k=5$, $\sigma=10^6$, and precision noise variance is 0.01. Here, the non-zero entries of the noise matrices
$\{\mathbf{E}_{i_a}\}$ are i.i.d. as $\mathcal{CN}(10,10^3)$. We present these experimental results in Fig.~\ref{fig:colluding}(a) and Fig.~\ref{fig:colluding}(b). The
plots confirm that the Byzantine worker nodes can use the results of
Proposition~\ref{prop:strog_coll} and Problem~\ref{prob:prop_weak} to
effectively design their base matrices when $\sigma^2_p>0$.

\textcolor{black}{Further, under the weakly colluding attack model, we obtain $p^{*}$ in closed form by solving the surrogate optimization problem in \eqref{eq:pval}. For the $(31,15)$ and $(27,15)$ settings, we have $v=\left\lfloor\frac{N-K}{2}\right\rfloor=8$ and $v=6$, respectively, and obtain $p^{*}=0.257$ and $p^{*}=0.3012$, respectively. As illustrated in Fig. \ref{fig:colluding}(a) and Fig. \ref{fig:colluding}(b), the worst-case accuracy for the $(31,15)$ and $(27,15)$ settings, using constraint lengths $8$ and $6$, respectively, occurs around the corresponding probability values. This confirms that the theoretically derived $p^{*}$ closely matches the empirical observations.}
From the viewpoint of the master node, the plots indicate that the
only way to alleviate the impact of the proposed attacks is to
increase the computational complexity, which may not be desirable.
Overall, the plots in Fig.~\ref{fig:colluding} confirm that in
contrast to the approach of injecting noise on every component of
their computations, the Byzantine worker nodes must carefully modify
their base matrices when $\sigma_{p}^2>0$.

\section{\textcolor{black}{Discussion on Worker Redundancy and Baseline Comparison}}
\label{sec:baseline_comparison}

\textcolor{black}{An important design consideration in Byzantine-resilient coded computing is the amount of worker redundancy required to tolerate adversarial workers while preserving exact computation. Since the proposed Robust ALCC framework provides resilience against Byzantine workers through a coding-theoretic decoder, it is important to distinguish the redundancy that is fundamentally required for Byzantine error correction from the redundancy introduced by a particular decoding algorithm. In this section, we first characterize the worker redundancy of Robust ALCC from first principles and then compare it with representative Byzantine-resilient coded-computing frameworks.}

\textcolor{black}{Recall that the worker responses in ALCC naturally form codewords of a $K$-dimensional DFT code of blocklength $N$, where $K=(k+t-1)D+1$ is determined by the polynomial degree, the number of data partitions, and the privacy parameter. Since DFT codes are Maximum Distance Separable (MDS) codes, they attain the Singleton bound with minimum distance $d_{\min}=N-K+1$. Therefore, correcting $A$ arbitrary Byzantine workers using unique decoding requires
\begin{align}
    d_{\min}\geq 2A+1,
\end{align}
which directly yields
\begin{align}
    N\geq K+2A.
\end{align}
Thus, the worker redundancy of Robust ALCC follows directly from the coding-theoretic requirement for exact correction of $A$ arbitrary Byzantine errors. It is not introduced by the proposed DFT-based decoder or by the joint error-localization algorithm. The above requirement corresponds to the conventional setting of unique decoding, where Byzantine errors are corrected from a received codeword without exploiting additional side information or error structure. More sophisticated decoding paradigms, such as Folded Lagrange Coded Computing (FLCC)~\cite{a6}, reduce worker redundancy by combining list decoding with a small amount of side information available at the master node. Nevertheless, irrespective of the decoding strategy, the code dimension $K$ increases with the polynomial degree $D$, implying that the required number of workers must also increase with $D$ to maintain resilience against a fixed number of Byzantine workers. The available side information only affects the rate at which this redundancy increases.}

\textcolor{black}{To place Robust ALCC in the context of existing literature, Table~\ref{tab:redundancy_comparison} compares its worker redundancy with representative Byzantine-resilient coded-computing frameworks. LCC, NSLCC, and Robust ALCC all perform exact polynomial computation and therefore exhibit the same worker-redundancy scaling under conventional unique decoding. FLCC achieves lower redundancy by adopting a different list-decoding framework with side information, whereas RBACC follows a fundamentally different recovery paradigm by allowing a tradeoff between approximation accuracy and Byzantine resilience through a tunable code dimension.}


\begin{table}[t]
\centering
\caption{\textcolor{black}{Comparison of worker redundancy with representative Byzantine-resilient coded computing approaches.}}
\label{tab:redundancy_comparison}
\renewcommand{\arraystretch}{1.15}
\begin{small}
\begin{tabular}{|p{2.3cm}|p{3.0cm}|p{5.0cm}|p{5.4cm}|}
\hline
\textbf{Framework} &
\textbf{Computation} &
\textbf{Byzantine-resilience approach} &
\textbf{Worker redundancy} \\
&
&
 & \\
\hline

LCC~\cite{b1} &
Exact polynomial &
Reed--Solomon unique decoding &
$N \geq K+2A$ \\
\hline

Proposed robust ALCC &
Exact polynomial &
DFT-based unique decoding &
$N \geq K+2A$\\
\hline

NSLCC~\cite{f1} &
Exact polynomial &
DCT-based error correction \cite{NCC} &
$N \geq K+2A$ \\
\hline

RBACC~\cite{rbacc} &
Approximate non-polynomial &
DCT-based error correction &
$K$ can be tuned, trading approximation accuracy for Byzantine resilience \\
\hline

FLCC~\cite{a6} &
Exact polynomial &
Folded Reed--Solomon list decoding with side information &
Lower redundancy than LCC. Approximately one additional worker per Byzantine worker \\
\hline
\end{tabular}
\end{small}
\end{table}

\section{\textcolor{black}{Summary and Future Directions}}
We have addressed an open problem raised by \cite{b2} and have identified how the nature of the computations returned by Byzantine worker nodes in ALCC allows the use of error correction algorithms from DFT codes. We have showed that our framework provides significantly higher accuracy than that of \cite{b2} in the presence of Byzantine worker nodes. Moreover, we have demonstrated that when the variance of the precision noise is negligible and when the number of Byzantine workers is small, our scheme achieves accuracy numbers close to that of the ALCC scheme without Byzantine worker nodes. We also discussed a variant of ALCC comprising both reliable and unreliable worker nodes. For this setting, we proposed a customized strategy for assigning evaluations to the unreliable worker nodes which further reduces the overall relative error of ALCC framework. As the final contribution of this work, we studied the vulnerabilities of \textcolor{black}{Robust}  ALCC against colluding attacks from the Byzantine worker nodes. Subsequently, we proposed two colluding attacks, which suggest a new sparsity structure on the noise matrices in contrast to the presumption that noise matrices should be fully non-zero.

\textcolor{black}{While the proposed joint-localization method improves localization accuracy by exploiting the common-support structure of Byzantine workers across multiple received codewords, its analysis assumes that the number of errors on each codeword is known. In Section~\ref{subsec:localization_estimated_errors}, we empirically showed that the proposed method continues to outperform independent localization even when the number of errors is estimated in the presence of precision noise. However, as the number of Byzantine workers approaches the error-correction capability, especially under larger precision-noise variances, inaccurate error-count estimation degrades reconstruction accuracy. Therefore, an important direction for future work is to jointly estimate the number of Byzantine errors and localize the adversarial workers, thereby improving the robustness of ALCC under finite-precision arithmetic.}

\pagenumbering{gobble}

\begin{appendix}
\subsection{Proof of Proposition \ref{prop:1}}
\label{proof:1}

For each pair $\bar{u} \in [u], \bar{h} \in [h]$, let $\mathbf{r}_{\bar{u}\bar{h}} = [\mathbf{R}_{1}(\bar{u}, \bar{h}) ~\mathbf{R}_{2}(\bar{u}, \bar{h}) ~\ldots~ \mathbf{R}_{N}(\bar{u}, \bar{h})]$ corresponds to a vector carved from the $(\bar{u}, \bar{h})$-th entry of each $\mathbf{R}_{i}$. In total, the master node receives $M=\bar{u}\times \bar{h}$ such vectors from the worker nodes as depicted in the following matrix $\mathbf{R}_{eff}\in \mathbb{R}^{M\times N}$, where each row captures a received vector $\mathbf{r}_{\bar{u}\bar{h}}$ of length $N$.

 \begin{small}
\begin{equation*}
\mathbf{R}_{eff} =
\begin{bmatrix}
\mathbf{R}_{1}(1,1) & \mathbf{R}_{{2}}(1, 1) & \ldots & \mathbf{R}_{{N}}(1, 1)\\
\mathbf{R}_{{1}}(1,2) & \mathbf{R}_{{2}}(1, 2) & \ldots & \mathbf{R}_{{N}}(1, 2)\\
\vdots & \vdots & \vdots & \vdots\\
\mathbf{R}_{{1}}(1,h) & \mathbf{R}_{{2}}(1, h) & \ldots & \mathbf{R}_{{N}}(1, h)\\
\mathbf{R}_{{1}}(2,1) & \mathbf{R}_{{2}}(2, 1) & \ldots & \mathbf{R}_{{N}}(2, 1)\\
\vdots & \vdots & \vdots & \vdots\\
\mathbf{R}_{{1}}(u,h) & \mathbf{R}_{{2}}(u, h) & \ldots & \mathbf{R}_{{N}}(u, h)\\
\end{bmatrix}.	
\end{equation*}
\end{small}

\noindent Note that, when $A = 0$, the set of vectors $\{\mathbf{r}_{\bar{u}\bar{h}}\}$ represents codewords from a $K$-dimensional Discrete Fourier Transform (DFT) code with blocklength $N$. To justify this claim, recall the encoding stage of ALCC, where the encoded matrix polynomial $l(z)\in \mathbb{R}^{m\times n}[z]$ is evaluated at the $N$-th roots of unity to distribute the set of evaluations $\{u(l_{i})\in \mathbb{R}^{m\times n}~|~ {i\in[N]\}}$ among $N$ worker nodes. Subsequently, each worker node computes the target function $f:\mathbb{R}^{m\times n}$ $\rightarrow$ $\mathbb{R}^{u\times h}$, and returns the result $f(u(l_{i}))\in \mathbb{R}^{u\times h}$ to the master node. Since the target function $f(\cdot)$ is a polynomial of degree $D$, applying it on each share $u(l_{i})\in \mathbb{R}^{m\times n}$, results in a polynomial evaluation $f(u(l_{i}))\in \mathbb{R}^{u\times h}$ of the matrix polynomial $f(l(z))\in \mathbb{R}^{u\times h}$, where the effective degree of $f(l(z))$ is $\tilde D=(k+t-1)D$. Let $\mathbf{m}_{\bar{u}\bar{h}}$ represent a row vector of length $K=\tilde D+1$ containing the coefficients of the scalar polynomial corresponding to the $(\bar{u},\bar{h})$-th entry of the matrix polynomial $f(l(z))$. Then, the set of vectors $\mathbf{r}_{\bar{u}\bar{h}}$ received during the reconstruction stage of ALCC can be represented as $\mathbf{r}_{\bar{u}\bar{h}} = \mathbf{m}_{\bar{u}\bar{h}}\mathbf{G}$, for each $\bar{u} \in [u], \bar{h} \in [h]$, where the matrix $\mathbf{G}$ is represented as

\begin{small}
\begin{equation}
\label{eq:Gmatrix}
\mathbf{G}=\begin{bmatrix}
1 & 1 & 1 & \cdots & 1 \\
1 & \omega_N^1 & \omega_N^2 & \cdots & \omega_N^{(N-1)} \\
1 & \omega_N^{2} & \omega_N^{4} & \cdots & \omega_N^{2(N-1)2} \\
\vdots & \vdots & \vdots & \ddots & \vdots \\
1 & \omega_N^{(K-1)} & \omega_N^{2(K-1)} & \cdots & \omega_N^{(K-1)(K-1)}
\end{bmatrix}_{K\times N},
\end{equation}
\end{small}

\noindent and $\omega_N = e^{-2\pi i/N}$ is the $N$-th root of unity. Note that, the above matrix $\mathbf{G}$ is identical to the matrix generated by taking the first $K$ rows from the DFT matrix $\mathbf{W}_{N}$ of order $N$ whose $(m,n)$-th element is represented as $\mathbf{W}_{N}(m,n)=\frac{1}{\sqrt{N}}\omega_N^{(m-1)(n-1)}$, for $1\leq m\leq N$ and $1\leq n\leq N$. This results from the fact that the generator matrix of an $(N,K)$ DFT code consists of any $K$ rows from the DFT matrix of order $N$ \cite{b3}. Therefore, in the absence of Byzantine worker nodes, the set of vectors, $\{\mathbf{r}_{\bar{u}\bar{h}}~|~ {\bar{u} \in [u], \bar{h} \in [h]}\}$ represents the codewords of an $(N,K)$ DFT code.
Furthermore, in the presence of $A$ Byzantine worker nodes, where $0<A<N$, the vector $\mathbf{r}_{\bar{u}\bar{h}}$ at the decoder of ALCC represents a noisy codeword of $(N,K)$ DFT code. Since, $(N,K)$ DFT codes have error-correction capability of at most $\lfloor \frac{N-K}{2}\rfloor$ errors, their corresponding error-correction algorithms can be applied on each noisy codeword $\mathbf{r}_{\bar{u}\bar{h}}$ as long as $A\leq \lfloor \frac{N-K}{2}\rfloor$. Finally, since the error vectors are real-valued, infinite precision on floating-point representation is necessary to perfectly localize and cancel these errors. This completes the proof.
 

\subsection{Proof of Theorem \ref{thm:corr}} 
\label{proof:2}

Recall that $\bar{g}(z)$ denotes error-locator polynomial of degree $A$, given by $\bar{g}(z)=g(z)+e(z)$. In particular, we write $g(z)=g_0+g_1z + \ldots + g_{A}z^{A}$ and $e(z)=e_0+e_1z+\ldots+e_{A}z^{A}$, where $\{g_0, g_1, \ldots, g_{A}$\} are the fixed coefficients of the error-locator polynomial, and $\{e_0, e_1,\dots, e_{A}$\} are the coefficients of $e(z)$, each distributed as $\mathcal{CN}(0, \sigma^2_p)$.

In the context of the theorem, let $X_{i_{a}}$ be a root of $g(z)$ and $X_{j_{b}}$ be another $N$-th root of unity such that $X_{i_{a}}\neq X_{i_{b}}$ and $g_{\bar{u}\bar{h}}(z)|_{z=X_{i_{b}}} \neq 0$. During the process of recovering the location of the errors, $X_{j_{b}}$ is incorrectly decoded as the root of $\bar{g}(z)$ if $||\bar{g}(X_{j_{b}})||^2 < ||\bar{g}(X_{i_{a}})||^2$. Therefore, to characterize this error event, we define $||\bar{g}(X_{i_{a}})||^2$ and $||\bar{g}(X_{j_{b}})||^2$ as two random variables and write 

\begin{equation}
\label{eq:pe interms e and e bar}
 PEP_{j_{b},i_{a}} =\text{Prob}(||c+\bar{e}||^2\leq ||e||^2),
 \end{equation}
where $c= g(z)|_{z=X_{j_{b}}}$,\hspace{0.2 mm} $\bar{e}=e(z)|_{z=X_{j_{b}}}$ and $e=e(z)|_{z=X_{i_{a}}}$. Expressing $X_{j_{b}}=e^{\iota \theta_{b}}$ such that $\theta_{b}=\frac{2 \pi j_{b}}{N}$, and $X_{i_{a}}=e^{\iota\theta_{a}}$ such that $\theta_{a}=\frac{2 \pi i_{a}}{N}$, the variables $\Bar{e}$ and $e$ in \eqref{eq:pe interms e and e bar} can be expanded as $\bar{e}=e_0+E_{\theta_{b}}$, $e=e_0+E_{\theta_{a}}$, 
where
\begin{equation}
\label{eq:error_theta}
E_{\theta_{b}} =e_1(e^{\iota\theta_{b}})+ e_2(e^{\iota 2\theta_{b}})+ \ldots +e_A(e^{\iota A\theta_{b}}),
\end{equation}
and 
\begin{IEEEeqnarray}{rcl}
\label{eq:error_theta_act}
E_{\theta_{a}} = e_1(e^{\iota\theta_{a}})+e_2(e^{\iota 2\theta_{a}})+\ldots +e_A(e^{\iota A\theta_{a}}).
\end{IEEEeqnarray}

\noindent After algebraic manipulation, we can rewrite \eqref{eq:pe interms e and e bar} as,
\begin{equation}
\label{eq:simpl_pairwise}
 PEP_{j_{b},i_{a}} =\text{Prob}(||c+e_0+E_{\theta_{b}}||^2\leq||e_0+E_{\theta_{a}}||^2).
\end{equation}

Although \eqref{eq:simpl_pairwise} has $A + 1$ complex random variables, we further reduce it to $A$ complex variables by marginalizing \eqref{eq:simpl_pairwise} over the random variable $e_{0}$. Continuing from \eqref{eq:simpl_pairwise}, we write all the complex variables in terms of their in-phase and quadrature components as $c = c_I +\iota c_Q$, $E_{\theta_{b}} = y_{I}+\iota y_{Q}$, $E_{\theta_{a}} =z_{I}+\iota z_{Q}$, and 
$e_0 = e_I + \iota e_Q$. Further, the condition $||c+e_0+E_{\theta_{b}}||^2\leq||e_0+E_{\theta_{a}}||^2$ can be rewritten as,
\[(c_I+y_{I}+e_I)^2+(c_Q+y_{Q}+e_Q)^2\leq(e_I+z_{I})^2+(e_Q+z_{Q})^2,\]
which can further be  simplified into
\begin{multline*}
2e_I(c_I+y_{I}-z_{I})+2e_Q(c_Q+y_{Q}-z_{Q})\leq
z_{I}^2+z_{Q}^2-(c_I+y_{I})^2-(c_Q+y_{Q})^2.
\end{multline*}
Fixing the random variables, $y_{I}, y_{Q}, z_{I}, z_{Q}$ to specific realizations, the scalar $2e_I(c_I+y_{I}-z_{I})+2e_Q(c_Q+y_{Q}-z_{Q})$ is a Gaussian random variable with zero mean and variance $2\sigma_p^2[(c_I+y_{I}-z_{I})^2+(c_Q+y_{Q}-z_{Q})^2]$. Using $r$ to denote the above random variable, the pairwise probability of error expression conditioned on $y_{I}, y_{Q}, z_{I}, z_{Q}$ can be written as,
\[\text{Prob}(r\leq z_{I}^2+z_{Q}^2-(c_I+y_{I})^2-(c_Q+y_{Q})^2),\] which can also be lower bounded as 
\[\text{Prob}(r\leq -(c_I+y_{I})^2-(c_Q+y_{Q})^2).\]
The above lower bound is valid since $z_{I}^2+z_{Q}^2$ is always non-negative. Further, since $r$ is zero mean, we rewrite the above as
\[\text{Prob}(r\geq (c_I+y_{I})^2+(c_Q+y_{Q})^2).\]
Using the standard definition of Q-function, we write the above expression as, 
\begin{equation*}
Q\left( \frac{(c_I+y_{I})^2+(c_Q+y_{Q})^2}{\sqrt{2\sigma_p^2\left((c_I+y_{I}-z_{I})^2+(c_Q+y_{Q}-z_{Q})^2\right)}}\right),
\end{equation*}
by using the variance of the random variable $r$. Subsequently, using the Chernoff bound approximation, the above expression can be written as, 
\begin{equation}
\label{eq:one_exp_before_bound}
e^{-\frac{((c_I+y_{I})^2+(c_Q+y_{Q})^2)^2}{4\sigma_p^2\left((c_I+y_{I}-z_{I})^2+(c_Q+y_{Q}-z_{Q})^2\right)}}.
\end{equation}
Finally, when taking expectation over the underlying random variables $e_{1}, e_{2}, \ldots, e_{A}$, we get a lower bound on \eqref{eq:one_exp_before_bound} as
\begin{equation}\label{eq:correlated lower bound} 
    \mathbb{E}_\{{e_1,e_2,\ldots e_A}\}\left\{ e^{-\frac{((c_I+y_{I})^2+(c_Q+y_{Q})^2)^2}{4\sigma_p^2\left((c_I+y_{I}-z_{I})^2+(c_Q+y_{Q}-z_{Q})^2\right)}}\right\}.
 \end{equation}
In the rest of the proof, we provide a lower bound on \eqref{eq:correlated lower bound}.

In \eqref{eq:one_exp_before_bound}, note that $y_{I}$ and $y_{Q}$ are independent Gaussian random variables with zero mean and variance $\frac{A\sigma^{2}_{p}}{2}$. This follows from \eqref{eq:error_theta}. Similarly, $y_{I} - z_{I}$ and $y_{Q} - z_{Q}$ are also independent Gaussian random variables with mean zero and variance 
\begin{equation}\label{variance_diff}
\sigma_{{diff}}^2= \sigma_p^2(A-(\mbox{cos}(\theta_d) + \mbox{cos}(2\theta_d)+\ldots +\mbox{cos}(A\theta_d))),
\end{equation}
such that $\theta_{d} = \theta_{a} - \theta_{b}$. When $\sigma^{2}_{p}$ is much smaller compared to the constants $c_{I}$ and $c_{Q}$, then we observe that we can choose a positive constant $\eta$ such that 
\begin{equation*}
\frac{\eta \left((c_I+y_{I}-z_{I})^2+(c_Q+y_{Q}-z_{Q})^2\right)}{\left((c_I+y_{I})^2+(c_Q+y_{Q})^2\right)} > 1,
\end{equation*}
with a probability close to one. Therefore, we can get a lower bound on $PEP_{j_{b},i_{a}}$ as,
\begin{IEEEeqnarray}{rcl}
\label{eq:lower bound final}
\geq \mathbb{E}_\{{e_1,e_2,\ldots e_A}\} \left\{e^{-\frac{\eta \left((c_I+y_{I}-z_{I})^2+(c_Q+y_{Q}-z_{Q})^2\right)}{4\sigma_p^2}}\right\}.
\end{IEEEeqnarray}
Given that $y_{I} - z_{I}$ and $y_{Q} - z_{Q}$ are statistically independent, \eqref{eq:lower bound final} can be rewritten as
\begin{IEEEeqnarray}{rcl}
\label{eqn:total prob correlated} 
PEP_{j_{b},i_{a}} \geq \mathbb{E}_{(y_{I}-z_{I})}\Bigg\{e^{-\frac{\eta(c_I+y_{I}-z_{I})^2}{4\sigma_p^2}}\Bigg\}  \times \mathbb{E}_{(y_{Q}-z_{Q})}\Bigg\{e^{-\frac{\eta(c_Q+y_{Q}-z_{Q})^2}{4\sigma_p^2}}\Bigg\}.
\end{IEEEeqnarray}
For each of the above two terms, we note that closed-form expressions can be computed to respectively obtain 
\begin{equation}\label{eqn:correlated1,3}   
\mathbb{E}_{(y_{I}-z_{I})}\{e^{-\frac{(c_I+y_{I}-z_{I})^2}{4\sigma_p}}\}=\frac{\sqrt{2} \sigma_p}{\sqrt{\sigma_{{diff}}^2}\sqrt{\eta +\gamma}}e^{-\frac{\eta c_I^2\gamma}{4\sigma_p^2(\eta+\gamma)}},
\end{equation}
and 
\begin{equation}\label{eqn:correlated2,4}   
\mathbb{E}_{(y_{Q}-z_{Q})}\{e^{-\frac{(c_Q+y_{I}-z_{I})^2}{4\sigma_p^2}}\}=\frac{\sqrt{2} \sigma_p}{\sqrt{\sigma_{{diff}}^2}\sqrt{\eta +\gamma}}e^{-\frac{\eta c_Q^2\gamma}{4\sigma_p^2(\eta+\gamma)}},
\end{equation}
where $\sigma_{diff}^2$ is as given in \eqref{variance_diff} and $\gamma = \frac{2\sigma^2_{p}}{\sigma_{{diff}}^2} $. Finally, substituting the above expressions in \eqref{eqn:total prob correlated}, we get 
\begin{equation} \label{eq:total correlated prob}
 PEP_{j_{b},i_{a}}\geq PL_{j_{b},i_{a}} \triangleq \frac{\kappa}{(1+\kappa)}e^{-\frac{\eta (c_I^2 +c_Q^2)\kappa}{4\sigma^2_p(1+\kappa)}},
\end{equation}
where $\eta, c_I$, $c_{Q}$ are constants which depends on $\mathbf{e}$ and the indices $j_{b}, i_{a}$, and $\kappa =\frac{2}{\eta \sum_{l=1}^{A} 1 - \mbox{cos}(l\theta_{d})}$, where $\theta_{d}=\theta_{a}-\theta_{b}$. This completes the proof.

\subsection{Proof of Proposition \ref{prop:all_one_matrix}}
\label{proof:3}
 

\textcolor{black}{We first consider the case $\sigma_p^2=0$. Since
$\mathbf{B}_{i_a}=\mathbf{J}_{u\times h}$ for every $a\in[v]$, all
$M=uh$ error-locator polynomials recovered at the master node are
identical and have degree $v=\lfloor \frac{N-K}{2}\rfloor$. Hence, in
the absence of precision noise, solving the roots of any one of these
polynomials is sufficient for error localization. Now consider the case when $\sigma_p^2>0$. Under the assumption
$\mathbf{B}_{i_a}=\mathbf{J}_{u\times h}$ for every $a\in[v]$, all
$M=uh$ error-locator polynomials correspond to the same set of $v$
error locations introduced by the $v$ Byzantine workers and have degree
$v$. Hence, the $M$ recovered error-locator polynomials are noisy versions of the same degree-$v$ error-locator polynomial. Since the coefficient perturbations of these polynomials are statistically independent and have variance
$\sigma_p^2$, the coefficient-wise averaging used to obtain
$\bar{g}_{\max}(z)$ results in an effective variance
of $\frac{\sigma_p^2}{M}$. Note that the lower bound on the pairwise localization error probability given in Theorem \ref{thm:corr} is non-decreasing with
respect to $\sigma_{p}^2$, as established in
Corollary \ref{lemma:sigma_beh}.
Since $\frac{\sigma_p^2}{M}<\sigma_p^2$ for $M>1$, the lower bound obtained
using $\bar{g}_{\max}(z)$ is no larger than that obtained using any one of $M$ error locator polynomials under independent localization. As a result, the proposed joint localization method offers smaller values of $P_{\mathrm{SLE}}$ in \eqref{eq:surrogate PL} compared to that obtained when independent localization is performed at the DFT decoder. This completes the proof.}


\subsection{Proof of Lemma \ref{lemma:inc_A}}
\label{proof:4}

In the expression on the lower bound in Theorem \ref{thm:corr} (which is the expression for $PL_{(i_{a}^{*},i_{a}^{*}+1,c)}$), we note that,
$$\frac{\kappa}{1 + \kappa} = \frac{2}{2 + \eta\sum_{l = 1}^{A_{c}} (1- \mbox{cos}(l\theta_{d}))},$$ where $\theta_{d}=\frac{2\pi}{N}(|i_{a}^*-(i_{a}^*+1)|)=\frac{2\pi}{N}$.
Since $(1-\mbox{cos}(l\theta_{d}))$ is a non-negative number for any $l \in [A_{c}]$ and $N$, the term $\sum_{l = 1}^{A_{c}} (1- \mbox{cos}(l\frac{2\pi}{N}))$ is a non-decreasing function of $A_{c}$. Therefore, $\frac{\kappa}{1 + \kappa}$ is a non-increasing function of $A_{c}$. In addition,  since $\frac{\kappa}{1 + \kappa}$ is in the exponent of $e^{-\frac{\eta (c_I^2 +c_Q^2)\kappa}{4\sigma^2_p(1+\kappa)}}$, it is evident that the term $e^{-{\frac{\eta(c_I^2 +c_Q^2)\kappa}{4\sigma^2_p(1+\kappa)}}}$ is a non-decreasing function of $A_{c}$. Overall, the lower bound is the product of the following three terms $A_{c}$, $\frac{\kappa}{1 + \kappa}$, $e^{-\{\frac{d\kappa}{\sigma^2_p(1+\kappa)}\}}$, for some constant $d$, which are non-decreasing, non-increasing, and non-decreasing functions of $A_{c}$, respectively. To study the behavior of this function with respect to $A_{c}$,  we apply $\mbox{log}_{e}$ on it to obtain $\mbox{log}_{e}(A_{c}) + \mbox{log}_{e}\frac{\kappa}{1 + \kappa} - \frac{d}{\sigma^{2}_{p}}\frac{\kappa}{1 + \kappa}$. Since $\mbox{log}_{e}\frac{\kappa}{1 + \kappa}$ is sub-linear, the term $\frac{d}{\sigma^{2}_{p}}\frac{\kappa}{1 + \kappa} + \mbox{log}_{e}(A_{c})$ dominates the lower bound when $\sigma^{2}_{p}$ is small, and therefore, the lower bound increases as $A_{c}$ increases from $1$ to $v$. This completes the proof.

\subsection{Proof of Proposition \ref{prop:strog_coll}}
\label{proof:5}

From Proposition \ref{prop:all_one_matrix}, $\mathbf{B^{\dagger}}_{eff}$ should not be the all-one matrix. Also, from Lemma \ref{lemma:inc_A}, each row of $\mathbf{B^{\dagger}}_{eff}$ must contain as many ones as possible. This implies that the number of ones in every row must be either $v$ or $v-1$. Let us suppose that out of the $M$ rows, $\Omega$ rows of $\mathbf{B^{\dagger}}_{eff}$ have all ones and the rest of the $M - \Omega$ rows have $v - 1$ ones at some positions. Under this situation, we prove that the optimal choice of $\Omega$ is one in order to maximize the objective function of Problem \ref{opt:B_{eff}}. With $\Omega$ rows of $\mathbf{B^{\dagger}}_{eff}$ containing all-ones, the master node can average them to obtain one error-locator polynomial, and the rest of the polynomials are in-tact.  We arrange the  polynomials such that the first polynomial captures the averaged degree $v$ polynomial, while the remaining are degree $v-1$ polynomials. With a total of $1 + M - \Omega$ polynomials, the master node proceeds to perform independent localization. Error event in the localization step occurs when either one of the polynomials experience errors. Therefore, with $M-\Omega+1$ total polynomials, the right-hand side of the expression in \eqref{eq: opt B_eff_modify} denoted by $PL$, can be expressed as

\begin{small}
\begin{IEEEeqnarray}{rcl}
\label{eq:total error prob strongly collude}
PL=\frac{1}{M - \Omega + 1}\left(v \times PL_{(i_{a}^{*},i_{a}^{*}+1,1)}\right)+ \frac{1}{M-\Omega + 1}  \left(\sum _{n=2}^{M-\Omega+1} (v-1) \times PL_{(i_{a}^{*},i_{a}^{*}+1,n)}\right),
\end{IEEEeqnarray}\end{small}

where  $PL_{(i_{a}^{*},i_{a}^{*}+1,1)}$ and  $PL_{(i_{a}^{*}, i_{a}^{*}+1,n)}$ denote the lower bound on pairwise error probability of averaged degree $v$ polynomial, and $n$-th degree $v-1$ polynomials, for $n\in[2:M-\Omega+1]$. Subsequently, using the derived expressions on lower bound in Theorem \ref{thm:corr}, the expression in \eqref{eq:total error prob strongly collude}(which is applicable for $\Omega \geq 1$), can be expressed as

\begin{small}
 \begin{multline}\label{eq:v-1 pe}
PL = \frac{1}{M - \Omega + 1}\times v \times\bigg( \frac{\kappa_{1}}{(1+\kappa_{1})}e^{-\frac{\Omega\eta(c_{I_{v}}^2 +c_{Q_{v}}^2)\kappa_{1}}{4\sigma^2_p(1+\kappa_{1})}}\bigg)+
\left(\frac{1}{M - \Omega + 1}\right)\times (v-1)\times \bigg( \sum_{n=1}^{M-\Omega}\frac{\kappa_{2}}{(1+\kappa_{2})}e^{-\frac{\eta(c_{I_{n}}^2 +c_{Q_{n}}^2)\kappa_{2}}{4\sigma^2_p(1+\kappa_{2})}}\bigg),
\end{multline}\end{small}

where $\kappa_{1} =\frac{2}{\eta\sum_{l = 1}^{v} 1 - \mbox{cos}(l\theta_{d}))}$ 
and  $\kappa_{2} =\frac{2}{\eta \sum_{l = 1}^{v-1} 1 - \mbox{cos}(l\theta_{d})}$, such that $\theta_{d} = \frac{2\pi}{N}$, and $c_{I_{v}}^2 +c_{Q_{v}}^2$, $c_{I_{n}}^2 +c_{Q_{n}}^2$ are the constants depend on $i_{a}^*$ for averaged degree $v$ polynomial and $n$-th degree $v-1$ polynomials. Notice the presence of $\Omega$ in the exponent of the first term of \eqref{eq:v-1 pe}. To prove that $\Omega = 1$ is the optimal choice, we need to show that
\begin{equation}
\label{eq:first_inequality}
PL|_{\Omega = 1} >PL|_{\Omega > 1},
\end{equation}
and
\begin{equation}
\label{eq:second_inequality}
PL|_{\Omega = 1} > PL|_{\Omega = 0}.
\end{equation}
Towards proving \eqref{eq:first_inequality} and \eqref{eq:second_inequality}, we replace all the pairwise error terms in the summation of second term in \eqref{eq:v-1 pe} by the maximum of of all $M-\Omega$ pairwise error terms i.e., $\max_{1 \leq n \leq M-\Omega}\bigg(\frac{\kappa_{2}}{(1+\kappa_{2})}e^{-\frac{\eta(c_{I_{n}}^2 +c_{Q_{n}}^2)\kappa_{2}}{4\sigma^2_p(1+\kappa_{2})}}\bigg)$. Therefore, instead of finding the optimal $\mathbf{B^{\dagger}}_{eff}$ which maximizes \eqref{eq:v-1 pe}, we find $\mathbf{B}_{eff}$ which aim to maximize the following expression which is an approximation of \eqref{eq:v-1 pe} denoted as $PL_{{approx}}$ given by

\vspace{-0.4cm}

\begin{small}
 \begin{multline}\label{eq:v-1 pe_approx}
PL_{{approx}} =\frac{1}{M - \Omega + 1}\times v \times\bigg( \frac{\kappa_{1}}{(1+\kappa_{1})}e^{-\frac{\Omega\eta(c_{I_{v}}^2 +c_{Q_{v}}^2)\kappa_{1}}{4\sigma^2_p(1+\kappa_{1})}}\bigg)+ \frac{M-\Omega}{M-\Omega +1}\times (v-1)\times \bigg(\frac{\kappa_{2}}{(1+\kappa_{2})}e^{-\frac{\eta(c_{I_{v-1}}^2 +c_{Q_{v-1}}^2)\kappa_{2}}{4\sigma^2_p(1+\kappa_{2})}}\bigg),
\end{multline}\end{small}

\vspace{-0.2cm}
where the second term denotes the maximum pairwise error probability term among $M-\Omega$ pairwise error terms of degree $v-1$ polynomials and $c_{I_{v-1}}^2 +c_{Q_{v-1}}^2$ in the exponent is a constant depends on the index $i_{a}^*$. Therefore, due to the approximation in \eqref{eq:v-1 pe_approx}, instead of proving \eqref{eq:first_inequality} and \eqref{eq:second_inequality}, we now prove for 
\begin{equation}
\label{eq:first_inequality_mod}
PL_{approx}|_{\Omega = 1} >PL_{approx}|_{\Omega > 1},
\end{equation}
\begin{equation}
\label{eq:second_inequality_mod}
PL_{approx}|_{\Omega = 1} > PL_{approx}|_{\Omega = 0}.
\end{equation}
In this context, when $\Omega = 0$, the remaining polynomials are of degree $v-1$. Therefore, we have
\begin{IEEEeqnarray}{rcl}
PL_{approx}|_{\Omega = 0}=(v-1)\times \bigg( \frac{\kappa_{2}}{(1+\kappa_{2})}e^{-\frac{\eta(c_{I_{v-1}}^2 +c_{Q_{v-1}}^2)\kappa_{2}}{4\sigma^2_p(1+\kappa_{2})}}\bigg).
\end{IEEEeqnarray}
On the other hand, when $\Omega = 1$, we have
\begin{small}
    \begin{multline}
\label{eq: strong coll.}
PL_{approx}|_{\Omega = 1} = \frac{1}{M}\times v \times\left(\frac{\kappa_{1}}{(1+\kappa_{1})}e^{-\frac{\eta(c_{I_v}^2 +c_{Q_v}^2)\kappa_{1}}{4\sigma^2_p(1+\kappa_{1})}}\right)+
\frac{M-1}{M}\times (v-1)\times\bigg(\frac{\kappa_{2}}{(1+\kappa_{2})}e^{-\frac{\eta(c_{I_{v-1}}^2 +c_{Q_{v-1}}^2)\kappa_{2}}{4\sigma^2_p(1+\kappa_{2})}}\bigg).
\end{multline}\end{small}
The above expression can be written as
\begin{IEEEeqnarray}{rcl}
PL_{approx}|_{\Omega = 1} = \frac{1}{M}\times v \times\left(\frac{\kappa_{1}}{(1+\kappa_{1})}e^{-\frac{\eta(c_{I_v}^2 +c_{Q_v}^2)\kappa_{1}}{4\sigma^2_p(1+\kappa_{1})}}\right)+
PL_{approx}|_{\Omega = 0}.
\end{IEEEeqnarray}
Finally, in the above expression, since the coefficient of $\frac{1}{M}$ is more than that of $\frac{M-1}{M}$ when $\sigma^{2}_{p}$ is small (due to Lemma 1), we can conclude $$PL_{approx}|_{\Omega = 1} > PL_{approx}|_{\Omega = 0}.$$
Towards proving \eqref{eq:first_inequality_mod}, we need to prove that $PL_{approx}$ in \eqref{eq:v-1 pe_approx} decreases as $\Omega$ increases in the range $[M]$. We immediately notice that the second term of \eqref{eq:v-1 pe} is a decreasing function of $\Omega$ for a given $N, v$ and $\sigma^{2}_{p}$. However, the first term of \eqref{eq:v-1 pe} is such that $\frac{1}{M - \Omega + 1}$ increases with $\Omega$, whereas the exponent term decreases with $\Omega$. Along the similar lines of the proof of Lemma 1, when $\sigma^{2}_{p}$ is small, the exponent term dominates the inverse linear term, and therefore, the first term of \eqref{eq:v-1 pe_approx} decreases with $\Omega$.  This completes the proof for \eqref{eq:first_inequality_mod}. Further, note that, because of using the approximation of original expression \eqref{eq:v-1 pe} in \eqref{eq:v-1 pe_approx}, the above strategy does not  yield the optimal $\mathbf{B^{\dagger}}_{eff}$, rather provides a near optimal choice, denoted as $\mathbf{B}_{eff}$. 
This completes the proof that only one row of $\mathbf{B}_{eff}$ must contain the all-one row, whereas the other rows must be filled with $v-1$ ones. Finally, in order to keep the choice of $\mathbf{B}_{eff}$ non-deterministic to the master node, the positions of ones on those rows that have $v-1$ must be chosen at random with uniform distribution. Thus, without loss of generality, in the statement of Proposition \ref{prop:strog_coll}, we recommend using the all-one row for the first row, and the rest of the rows to be filled with $v-1$ rows. This completes the proof.

\subsection{Rationale behind Problem \ref{prob:prop_weak}}
\label{proof:6}
Due to Lemma \ref{lemma:inc_A}, when $\sigma^{2}_{p}$ is small, the lower bound on error rate of localization of an error locator polynomial is an increasing function of $A_{c}$, where $A_{c}$ denotes the degree of the error locator polynomial, which also captures the number of non-zero entries of a row of $\mathbf{B}_{eff}$. As a consequence, $\mathbf{B}_{eff}$ must be such that the average number of zeros (captured by $\mathbb{E}[N_{0}] = Mvp$) in it should be minimized. Also due to Proposition \ref{prop:all_one_matrix}, given that the master node can perform averaging of all the degree $v$ polynomials, the average number of rows that have all ones in it, denoted by $\mathbb{E}[N_{\textbf{1}}] = M(1-p)^{v}$, should be minimized. Thus, combining the two objectives, we propose $p^{*}$ in \eqref{eq:pval}, which can be obtained in closed form as

\begin{equation}
p^{*}=1-\sqrt[(v-1)]{\frac{1}{v}}.
\label{eq:p*}
\end{equation}
Therefore, $p^{*}$ can be obtained for a given choice of $v= \lfloor \frac{N-K}{2}\rfloor$, using \eqref{eq:p*}. Finally, this value of $p^{*}$ can be used to generate $\mathbf{B}_{eff}$ at the Byzantine worker nodes. This completes the proof.

\end{appendix}

\end{document}